\documentclass[aps, 10pt, notitlepage, twocolumn, superscriptaddress,
nofootinbib,longbibliography]{revtex4-2}

\usepackage{dcolumn}% Align table columns on decimal point
\usepackage{bm}% bold math
\usepackage{ifpdf}
\usepackage{hyperref}
\usepackage{bm}
\usepackage{xcolor,color,graphicx,graphics}
\usepackage[spanish,english]{babel}%, portuguese
\usepackage[latin1]{inputenc}
\usepackage[OT1]{fontenc}
\usepackage{latexsym,amssymb,amsmath,amsfonts}
\usepackage{makeidx}
\usepackage{epsfig,subfigure}
\usepackage{natbib}
\usepackage{epstopdf}
\usepackage{mathrsfs}
\hypersetup{colorlinks=true, linkcolor=blue, citecolor=green}
\usepackage{enumerate}
\usepackage{xcolor}
 \usepackage{multirow}
\definecolor{red}{rgb}{1,0,0}

\def\+{^\dagger}

\def\<{\leftarrow}
\def\>{\rightarrow}

\def\({\left(}
\def\){\right)}

\def\arcsinh{\mathop{\rm arcsinh}\nolimits}

\newcommand{\bi}{\begin{itemize}} 				\newcommand{\ei}{\end{itemize}}
\newcommand{\benu}{\begin{enumerate}} 		\newcommand{\enu}{\end{enumerate}}
\newcommand{\bd}{\begin{dinglist}{0}}     \newcommand{\ed}{\end{dinglist}}
\newcommand{\bfig}{\begin{figure}[htbp]}  \newcommand{\efig}{\end{figure}}
        			
\newcommand{\bc}{\begin{center}} 				  \newcommand{\ec}{\end{center}}
\newcommand{\be}{\begin{equation}} 				\newcommand{\ee}{\end{equation}}
\newcommand{\bsub}{\begin{subequations}}  \newcommand{\esub}{\end{subequations}}
\newcommand{\ben}{\begin{eqnarray}} 			\newcommand{\een}{\end{eqnarray}}
\newcommand{\ba}[1]{\begin{array}{#1}} 		\newcommand{\ea}{\end{array}}
\newcommand{\bea}{\begin{equation}\begin{array}{rcl}}
\newcommand{\eea}{\end{array}\end{equation}}

\begin{document}
\title{Optical appearance of black holes surrounded by a dark matter halo}

\author{Caio F. B. Macedo} \email{caiomacedo@ufpa.br}
\affiliation{Faculdade de F\'isica, Campus Salin\'opolis, Universidade Federal do Par\'a, 68721-000, Salin\'opolis, Par\'a, Brazil}
\author{Jo\~{a}o Lu\'{i}s Rosa} \email{joaoluis92@gmail.com}
\affiliation{University of Gda\'{n}sk, Jana Ba\.{z}y\'{n}skiego 8, 80-309 Gda\'{n}sk, Poland}
\affiliation{Institute of Physics, University of Tartu, W. Ostwaldi 1, 50411 Tartu, Estonia}
\author{Diego Rubiera-Garcia} \email{drubiera@ucm.es}
\affiliation{Departamento de F\'isica Te\'orica and IPARCOS,
	Universidad Complutense de Madrid, E-28040 Madrid, Spain}

\date{\today}
\begin{abstract}
Black holes in General Relativity are described by space-time metrics that are simpler in comparison to non-vacuum compact objects. However, given the universality of the gravitational pull, it is expected that dark matter accumulates around astrophysical black holes, which can have an impact in the overall gravitational field, especially at galactic centers, and induce non-negligible effects in their observational imprints. In this work we study the optical appearance of a spherically symmetric black hole both when orbited by isotropically emitting light sources and when surrounded by a (geometrically and optically thin) accretion disk, while immersed in a dark matter halo. The black hole geometry plus the dark matter halo come as a solution of Einstein's field equations coupled to an anisotropic fluid whose density component follows a Hermquist-type distribution. Even in situations in which the geodesic description differs profoundly from the isolated black hole case, we find minor modifications to the primary and secondary tracks of the isotropic orbiting sources, and to the width, location, and relative luminosity of the corresponding photon rings as compared to the Schwarzschild black hole at equal black hole mass and emission models. This fact troubles distinguishing between both geometries using present observations of very-long baseline interferometry. 
\end{abstract}

\maketitle

\section{Introduction}

The recent experimental advances in the field of gravitational physics have opened up a new era in the employ of electromagnetic radiation to investigate the nature of ultra-compact bodies including black holes. Most notably are the groundbreaking discoveries of the Event Horizon Telescope (EHT), which detected the electromagnetic radiation emitted by the super-heated plasma in orbit around the supermassive object at the center of both the M87 and Milky Way galaxies \cite{EventHorizonTelescope:2019dse,EventHorizonTelescope:2021bee,EventHorizonTelescope:2022wkp}, and the GRAVITY instrument, with the detection of infrared flares in the vicinity of our own galactic center \cite{GRAVITY:2020lpa,GRAVITY:2023avo}. This new window to peer into the Universe is allowing the gravitational community to put to observational test the imaging of black holes via General Relativistic Magneto-HydroDynamic (GRMHD) simulations of the plasma orbiting it \cite{Johnson:2019ljv,Gold:2020iql}.

The Kerr metric is the cornerstone of several theorems developed within General Relativity (GR) guaranteeing that it is the unique stationary, axisymmetric solution of the Einstein field equations in vacuum possessing Killing horizons~\cite{Mazur:2000pn}. Indeed, the experimental achievements outlined above are in agreement with the theoretical predictions of the Kerr hypothesis \cite{Will:2014kxa,Yagi:2016jml}, according to which the complete gravitational collapse of a body in an appropriate astrophysical setup results in a spinning and electrically neutral black hole \cite{Kerr:1963ud,Penrose:1964wq}. Such a hypothesis has thus become the benchmark against which every observation is compared, and alternative space-times describing exotic compact objects are challenged to fit the observations better \cite{Bambi:2011mj,Cardoso:2019rvt}.

The Kerr space-time features unstable bound null orbits, also known as photon shells \cite{Perlick:2004tq}, which degenerate into the photon sphere in the Schwarzschild limit. Consequently, photons with near-critical impact parameters to such orbits can wind several times around the black hole before being released to the asymptotic observer or absorbed by the event horizon. Thus, GRMHD simulations produce a bright ring of radiation enclosing a central brightness depression in the observer's screen \cite{Broderick:2020wda}. Both features are tightly tied to the critical curve, i.e., the projection of the photon sphere/shell in the observer's screen. Higher-order images of the plasma surrounding the black hole and which asymptotically approach the critical curve are known as {\it photon rings} \cite{Gralla:2019drh}, and the {\it black hole shadow} marks the boundary at which there is a sharp decrease in the observed luminosity \cite{Falcke:1999pj}. Many works in the field provide refinements to such a picture \cite{Gralla:2019xty,Narayan:2019imo,Gralla:2020srx,Vincent:2022fwj,Paugnat:2022qzy,Broderick:2023jfl} or directly contest it via a supply of alternative black hole models (e.g. supported by additional matter fields or proposed within modifications of GR \cite{Wielgus:2021peu,Staelens:2023jgr,Olmo:2023lil,Guerrero:2022qkh,Guerrero:2021ues,Asukula:2023akj,Hou:2022eev}) or instead using horizonless compact objects \cite{Vincent:2020dij,Ishkaeva:2023xny,Rosa:2023hfm,Rosa:2022tfv,Rosa:2023qcv,Guerrero:2022msp,Olmo:2021piq,Wang:2023nwd,Rosa:2022toh,Tamm:2023wvn,Rosa:2024bqv}, providing a fertile playground to describe both black hole geometries and accretion disk physics via the images of ultra-compact objects \cite{EventHorizonTelescope:2020qrl,Lara:2021zth,Vagnozzi:2022moj,Eichhorn:2022oma}.

Astrophysical black holes are not isolated objects. In addition to their accretion disks, they are also immersed in the dark matter (DM) halo that supposedly engulfs the whole galaxy. Photons emitted outwards by matter in the vicinity of the black hole are also lensed by the dark matter halo, potentially leaving additional imprints in the resulting optical appearance. The main aim of this work is to determine the optical appearance of such a recently proposed spherically symmetric geometry \cite{Cardoso:2021wlq} (other models have been analyzed in e.g. \cite{Zeng:2020vsj,Zeng:2021mok,Ovgun:2023wmc}). In an adequate astrophysical context, this geometry is compact enough to hold a photon sphere, and thus it also produces both the photon ring and central brightness depression, which allows for a comparison with the canonical Schwarzschild black hole images. The critical impact parameter of the photon sphere in this geometry is modified, while the radius of the horizon remains unchanged (at equal black hole mass). These features provide an interesting scenario allowing to explore the images cast by background geometries with similar profiles for the emission of the accretion disk but different radii of the shadow.

In this work we analyze two different scenarios. First we study the images produced by the orbits of isotropically emitting light sources (hot spots), and second we employ a geometrically and optically thin accretion disk. In the former, we recur to the well-known ray-tracing software GYOTO \cite{Vincent:2011wz,Grould:2016emo}, which was proven useful in the imaging of diverse compact objects in an astrophysical context \cite{Vincent:2020dij,Lamy:2018zvj,Vincent:2016sjq,Vincent:2012kn,Rosa:2023qcv,Rosa:2022toh,Tamm:2023wvn,Rosa:2024bqv}, and perform an astrometric analysis of several observables, namely the integrated flux, magnitude, and centroid of the observation. In the latter, we recur to our own Mathematica-based ray-tracing code, previously used in several other publications \cite{Rosa:2023hfm,Rosa:2022tfv,Rosa:2023qcv,Olmo:2023lil,Guerrero:2022msp,Guerrero:2022qkh,Olmo:2021piq,Guerrero:2021ues,Asukula:2023akj,Rosa:2024bqv}, to model a monochromatic emission in the disk frame described by suitable adaptations of the Standard Unbound distribution previously employed in semi-analytical analysis to emulate the results of specific scenarios of GRMHD simulations \cite{Gralla:2020srx}, and we analyze the observed photon ring structure and central brightness depression. Using these two techniques, we compare the obtained results with those of a Schwarzschild black hole, and we discuss the possibilities for their disentanglement in future observations. Our analysis complements and significantly extends the one of \cite{Xavier:2023exm} in the photon rings and shadow properties of these geometries.

This paper is organized as follows. In Sec.~\ref{sec:framework} we describe the background geometry, the effective potential for null geodesics, and split the configurations into two different regimes, in which the DM distribution have low and high compactness, discussing the relevant features for the generation of images. In Sec.~\ref{sec:hotspots} we consider the orbits of isotropically emitting sources and construct the associated astrometric observables. In  Sec.~\ref{sec:disks} we present the ray-tracing procedure, the observation constraints of the black hole shadow, the accretion disk models and use all these ingredients to generate images and characterize the features of the photon rings and central brightness depression for several configurations of the background geometries and emission models of the disk. Finally, in Sec.~\ref{sec:conclusion} we present our final remarks. In what follows we adopt a system of geometrized units such that $G=c=1$, where $G$ is the gravitational constant and $c$ is the speed of light.

\section{Theory and framework} \label{sec:framework}

\subsection{Background geometry and effective potential}

Let us start by introducing the background geometry to be analyzed. To model the dark matter halo, we consider a Hernquist-type radial density distribution of the form \cite{Hernquist:1990be}
\begin{equation}
\rho=\frac{M_{\rm DM}a_0}{2\pi r(r+a_0)^3},
\end{equation}
where $\rho$ denotes the energy density, accounting for the galactic profiles of elliptical galaxies, and characterized by both the mass $M_{\rm DM}$ of the halo, and by a new constant $a_0$ which sets its typical length scale. With this matter distribution, a solution of the Einstein equations coupled to an anisotropic fluid was found in \cite{Cardoso:2021wlq} and can be suitably written in the usual spherical coordinates $\left(t,r,\theta,\phi\right)$ as follows
\begin{equation} 
ds^2=-A\left(r\right)dt^2+\frac{dr}{B(r)}+r^2 (d\theta^2 +\sin^2 \theta d\phi^2), \label{eq:le}
\end{equation}
where the metric functions are
\begin{align}
A(r)&=\left(1-\frac{2M_{\rm BH}}{r}\right)  e^{\Upsilon}, \\
B(r)&= 1-\frac{2m(r)}{r}, \\
m(r)&=M_{\rm BH} + \frac{M_{\rm DM}r^2}{(a_0+r)^2} \left(1-\frac{2M_{\rm BH}}{r} \right)^2,
\end{align}
and the parameters
\begin{align}
 \Upsilon &=-\pi \sqrt{\frac{M_{\rm DM}}{\xi}}\nonumber \\
&+ 2\sqrt{\frac{M_{\rm DM}}{\xi}} \text{arctan} \left(\frac{r+a_0-M_{\rm DM}}{\sqrt{M_{\rm DM}\xi}} \right), \\
\xi &=2a_0-M_{\rm DM}+4M_{\rm BH}.
\end{align}
This geometry describes a black hole of source mass $M_{\rm BH}$ surrounded by a dark matter halo of mass $M_{\rm DM}$, the full system hereafter dubbed as dark matter halo black hole, DMHBH. Consequently, the ADM mass $M$ of the space-time is given by $M=M_{BH}+M_{\rm DM}$. Interestingly, the horizon radius in this geometry turns out to coincide with the usual Schwarzschild radius of a black hole of similar source mass, i.e., $r_h=2M_{BH}$, a feature of interest for the analysis of the observational properties of these configurations. It is typically assumed that a hierarchy of scales applies to these DMHBHs, that is, $M_{BH} \leq M \ll a_0$, which also guarantees that the only singularity present in the geometry is the usual central one at $r=0$. In this work we shall study a few configurations of interest that fall outside of such assumption, but we guarantee nevertheless that only the black hole singularity is present in the space-time. Note that one can quantify the compactness of the dark matter halo via the introduction of a compactness parameter of the form 
\begin{equation}
\mathcal{C} \equiv \frac{M_{\rm DM}}{a_0} 
\end{equation}
which is bounded by galactic dynamics to $\mathcal{C} >10^{-4}$
\cite{Navarro:1995iw}. The parameter $\mathcal C$ is also a quantity of interest in the analysis of the observational properties of these models.

Before we proceed with the analysis of the DMHBH models, let us define a few parameters for later convenience. As noted previously, the metric in Eq.~\eqref{eq:le} is given in terms of three independent parameters, namely the mass of the black hole $M_{\rm BH}$, the mass of the DM halo $M_{\rm DM}$, and the typical length scale of the DM halo $a_0$. Since for observational purposes the quantity of interest is the ADM mass $M$, we find it convenient to express the mass of the DM halo in terms of a factor $n$ of the mass of the black hole, i.e., $M_{\rm DM}=n M_{\rm BH}$, from which one obtains $M=\left(n+1\right)M_{\rm BH}$. Normalizing all quantities of interest with respect to the ADM mass of the space-time, one can thus replace the dependency of the metric in the two masses $M_{\rm BH}$ and $M_{\rm DM}$ by a dependency on the ADM mass $M$ and the factor $n$ as
\begin{equation}
M_{\rm BH}=\frac{1}{n+1}M, \qquad M_{\rm DM}=\frac{n}{n+1}M,
\end{equation}
where the parameter $n$ ranges from $n=0$, corresponding to the case for which the DM halo is absent, to $n\to\infty$, corresponding to the limit for which the entire mass is contained in the DM halo. To simplify the notation and compactify the range of the parameter $n$, we define $k=\frac{n}{n+1}$ (or, conversely, $n=\frac{k}{1-k}$), such that if $k=0$ the mass of the DM halo vanishes and if $k=1$ the entire mass is contained in the DM halo. Furthermore, we introduce the normalized dimensionless length scale of the DM halo $\bar a_0$ as $a_0=\bar a_0 M$. In summary, these definitions allow one to fully describe each DMHBH model in terms of two dimensionless free parameters $k$ and $\bar a_0$ with
\begin{equation}
\frac{M_{\rm BH}}{M}=\left(1-k\right),\quad \frac{M_{\rm DM}}{M}=k , \quad \bar a_0=\frac{a_0}{M}.
\end{equation}

For the purpose of generating observable quantities in the upcoming sections, it is necessary to analyze the equations of geodesic motion. These equations can be obtained from the extrema of the Lagrangian $2{\cal L}=g_{\mu\nu}\dot{x}^\mu\dot{x}^\mu$, where dots represent derivatives with respect to the affine parameter along the geodesics, subjected to the constraint $2{\cal L}=-\delta$, with $\delta=0$ for null (massless) particles and $\delta=1$ for timelike (massive) particles~\cite{Jusufi:2020mmy}. Due to the spherical symmetry of the space-time under study, described by the line element in Eq.~\eqref{eq:le}, one can restrict this analysis to the equatorial plane $\theta=\pi/2$ without loss of generality. The resultant radial geodesic equation resembles the form of the scattering of a particle in an effective potential according to the equation
\begin{equation} \label{eq:effp}
\frac{A}{B}\dot{r}^2=E^2-V_{\rm eff}(r)
\end{equation}
where we have defined the conserved quantity $E=-A\dot{t}$ as the specific energy of the particle. The effective potential $V_{\rm eff}(r)$ reads as
\begin{equation}\label{eq:potential}
V_{\rm eff}(r)={A}\left(\delta+\frac{L^2}{r^2}\right),
\end{equation}
where we have defined the conserved quantity $L=r^2\dot\phi$ as the specific angular momentum of the particle. The trajectories of photons can thus be obtained by taking $\delta=0$ and performing the numerical integration of Eq.~\eqref{eq:effp}. By analyzing the equation above, one finds that for values of the impact parameter $b \equiv L/E$ such that $b<b_c$, where $b_c$ corresponds to those trajectories fulfilling Eq.~\eqref{eq:effp} and which are extrema of the effective potential, i.e., 
\begin{equation} \label{eq:ps}
b_c= V_{\rm eff}^{-1/2}(r_{ps}); \quad V_{\rm  eff}'(r_{ps})=0; \quad V_{\rm  eff}'' (r_{ps})>0 
\end{equation}
where a primes denotes derivatives with respect to the radial coordinate, then such photons are absorbed by the event horizon of the central black hole. Similarly, those with $b>b_c$ are scattered off the central potential back to asymptotic infinity, and those with $b \gtrsim b_c$ may perform several revolutions around the black hole lingering there for longer times before being released to the asymptotic observer.  Additionally, some properties of accretion disks can be inferred from analyzing the time-like circular geodesics, so we shall also highlight some of these features in our discussion below. For clarity of our analysis, we separate the discussion between low and high compactness regimes.

%%%%%%%%%%%%%%%%%%%%%%%%%%%%%%%%%%%%%%%%%%%%%%%%%%%%%%%%

\subsection{Low-compactness configurations}

In the low-compactness regime, the geodesic structure resembles the one of the Schwarzschild space-time, with small deviations due to the presence of the DM halo. These deviations are parametrically connected to the Schwarzschild case in the limit of ${\cal C}\to 0$ and, therefore, they can be described through a perturbative expansion.

\begin{figure}
	\includegraphics[width=\columnwidth]{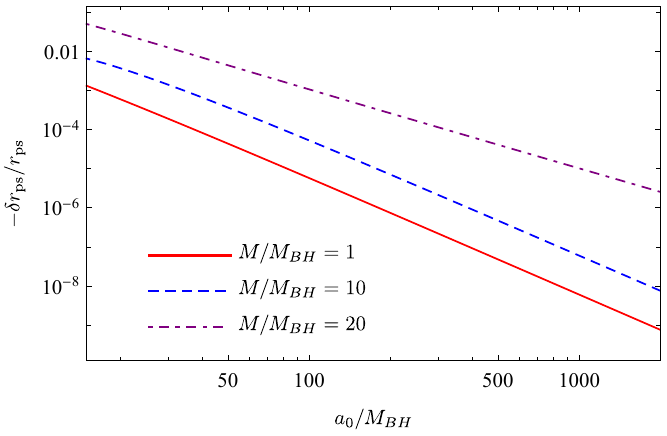}
	\caption{Normalized deviation of $r_{ps}$ between the numerical value and the analytical approximation obtained through Eq.\eqref{eq:light_a} as function of $a_0$.}\label{fig:light}
\end{figure}

The photon sphere (the locus of unstable bound geodesics) corresponds to the critical (unstable) points of the effective potential in Eq.\eqref{eq:potential} with $\delta=0$, which can be found via the fulfilment of the equation 
\begin{equation} \label{eq:rm}
	r_{ps}=\frac{2A_{ps}}{A_{ps}'}
\end{equation}
where the subindex $ps$ in the metric components denotes the evaluation at the photon sphere radius, that is, $A_{ps} \equiv A(r_{ps})$, as provided by the fulfilment of Eq.~(\ref{eq:ps}). In general, the photon sphere equation (\ref{eq:rm}) must be solved numerically to obtain $r_{ps}$. Considering the DMHBH and for small compactness values, the above equation can be solved as 
\begin{equation}
	r_{ps} \approx r_{ps}^S \left(1+\frac{\mathcal{C}M_{\rm BH}}{a_0}  \right) + \mathcal{O}\left(\frac{1}{a_0^3} \right)\label{eq:light_a}
\end{equation}
where $r_{ps}^S=3  M_{\rm BH}$ is the radius of the photon sphere for the Schwarzschild black hole. In Fig.~\ref{fig:light} we show the deviation of Eq.~\eqref{eq:light_a} from the corresponding numerical value, confirming its validity in the low-compactness scenario. Light rays that asymptotically approach this curve produce, in the observer's screen, a photon ring composed of trajectories that have winded $n$ half-times around the black hole before being released to the observer's screen. To achieve this in our case, the impact parameter must be equal to the critical value
\begin{equation} \label{eq:bc}
	b_c \approx b_c^S \left[1+\mathcal{C} + \frac{\mathcal{C}^2}{6} \left(5-18\frac{M_{\rm BH}}{M_{\rm DM}} \right) \right] + \mathcal{O}\left(\frac{1}{a_0^3} \right),
\end{equation}
where $b_c^S = 3 \sqrt{3} M_{\rm BH} \approx 5.1916M_{\rm BH}$ is the critical impact parameter in the Schwarzschild space-time. From these expressions we infer two general properties of DMHBHs in the low-compactness regime. First, the expressions for the photon sphere and critical impact parameter are modified outwards and upwards, respectively, from their Schwarzschild values. Second, they indicate a stronger sensibility of $b_c$ (in comparison with $r_m$) to variations in the compactness,  given the fact that the latter is additionally suppressed by a factor $M_{\rm BH}/a_0 \ll 1$. Note, however, that these two results are modified in the high compactness case, as a richer structure appears when two photon spheres are present.

\begin{figure*}
	\includegraphics[width=1\columnwidth]{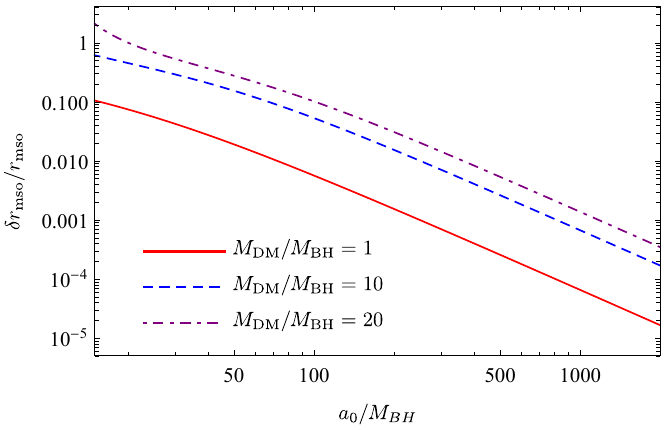}\includegraphics[width=1\columnwidth]{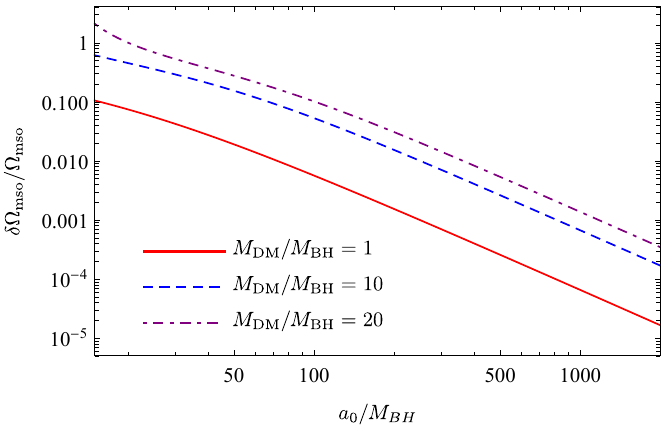}
	\caption{Normalized deviation of $r_{\rm MSO}$ and $\Omega_{\rm MSO}$ between the numerical value obtained with a numerical root finder and the analytical approximations obtained through Eqs.~\eqref{eq:isco} and~\eqref{eq:wisco}, respectively, as function of $a_0$. }
	\label{fig:isco_wisco}
\end{figure*}

Concerning circular time-like geodesics, for low-compactness the structure is similar to the one in the Schwarzschild space-time. Considering the potential in Eq.~\eqref{eq:potential} with $\delta=1$ and respective radial derivatives, and under the restriction $\dot r = \ddot r = 0$ for circular orbits, one obtains a system of two equations that can be solved for the specific energy and angular momentum. Furthermore, the stability of those orbits can be studied via the analysis of the second-order radial derivative of the effective potential in Eq.~\eqref{eq:potential}. Indeed, the radii for which $V_{\rm eff}''=0$ correspond to the ones at which a transition between stable and unstable orbits occurs. Circular orbits at those radii are called Marginally Stable Orbits (MSO). For the Schwarzschild space-time, there is only one MSO at $r_{\rm MSO}=6M_{\rm BH}$ which corresponds to the so-called Innermost Stable Circular Orbit (ISCO) as for any radius below the radius of the ISCO only unstable orbits exist. In most accretion models, the ISCO corresponds to the inner edge of the disk. An analysis of the second radial derivative of the potential for circular orbits in our case shows that there is a MSO in the DMHBH model that is parametrically connected to the Schwarzschild ISCO. The radius of this MSO can be obtained perturbatively in powers of $1/a_0$ as
\begin{equation}
	r_{\rm MSO} \approx r_{\rm ISCO}^S \left(1-\frac{32\mathcal{C}M_{\rm BH}}{a_0}+{\cal O}(1/a_0^2) \right),\label{eq:isco}
\end{equation}
where $r_{\rm ISCO}^S=6M_{\rm BH}$ is the radius of the ISCO for the Schwarzschid black hole. Another important quantity is the orbital frequency $\Omega$, which can be obtained through $\Omega=\sqrt{A(r_o)/2r_o}$ and is given perturbatively by 
\begin{equation}
	\Omega_{\rm MSO,0}=\Omega_{\rm ISCO}^S\left(1-\frac{M_{\rm DM}}{a_0}+{\cal O}(1/a_0^2)\right),\label{eq:wisco}
\end{equation}
where $\Omega_{\rm ISCO}^S=1/(6\sqrt{6} M_{\rm BH})$ is the orbital frequency at the ISCO for the Schwarzschild space-time. We compare the results obtained through the analytical approximations for the radius and orbital frequency of the MSO with the numerical values obtained through a numerical root finder in Fig.~\ref{fig:isco_wisco}. We observe a better agreement in the low-compactness regime (larger $a_0/M_{BH}$), as expected. From Eqs.\eqref{eq:isco} and \eqref{eq:wisco} one verifies that, although the radius of the inner edge of the disk decreases, the corresponding orbital frequency also decreases. This result is somewhat counter-intuitive, as one usually expects the orbital frequency to increase when the orbital radius decreases.

\subsection{High-compactness configurations}
\subsubsection{Stability window for circular geodesic and accretion disks}

While the configurations of physical interest are usually in the low-compactness regime, if one considers the possibility of the accumulation of dark matter in certain regions closer to the black hole, these could lead to the formation of denser compact-like objects. These configurations are usually called dark matter spikes \cite{Merritt:2002vj}. Furthermore, in the axionic dark matter scenario, the compactness of these configurations may be large enough to induce photon-ring formation~\cite{Guerra:2019srj}. Therefore, in this subsection we analyze the geodesic structure and non-perturbative corrections for photon rings and stable circular orbits in the high-compactness regime.

\begin{figure}
	\includegraphics[width=\columnwidth]{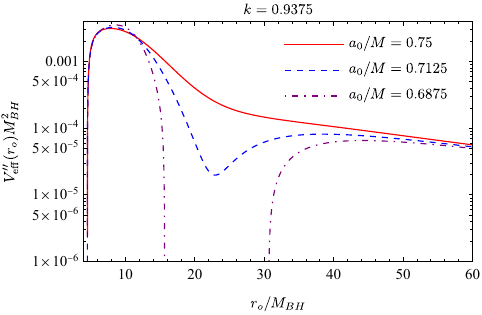}
	\caption{Transitions between stable and unstable orbits illustrated by changes in the second derivative of the effective potential. For a fixed value of $k$, there is a critical value of $a_0/M$ for which a pair of MSOs appear. 
	%A second peak develops at a given value of $k$, which can bring new features to the disk profile.
		}\label{fig:ddpot}
\end{figure}

For the majority of the parameter space, the MSOs correspond to ISCOs, and the circular geodesic structure closely resembles the one in the Schwarzschild space-time. However, in the high-compactness regime a more complex geodesic structure emerges, featuring additional transition points between stable and unstable orbits. This feature is illustrated in Fig.~\ref{fig:ddpot}, where we plot the second derivative of the effective potential computed at different orbital radius $r_o$ considering $M_{\rm DM}/M_{\rm BH}=15$ ($k=0.9375$) and varying $a_0/M$. The critical value for the compactness above which the additional transitions arise depends on the value of $M_{\rm DM}$, and takes approximately the value ${\cal C}=1.26$ in the limit $M_{\rm DM}\gg M_{\rm BH}$.

\begin{figure}
	\includegraphics[width=\columnwidth]{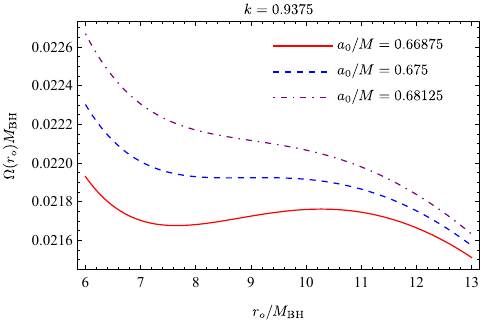}
	\caption{Orbital frequency of time-like circular geodesics around DMHBH systems as a function of the orbital radius. For a given value of the compactness a maximum in the frequency appears at $r_o/M_{BH}\approx 10.2$ (for this particular value of $k$).}\label{fig:orb_fre}
\end{figure}

The existence of a stability window for circular geodesics can (possibly) modify the structure of accretion disks around these objects and, therefore, present a distinctive feature for high-compactness configurations. These can be obtained via GRMHD simulations, for instance, which is beyond the scope of this paper. However, it has been noticed in other scenarios that changes in the inner edge of accretion disks can present themselves whenever the orbital velocity as a function of the radial coordinate $r$ features a maximum at some finite $r>0$~\cite{Olivares:2018abq}. We observe that in high-compactness scenarios this is indeed the case, as it can be seen in Fig.~\ref{fig:orb_fre}. The appearance of the maximum occurs near (but not exactly) the critical value for $a_0$ described above, such that is can occurs even when additional MSOs are not present. The location of the maximum also depends on $k$. Therefore, we can expect a discontinuity in the disk structure and in the image of the black hole as one explores the parameter space of the solution.

\subsubsection{Photonspheres, critical impact parameter and trapped regions}

Another consequence of high-compactness configurations is the formation of additional photon spheres. This feature was already pointed out in Ref.~\cite{Xavier:2023exm}. In horizonless compact objects, photon rings come in pairs~\cite{Cunha:2017qtt}, i.e., the total number of photon rings should be an even number, except for possibly degenerate meta-stable pairs. For the DMHBH model, due to the existence of an event horizon, the total number of photon rings is an odd number. In the low-compactness region, there is a photon sphere that is parametrically connected to the Schwarzschild one. Similarly to what we previously described for time-like circular geodesics, as the compactness increases, another non-perturbative pair of photon spheres develops, one of which is stable and the other is unstable. Our results indicate that the additional pair of photon spheres appear for a compactness of ${\cal C}>1.35$ in the limit $M_{\rm DM}\gg M_{\rm BH}$.

\begin{figure}
	\includegraphics[width=\columnwidth]{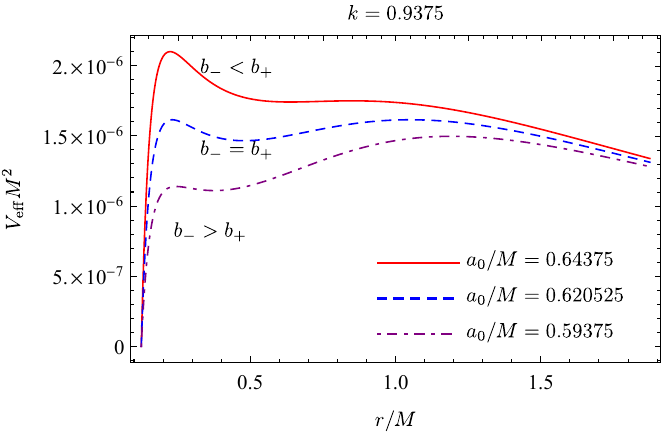}
	\caption{Null geodesic potential illustrating the behavior of the impact parameters $b_\pm$.}\label{fig:potential}
\end{figure}

Similarly to low-compactness configurations, the radii of the photon spheres can be found through Eq.\eqref{eq:rm}, which describes the extrema of the effective potential. The potential is also related to the critical impact parameter, as for the photon to be observed it must be capable of reaching asymptotic infinity from the photon sphere. To illustrate this, we show the effective potential in Fig.~\ref{fig:potential} for a fixed $M_{\rm DM}=15M_{\rm BH}$, and varying the value of $a_0$. These results imply the existence of two different impact parameters, each corresponding to unstable photon orbits (located at the maxima of the potential), namely, $b_-$ for the internal one and $b_+$ for the external one. The minimum of the potential between these two maxima corresponds to the \textit{stable} photon orbit. The critical impact parameter corresponds therefore to\footnote{A similar structure appears in the case of black holes surrounded by spherical shells \cite{Macedo:2015ikq}.}
\begin{equation}
	b_c={\rm min}(b_-,b_+).
\end{equation}  
As the critical impact parameter is related both to the photon rings and central brightness depression features of the image, we have a scenario in which the addition of the photonspheres due to the increase in compactness can drastically change the observational properties of the highly compact DMHBH configurations, inducing a discontinuity in the size of the shadow as a function of compactness.

\begin{figure}
	\includegraphics[width=\columnwidth]{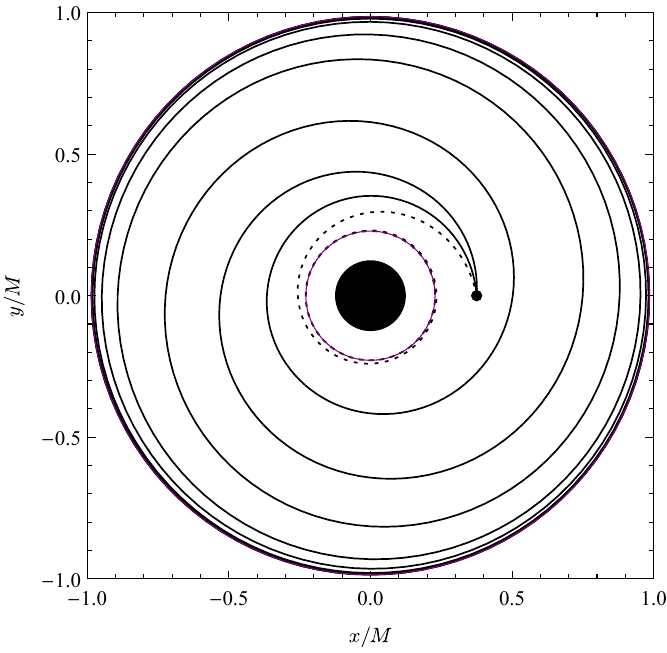}
	\caption{Null geodesics emitted from a point-like source located at $(x,y)=(6,0)M_{\rm BH}$, for a configuration with $M_{\rm DM}=15M_{\rm BH}$ and $a_0=0.63M$. The black lines indicate orbital motion with $b=b_+$, emitted in the inward and outward directions. Light rays emitted within the angle defined by the black lines are trapped in between the potential barriers. The dashed line considers a light ray with $b=b_-$.}\label{fig:trapped}
\end{figure}

Finally, we note the possibility of having emitting matter in between the two photon spheres, due to the fact that time-like circular geodesics with radii between the stable photon sphere and the inner unstable photon sphere are stable orbits. As these orbits are highly energetic, one can expect a considerable amount of power being emitted from this region. However, as noticed from standard electrodynamics, highly energetic particles emit in a narrow beam in the direction of their velocities, with a typical width of $\sim({dt}/{d\tau})^{-1}$ \cite{Misner:1973prb}. Consequently, many of these light rays -- if not all -- get trapped in between the two peaks of the potential, thus being inaccessible to the external observer. To illustrate this effect, in Fig.~\ref{fig:trapped} we depict light rays emitted upwards from a point in a region between the stable and the inner unstable photon sphere, located at $(x,y)=(6,0)M_{BH}$. Note that this position would correspond to a stable time-like circular orbit with $(dt/d\tau)^{-1}\sim 0.005$. The black solid lines represent light rays that have $b=b_+$, emitted in the inward and outward directions, out-spiralling asymptotically towards the outer photon sphere, and orbiting there in an unstable motion. Geodesics emitted in the cone within the black lines as boundaries are trapped in an orbital motion dictated by the effective potential. The trapped light rays have an impact parameter $b>b_+$ such that, as expected, they cannot escape the outer potential barrier. Likewise, when projecting a light ray inwards, one can orbit the inner unstable photon sphere for an arbitrarily large amount of time if $b=b_-$, represented by the black dashed line, and light rays with $b<b_-$ are absorbed by the black hole.

%%%%%%%%%%%%%%%%%%%%%%%%%%%%%%%%%%%%%%%%%%%%%%%%%%%%

\section{Hot spots and astrometry}\label{sec:hotspots}

\subsection{Setup and astrometric observables}

Consider now an astrophysical setting for which the DMHBH is orbited by some isotropically emitting light source. In this situation, one could recur to the analysis of astrometric quantities to assess the validity of the model. For this purpose, we recur to the well-known open-source ray-tracing software GYOTO \cite{Vincent:2011wz}, where we simulate the orbits of a spherical source with a radius of $r_s=0.5M$ on the equatorial plane $\theta=\pi/2$. We take the total ADM mass of the space-time to be given by the mass of Sgr A*, i.e., $M_=4.26\times 10^6 M_\odot$, where $M_\odot$ is the solar mass. The value for the orbital radius is chosen to be $r_o=8M$, to be consistent with the observed radius at which the GRAVITY instrument detected the orbit of infrared flares at Sgr A* \cite{GRAVITY:2020lpa}. Furthermore, the distance between the observer and the center of the DMHBH configuration is chosen to be the distance between the Sun and Sgr A*, i.e., $d=8.23 \text{kpc}$. We select nine DMHBH models to analyze, described by the parameters $k=\{0.3,0.6,0.9\}$ and $\bar a_0=\{1, 10, 100\}$, plus an additional high-compactness configuration described by the parameters $k=0.9375$ and $\bar a_0=0.620525$.

The GYOTO software is run under the assumptions outlined in the previous paragraph for a given DMHBH model and outputs a 2-dimensional matrix of specific intensities $I^\nu_{lm}$ for a given time instant $t_k$, which can be interpreted as the observed image, where each pixel $\{l,m\}$ is associated with a given observed intensity. The simulations are repeated for a total of 180 equally spaced time instants in the range $t_k\in\left[0,T\right[$, where $T$ is the orbital period of the source. As a result, one obtains a cube of data $I_{klm}=\delta\nu I^\nu_{lm}$, where $\delta\nu$ is the spectral width. This process is then repeated for every DMHBH model considered, thus resulting in a cube of data for each of the models under analysis.

The cubes of data produced through the process summarized in the previous paragraph can then be used to produce three observable quantities, namely the time integrated fluxes $\left<I\right>_{lm}$, the temporal fluxes $F_k$, and the temporal centroids $\vec{c}_k$, which are defined respectively as:
\begin{equation}
\left<I\right>_{lm}=\sum_k I_{klm},
\end{equation}
\begin{equation}
F_k=\sum_l\sum_m\Delta\Omega I_{klm},
\end{equation}
\begin{equation}
\vec{c}_k=\frac{1}{F_k}\sum_l\sum_k\Delta\Omega I_{klm}\vec{r}_{lm},
\end{equation}
where $\Delta\Omega$ represents the solid angle of a single pixel and $\vec{r}_{lm}$ denotes the displacement vector of the pixel $\{l,m\}$ with respect to the center of the observed image. Furthermore, from the temporal fluxes $F_k$, one can construct a more useful astronomical observable known as the temporal magnitude, defined as
\begin{equation}
m_k=-2.5 \log \left(\frac{F_k}{\min F_k}\right).
\end{equation}

\begin{figure*}[t!]
\includegraphics[scale=0.35]{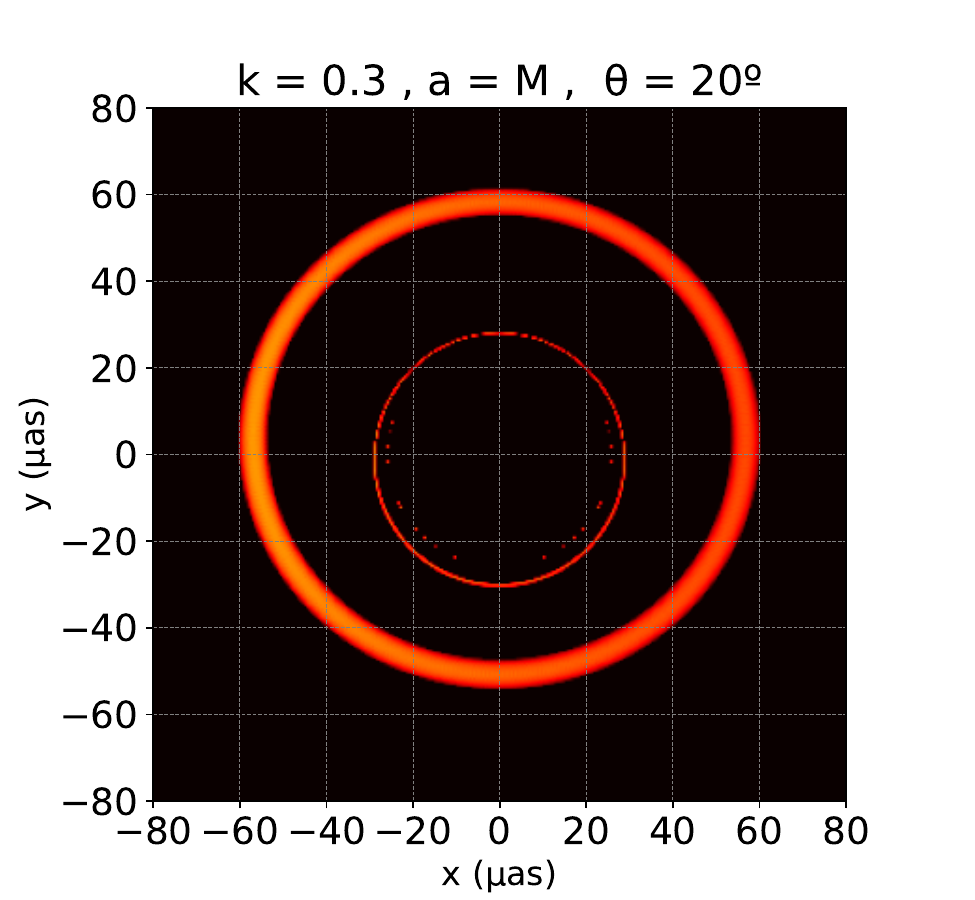}
\includegraphics[scale=0.35]{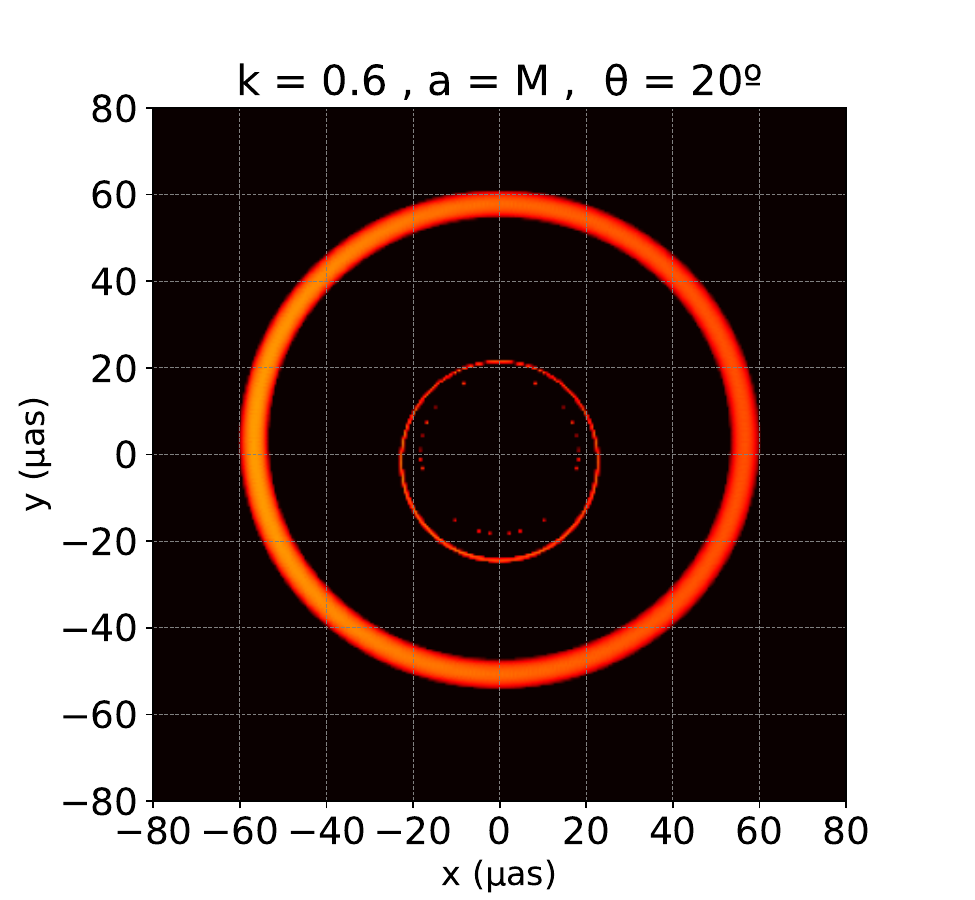}
\includegraphics[scale=0.35]{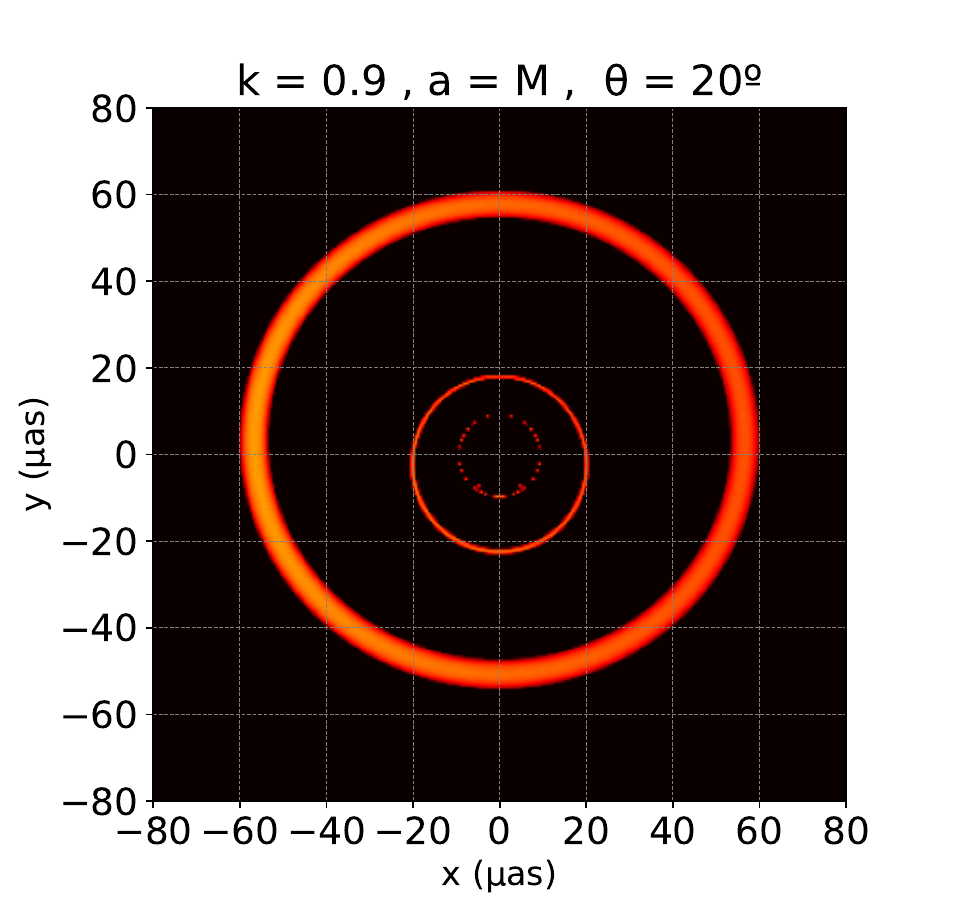}\\
\includegraphics[scale=0.35]{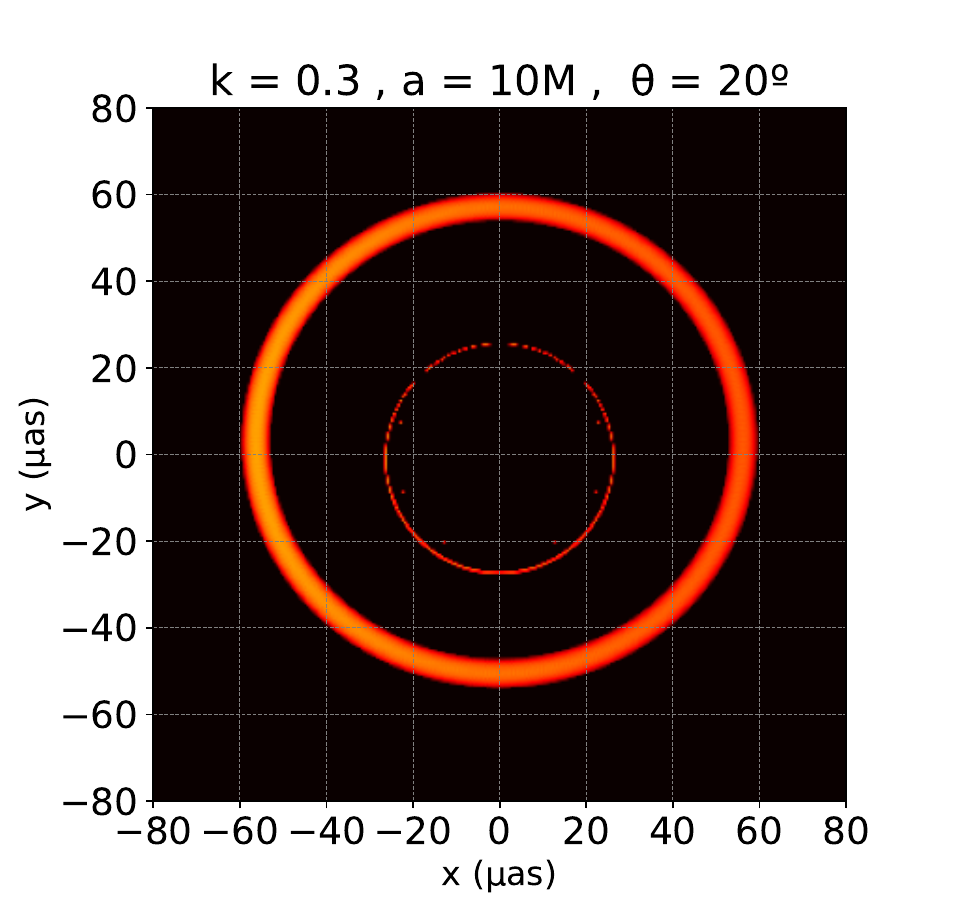}
\includegraphics[scale=0.35]{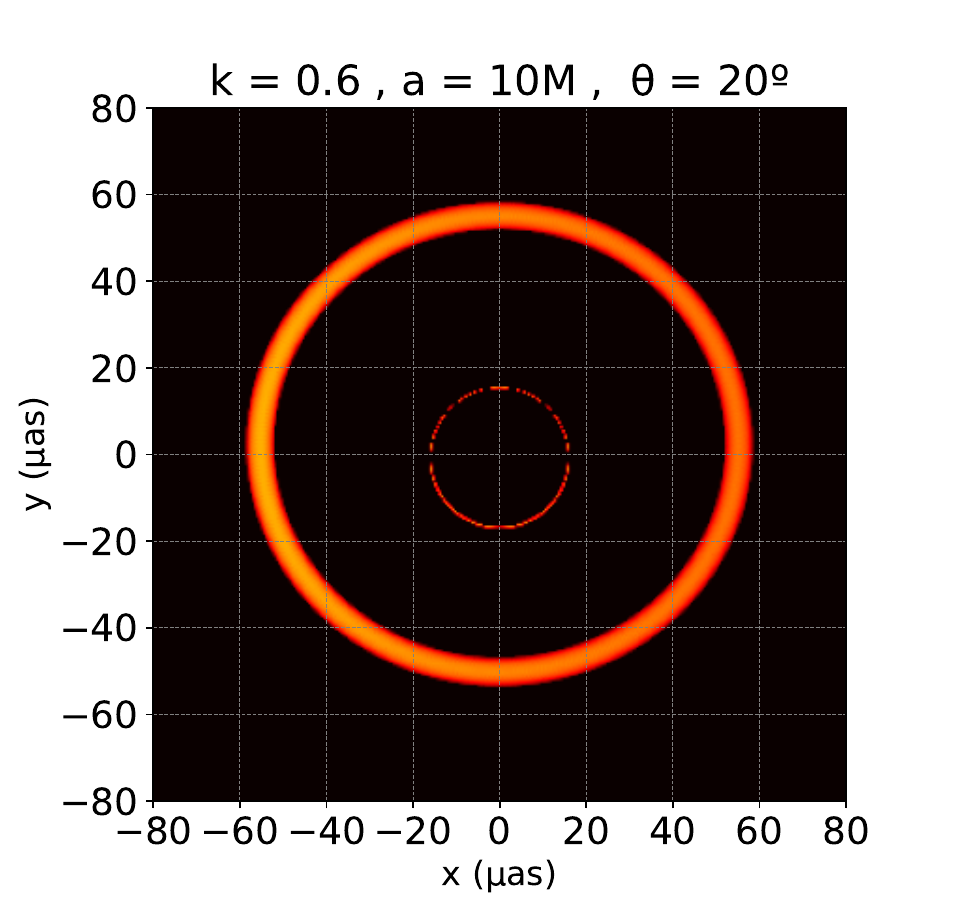}
\includegraphics[scale=0.35]{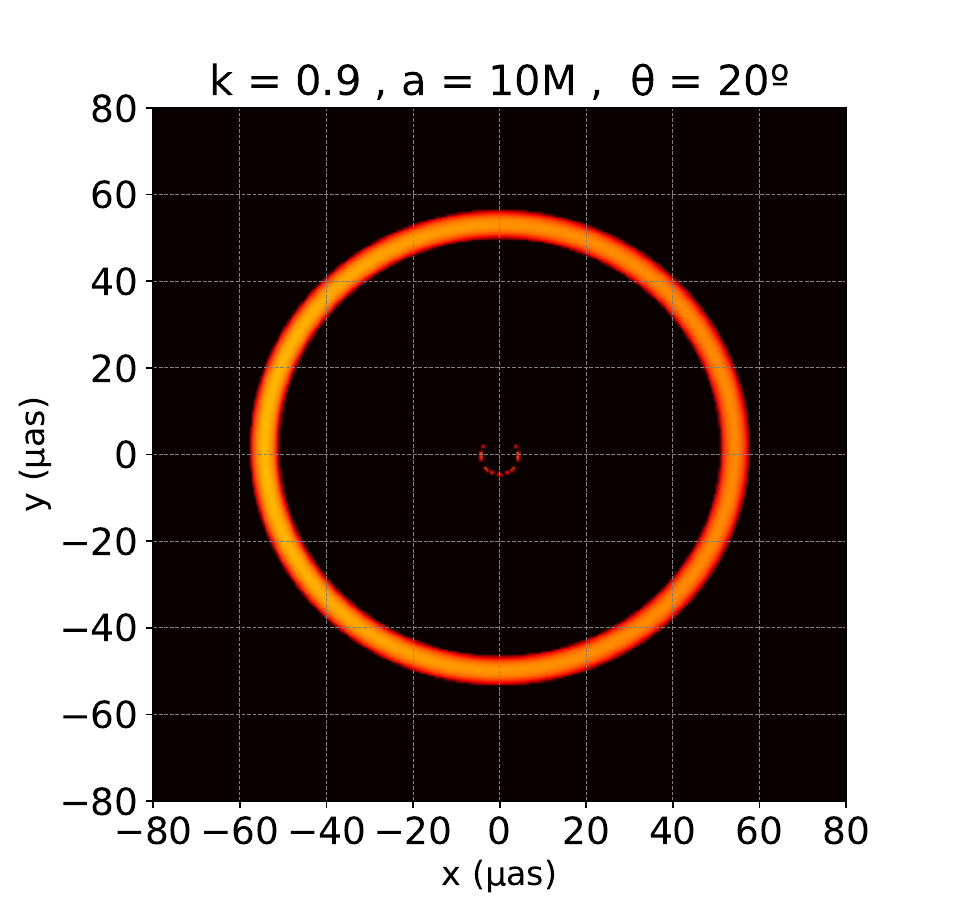}\\
\includegraphics[scale=0.35]{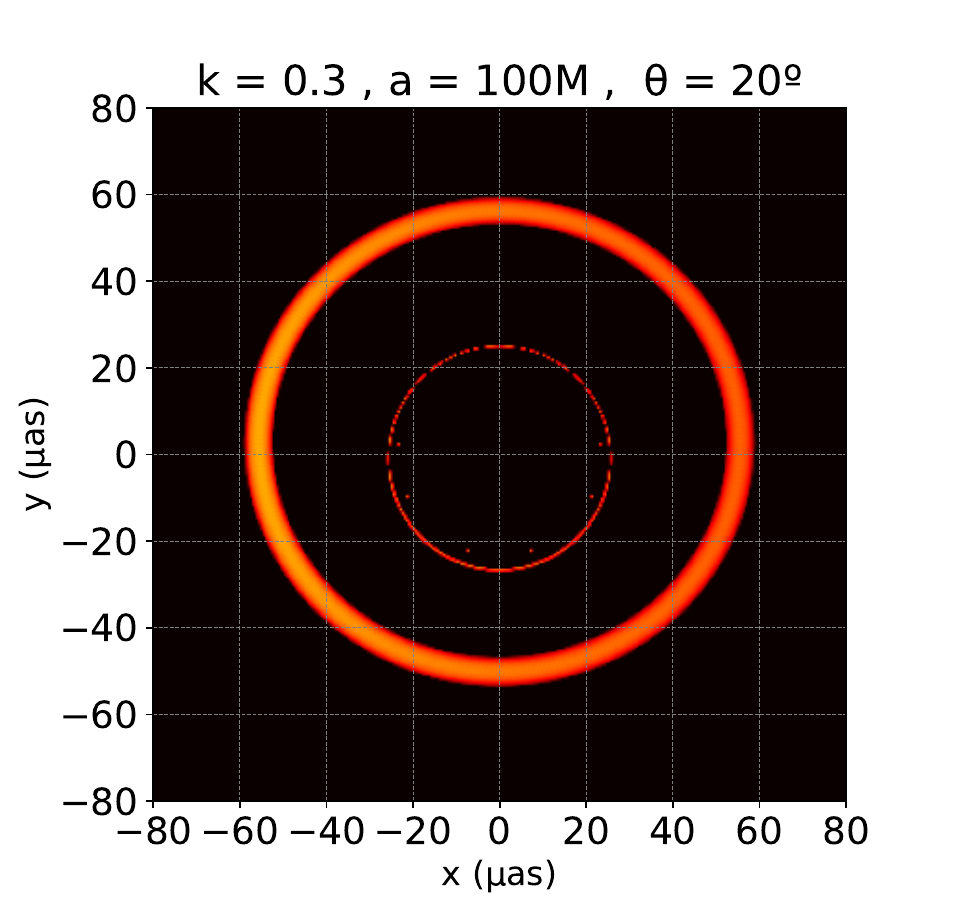}
\includegraphics[scale=0.35]{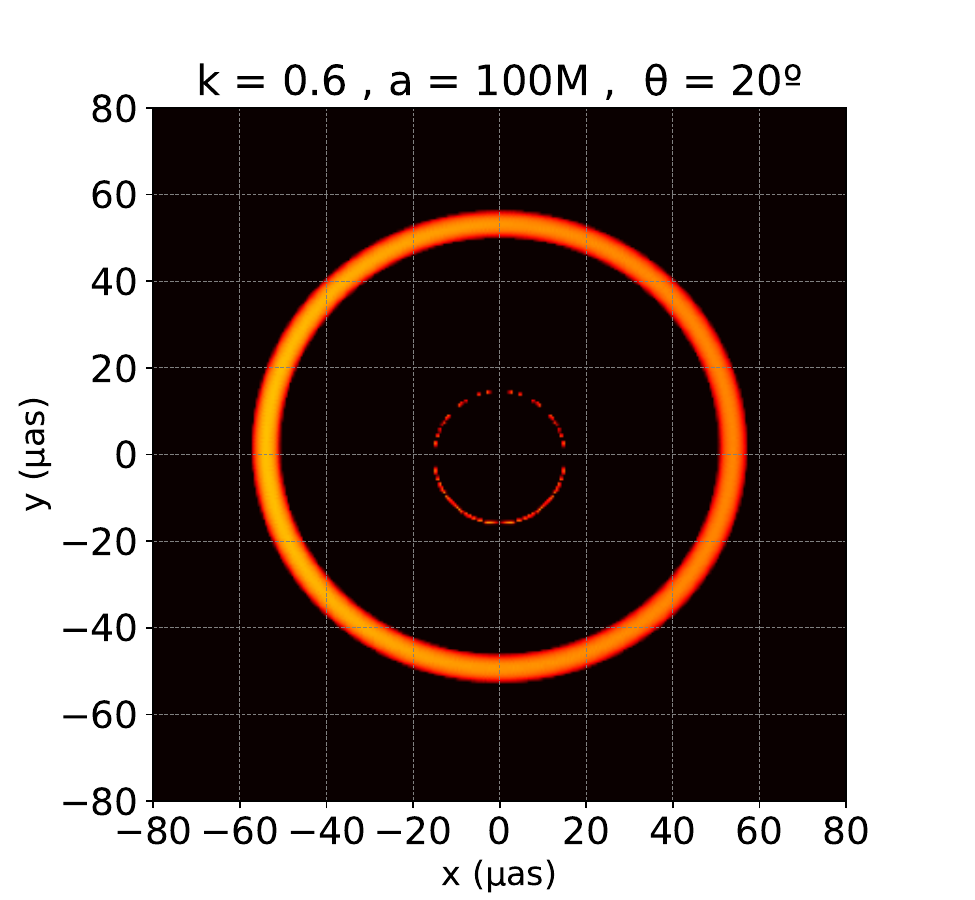}
\includegraphics[scale=0.35]{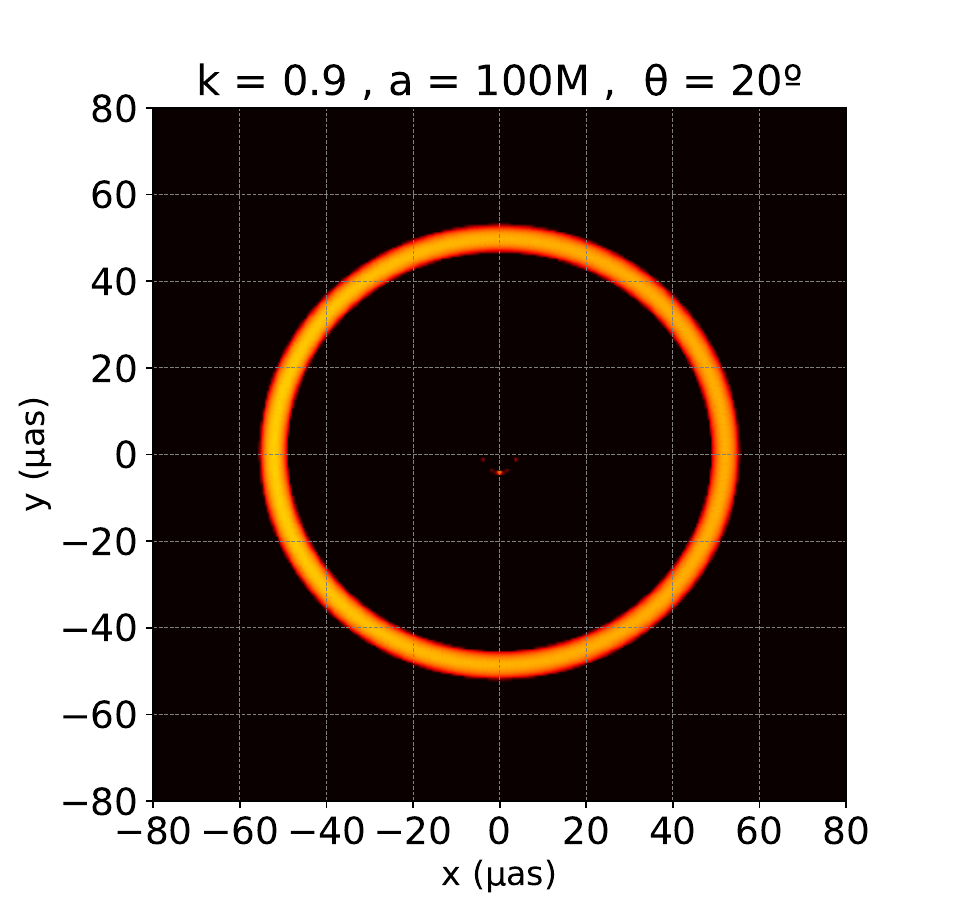}
\caption{Time integrated fluxes $\left<I\right>_{lm}$ for an observation angle of $\theta=20^\circ$ for the DMHBH models with $a=1M$ (top row), $a_0=10M$ (middle row), and $a_0=100M$ (bottom row), and with $k=0.3$ (left column), $k=0.6$ (middle column), and $k=0.9$ (right column).}
\label{fig:flux20deg}
\end{figure*}

\begin{figure*}[t!]
\includegraphics[scale=0.35]{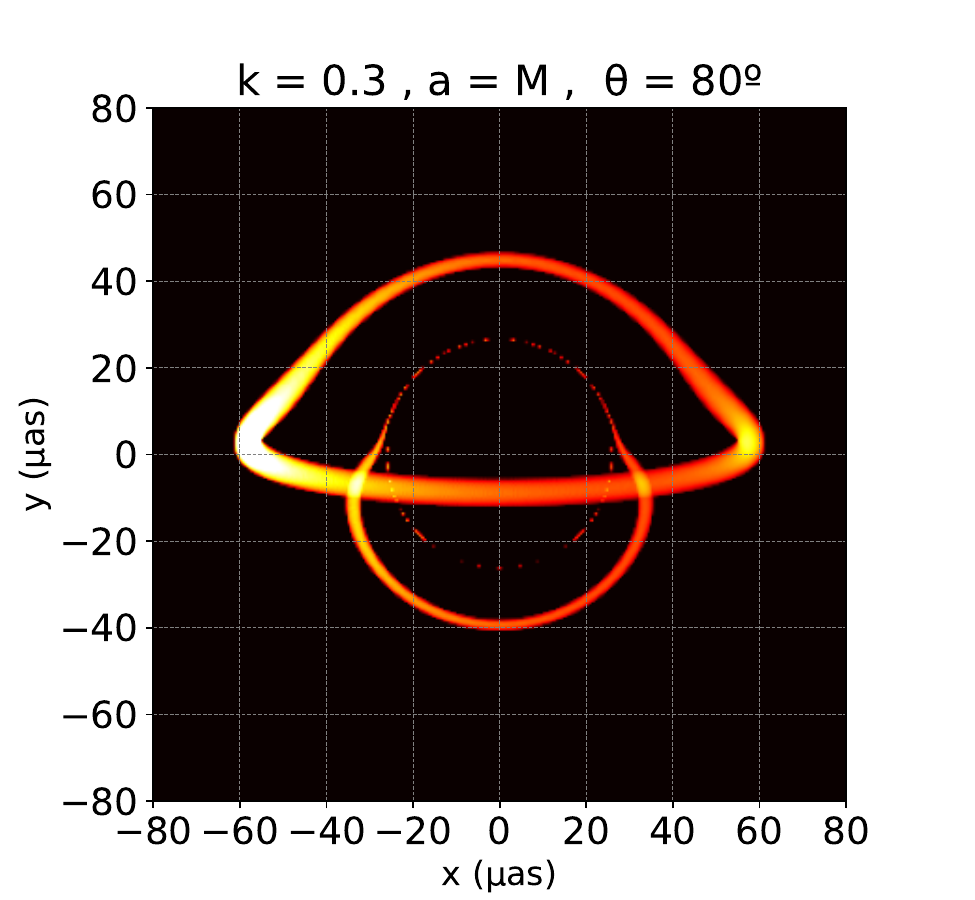}
\includegraphics[scale=0.35]{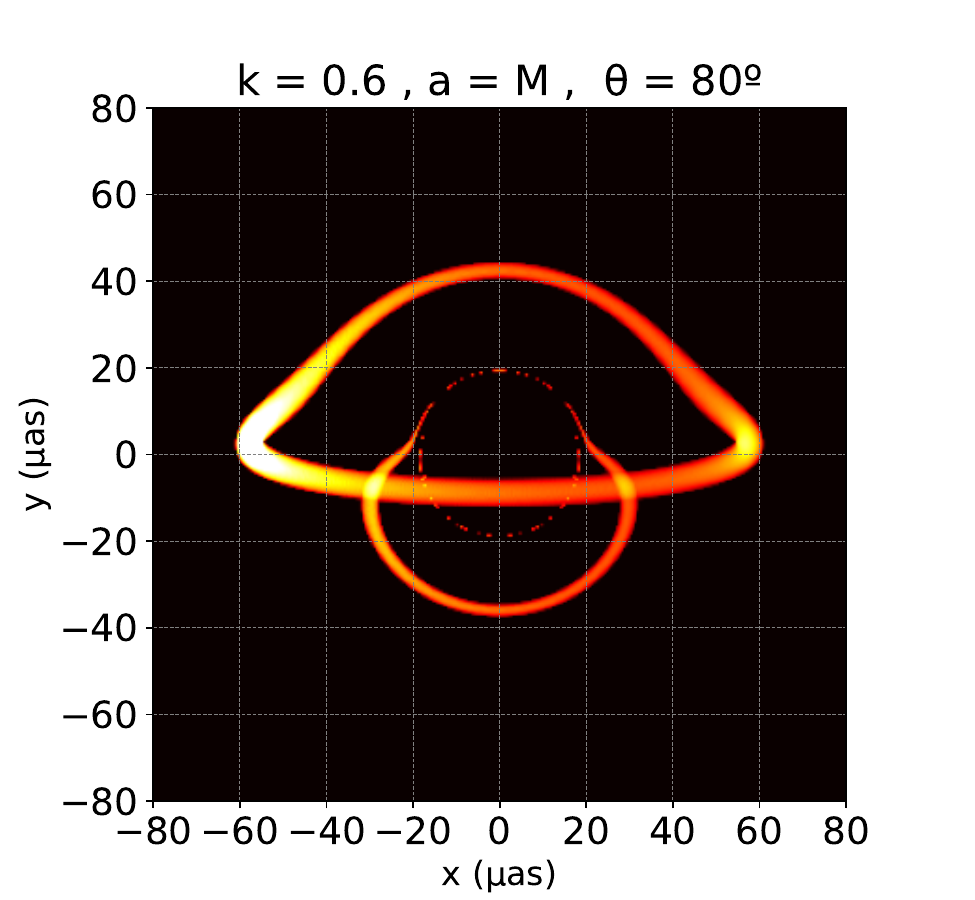}
\includegraphics[scale=0.35]{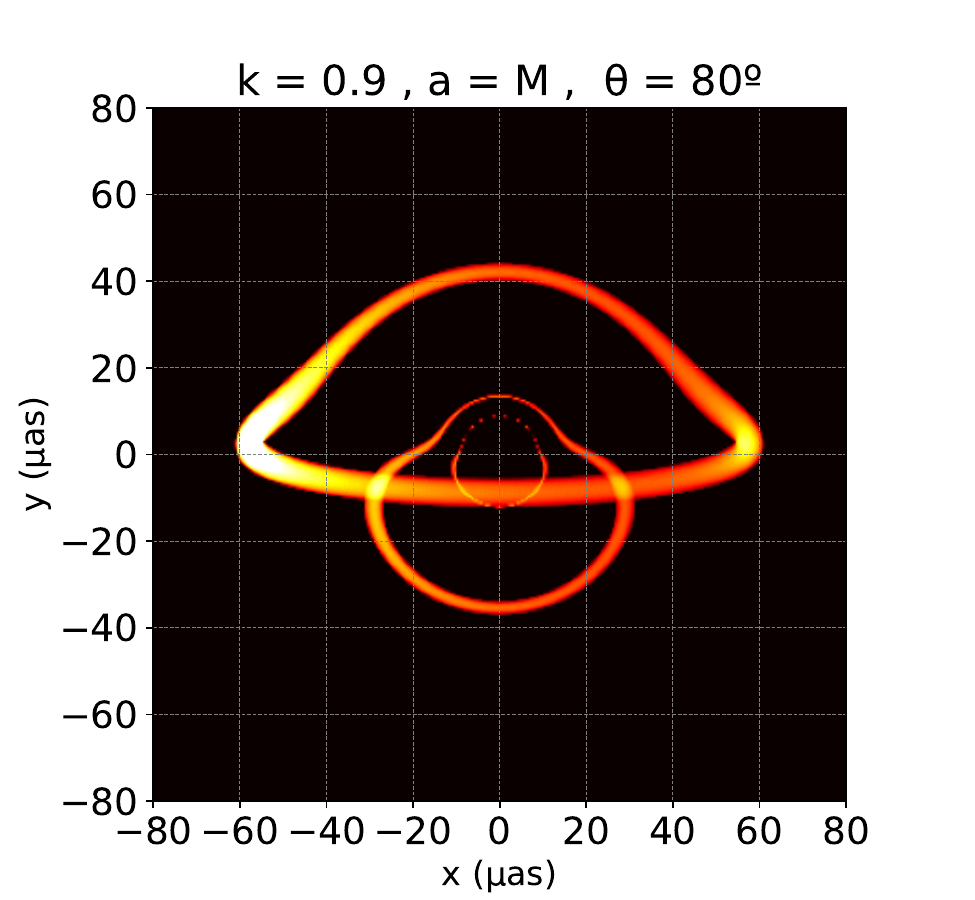}\\
\includegraphics[scale=0.35]{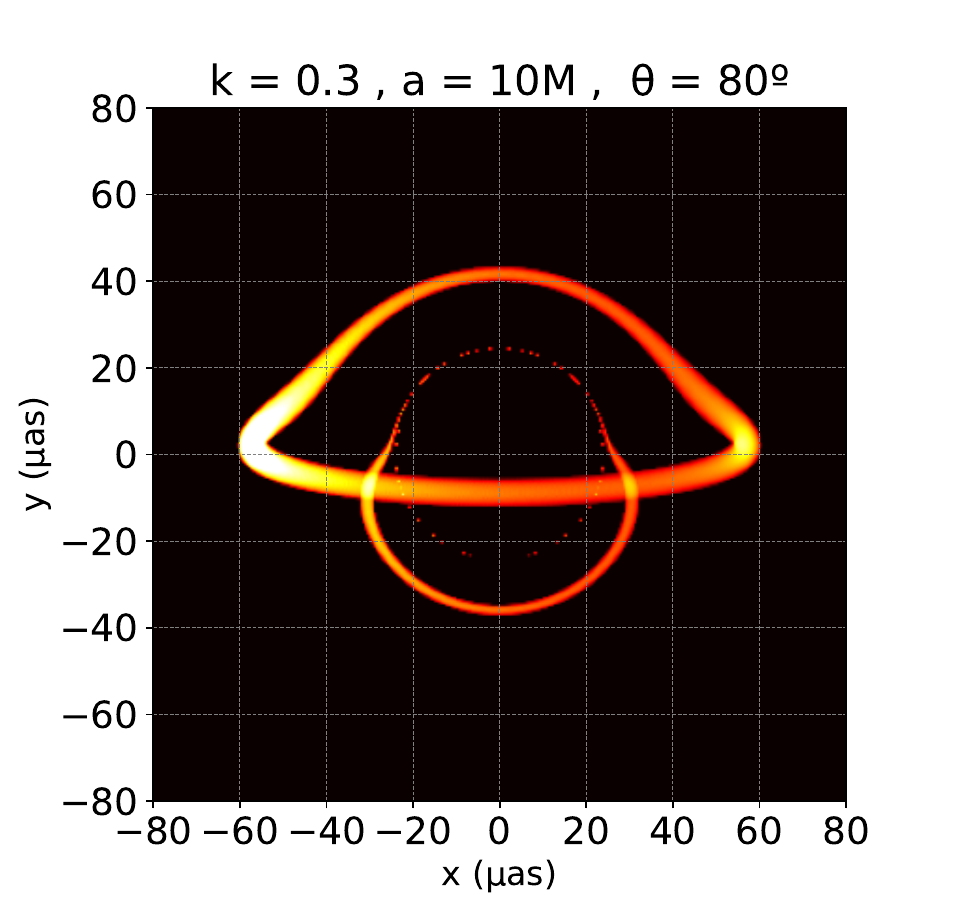}
\includegraphics[scale=0.35]{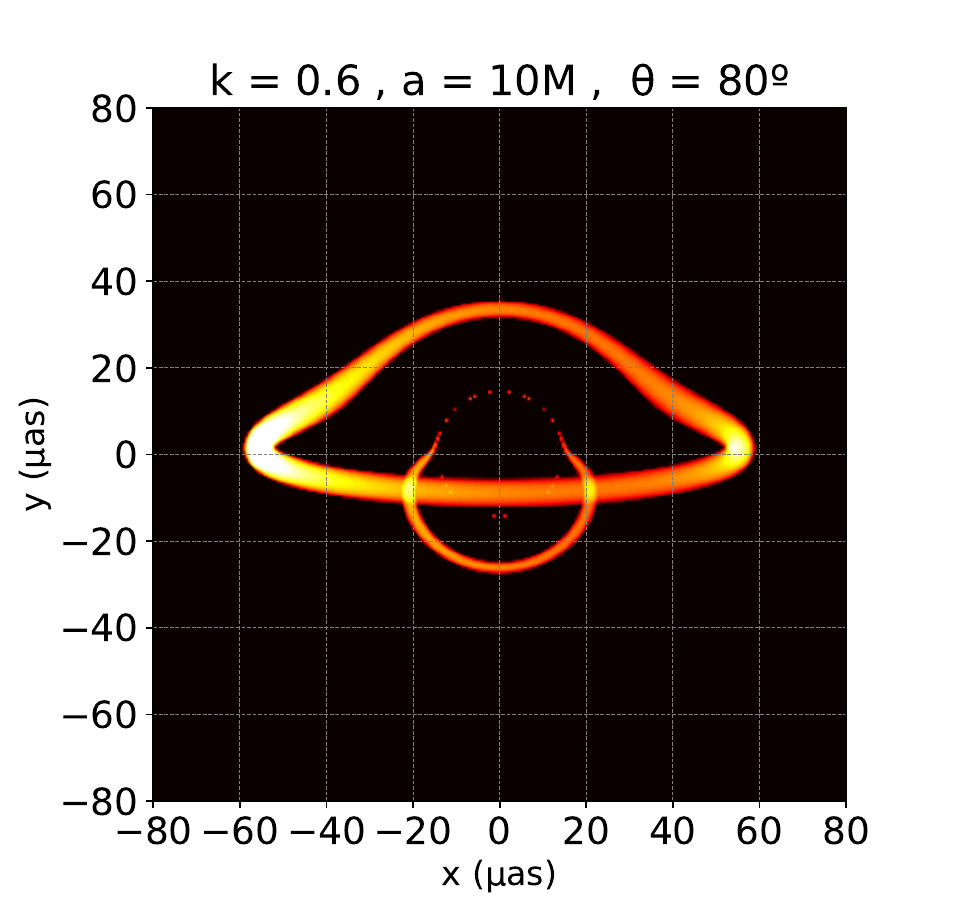}
\includegraphics[scale=0.35]{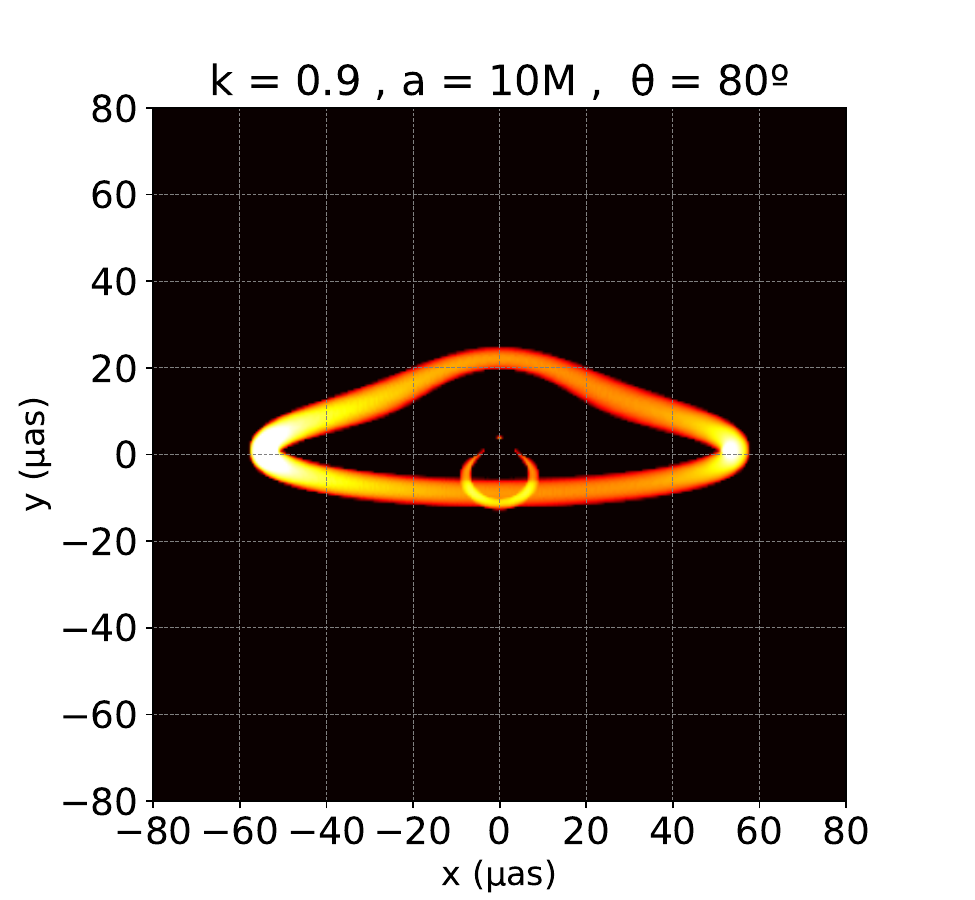}\\
\includegraphics[scale=0.35]{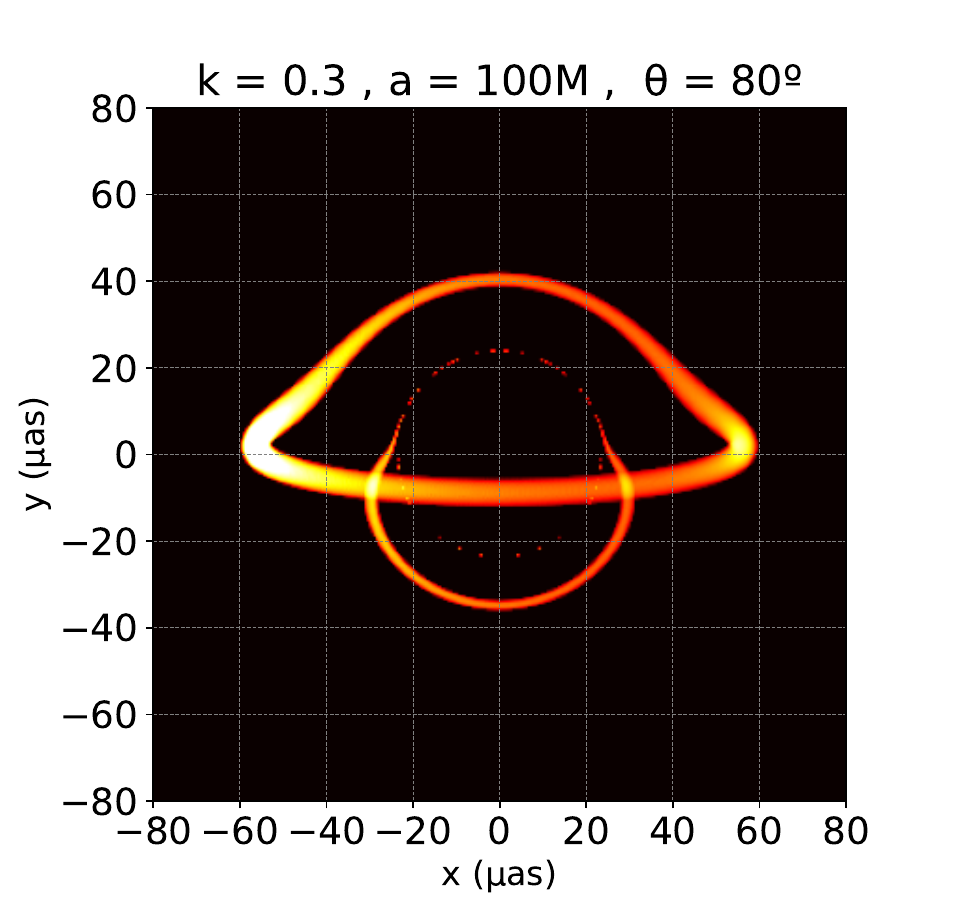}
\includegraphics[scale=0.35]{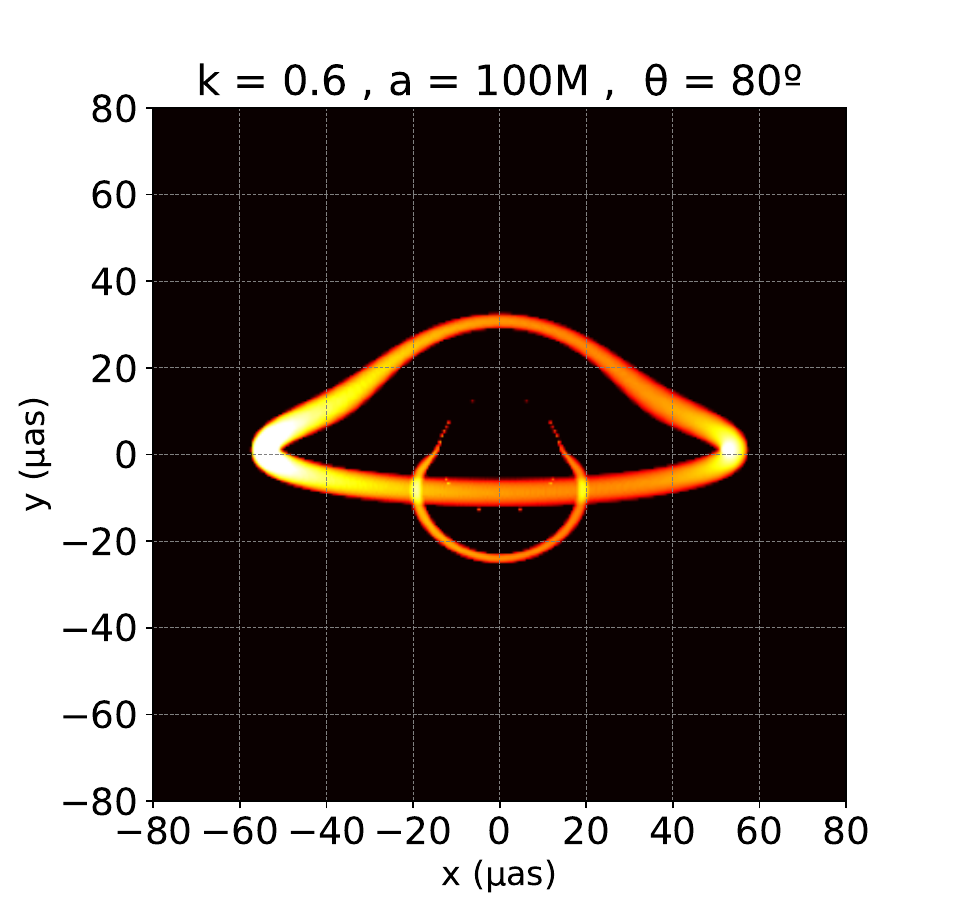}
\includegraphics[scale=0.35]{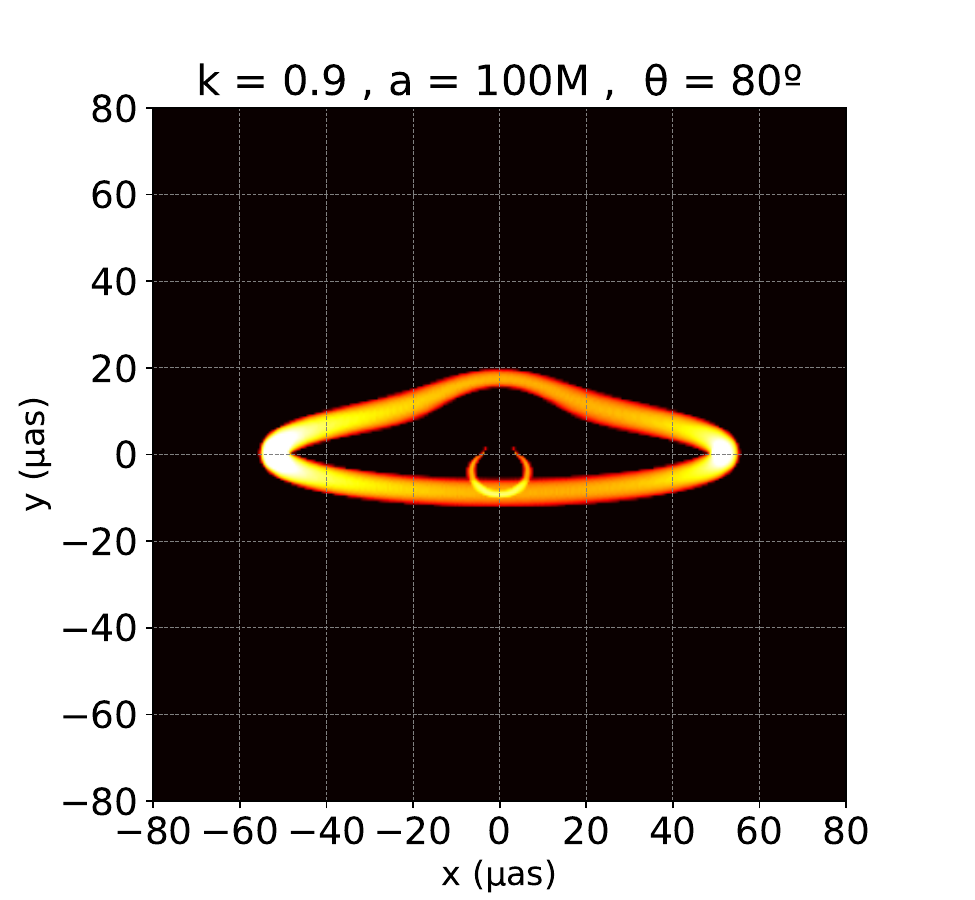}
\caption{Time integrated fluxes $\left<I\right>_{lm}$ for an observation angle of $\theta=80^\circ$ for the DMHBH models with $a=1M$ (top row), $a_0=10M$ (middle row), and $a_0=100M$ (bottom row), and with $k=0.3$ (left column), $k=0.6$ (middle column), and $k=0.9$ (right column).}
\label{fig:flux80deg}
\end{figure*}

In the following, we analyze both the time integrated fluxes and the temporal centroids separately.

\subsection{Time integrated fluxes}

The observed time integrated fluxes for an observation angle of $\theta=20^\circ$ and $\theta=80^\circ$ are given in Figs. \ref{fig:flux20deg} and \ref{fig:flux80deg}, respectively. For low inclination, one observes that the time integrated flux consists solely of two components, a wider primary track corresponding to the trajectory of the primary image in the observer's screen, and a narrower approximately circular component corresponding to the superposition of the secondary and photon ring tracks, associated with the trajectory of the secondary image and location of the critical curves. When the inclination of the observer increases, one observes a flattening of the primary track and a clear separation between the secondary and photon ring tracks, similarly to what is expected from a black hole space-time. 

The impact of the two free parameters of the model, namely $k$ and $a_0$, is fundamentally different. Keeping the parameter $k$ constant and increasing the value of $a_0$ from $a_0=10M$ to $a_0=100M$, one observes a slight, barely noticeable, decrease in the light deflection angles of the primary and secondary tracks, while the radius of the photon ring track remains constant. This happens because the mass of the central black hole is kept constant through this variation in $a_0$, and thus the radius of the photon ring remains unchanged. However, if the value of $a_0$ is smaller, i.e., of order $a_0 = M$, a variation in $a_0$ induces more evident modifications in the integrated fluxes, including large variations in the deflection angles of the primary and secondary images, as well as the appearance of an additional secondary for $k=0.9$. On the other hand, a variation of $k$ while keeping $a_0$ constant leads to strong qualitative changes in the observed time integrated fluxes, especially for large inclinations. Indeed, an increase in $k$ implies that the mass of the black hole decreases and the mass of the DM halo increases. As a consequence, the radius of the photon ring of the configuration decreases, a feature that is clearly observable in the time integrated fluxes as a decrease in the radius of the secondary and photon ring tracks. Furthermore, the lower compactness of the configuration also affects the magnitude of the light deflection, which decreases abruptly for large $a_0$.

\subsection{Temporal centroids and magnitudes}

\begin{figure*}[t!]
\includegraphics[scale=0.28]{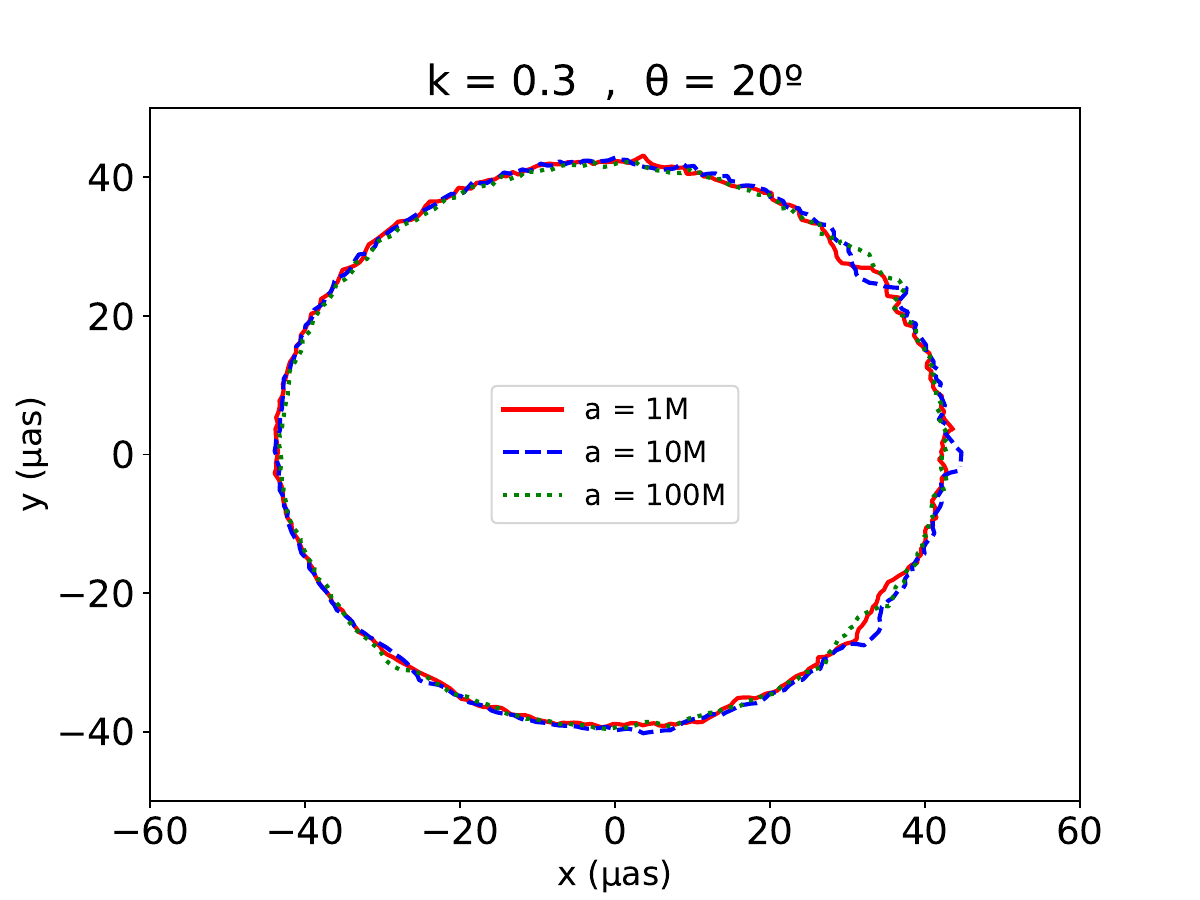}
\includegraphics[scale=0.28]{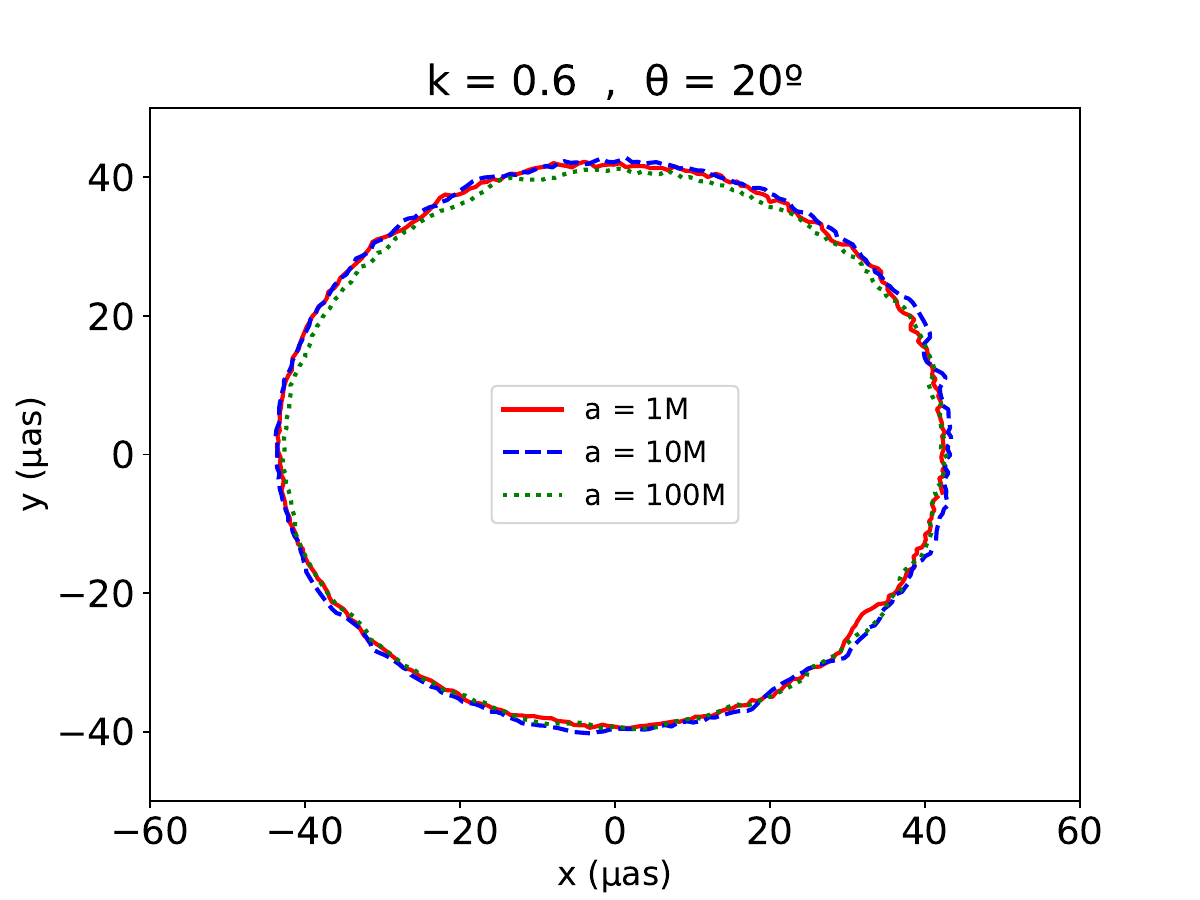}
\includegraphics[scale=0.28]{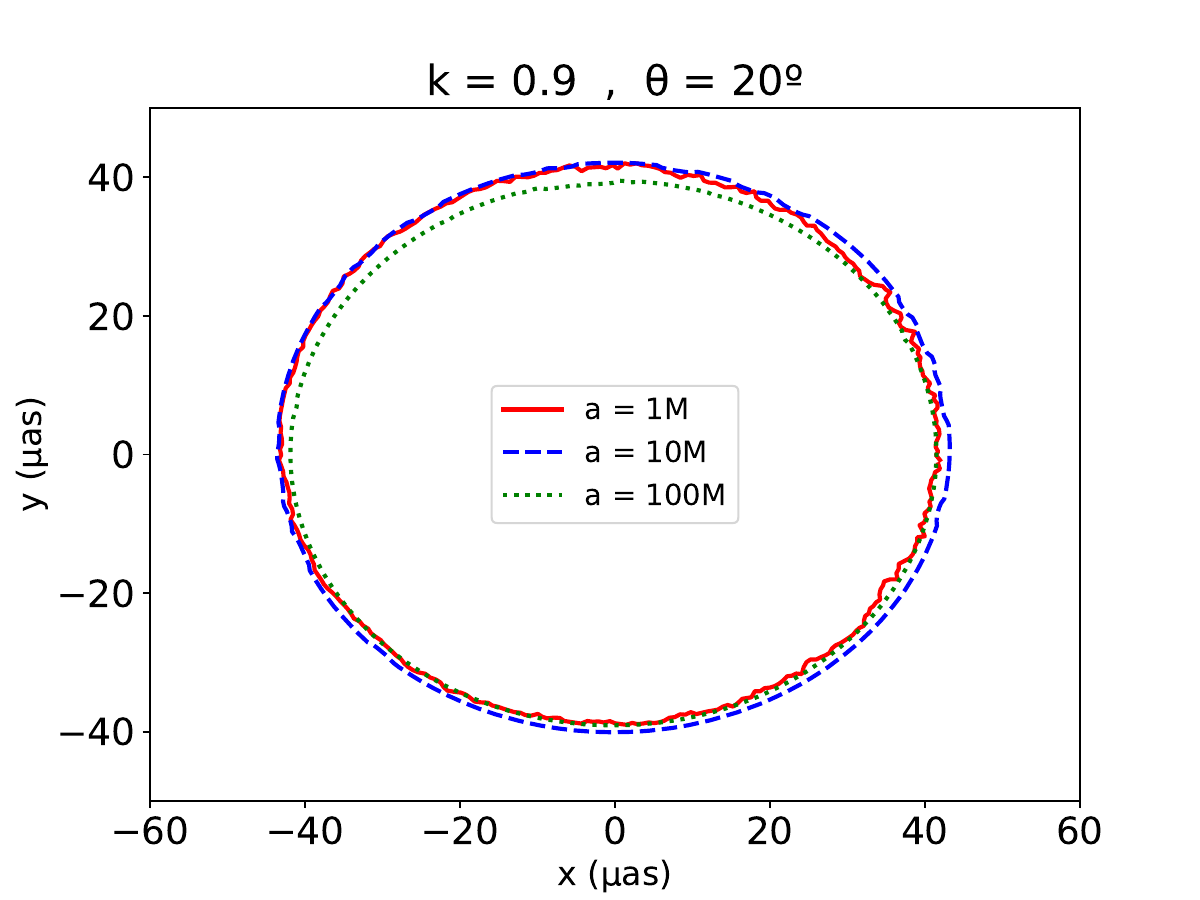}\\
\includegraphics[scale=0.28]{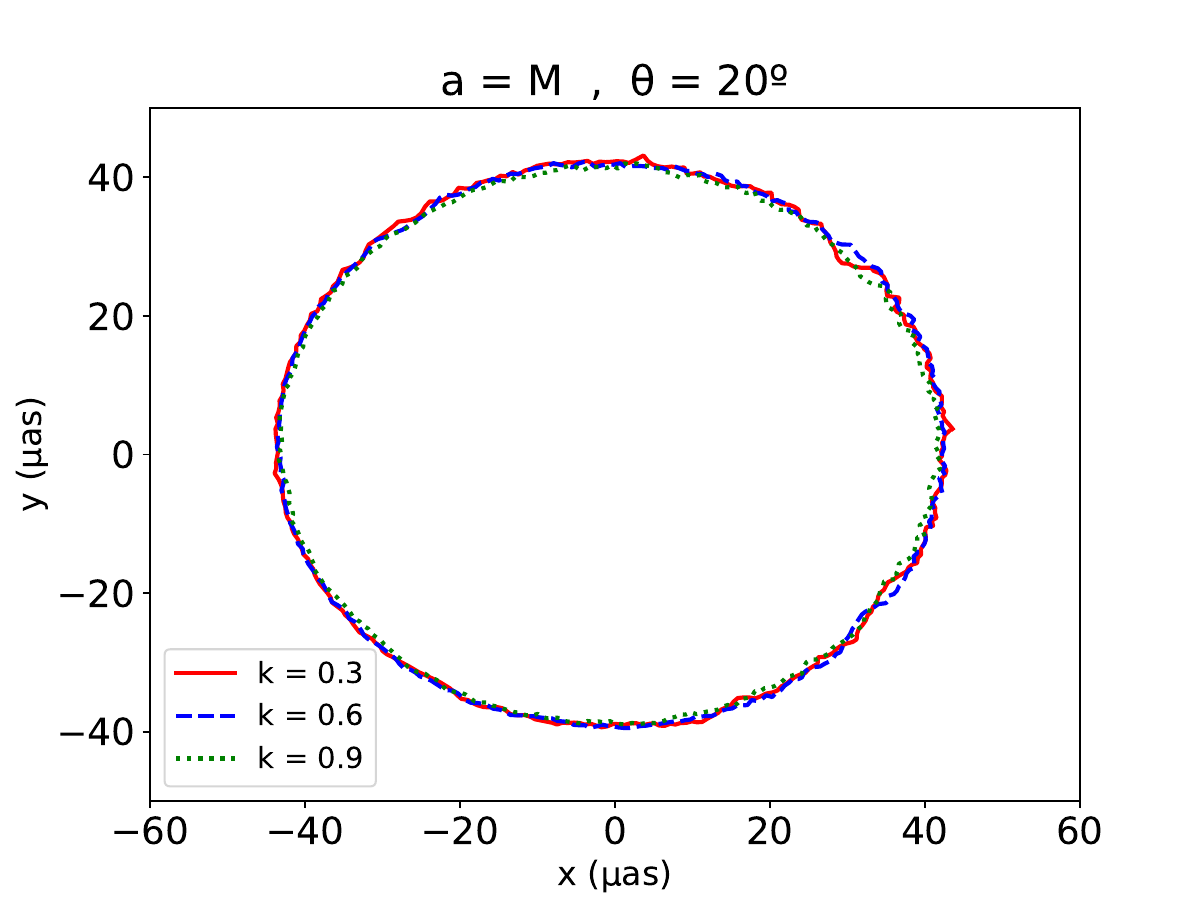}
\includegraphics[scale=0.28]{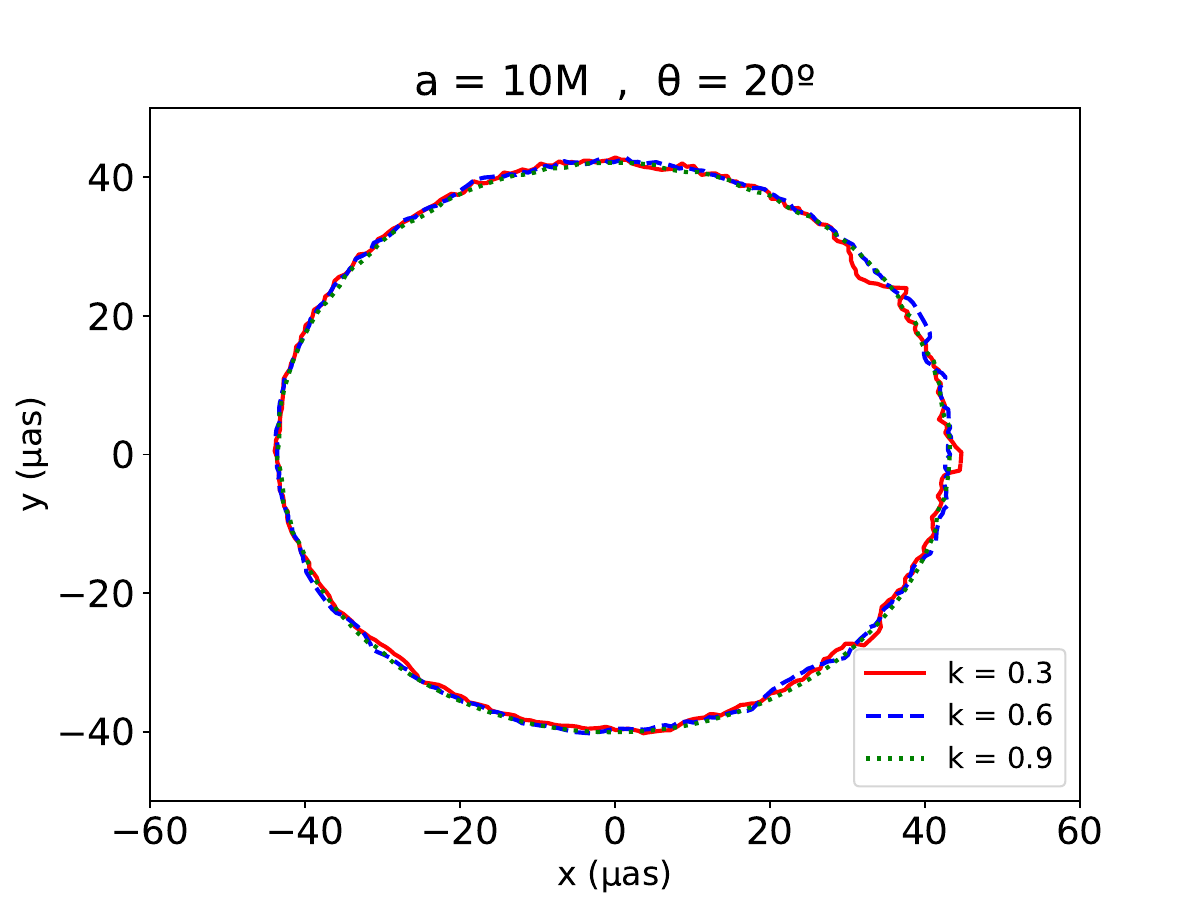}
\includegraphics[scale=0.28]{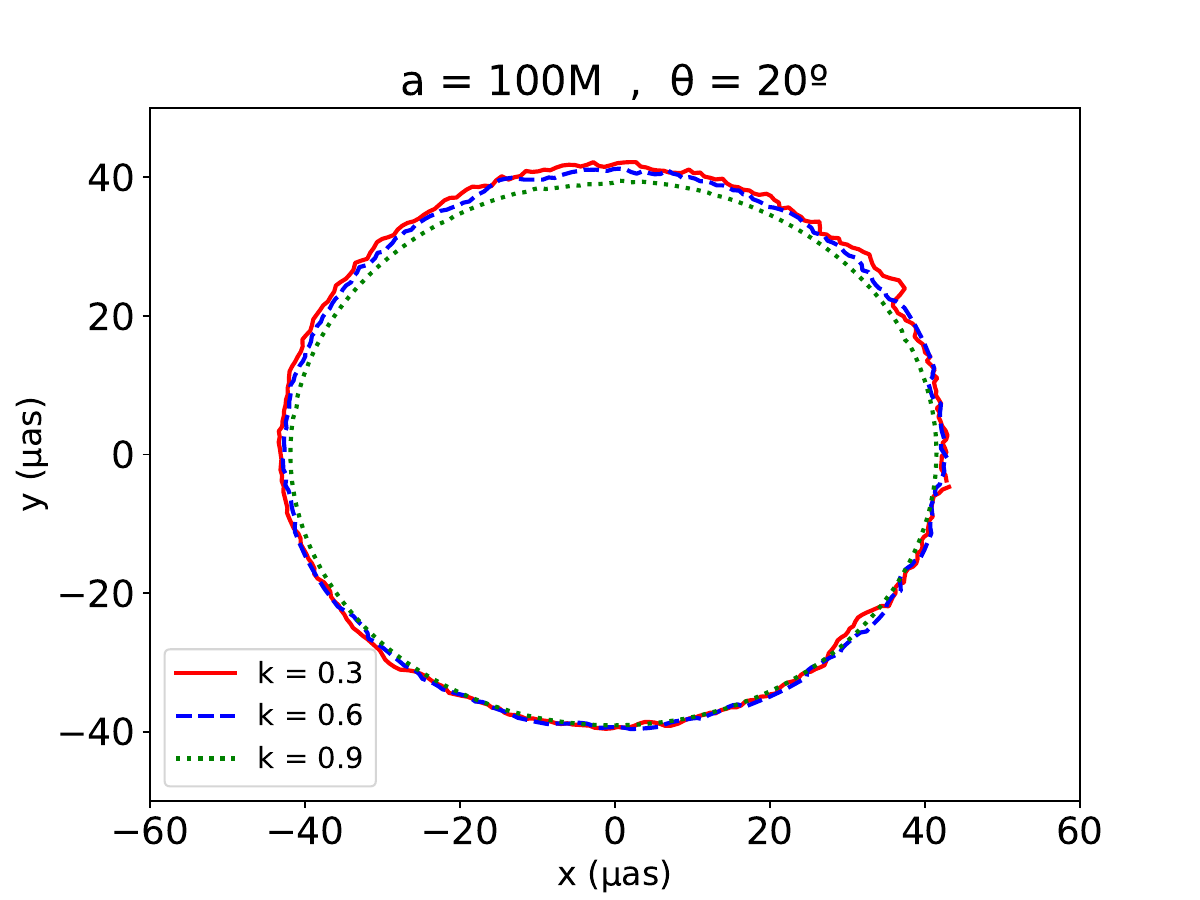}\\
\includegraphics[scale=0.28]{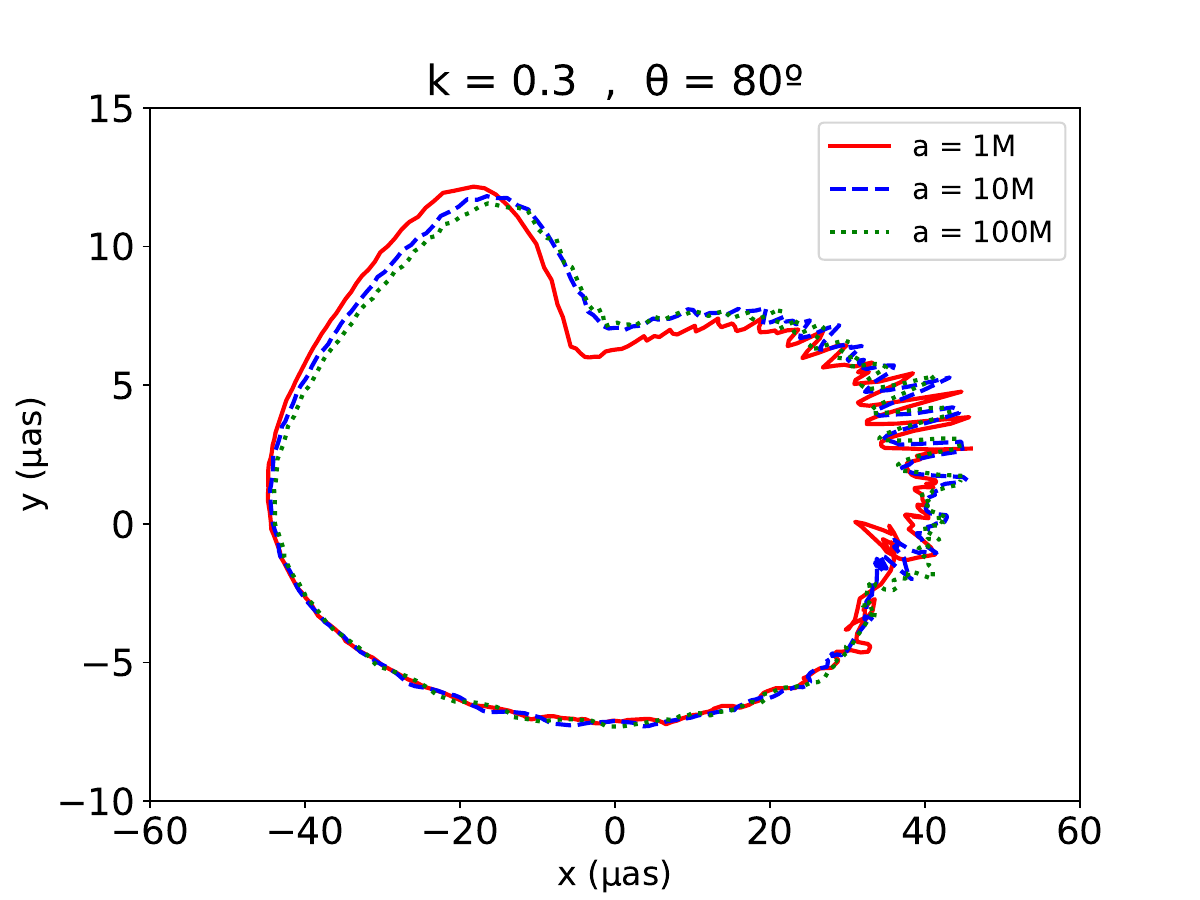}
\includegraphics[scale=0.28]{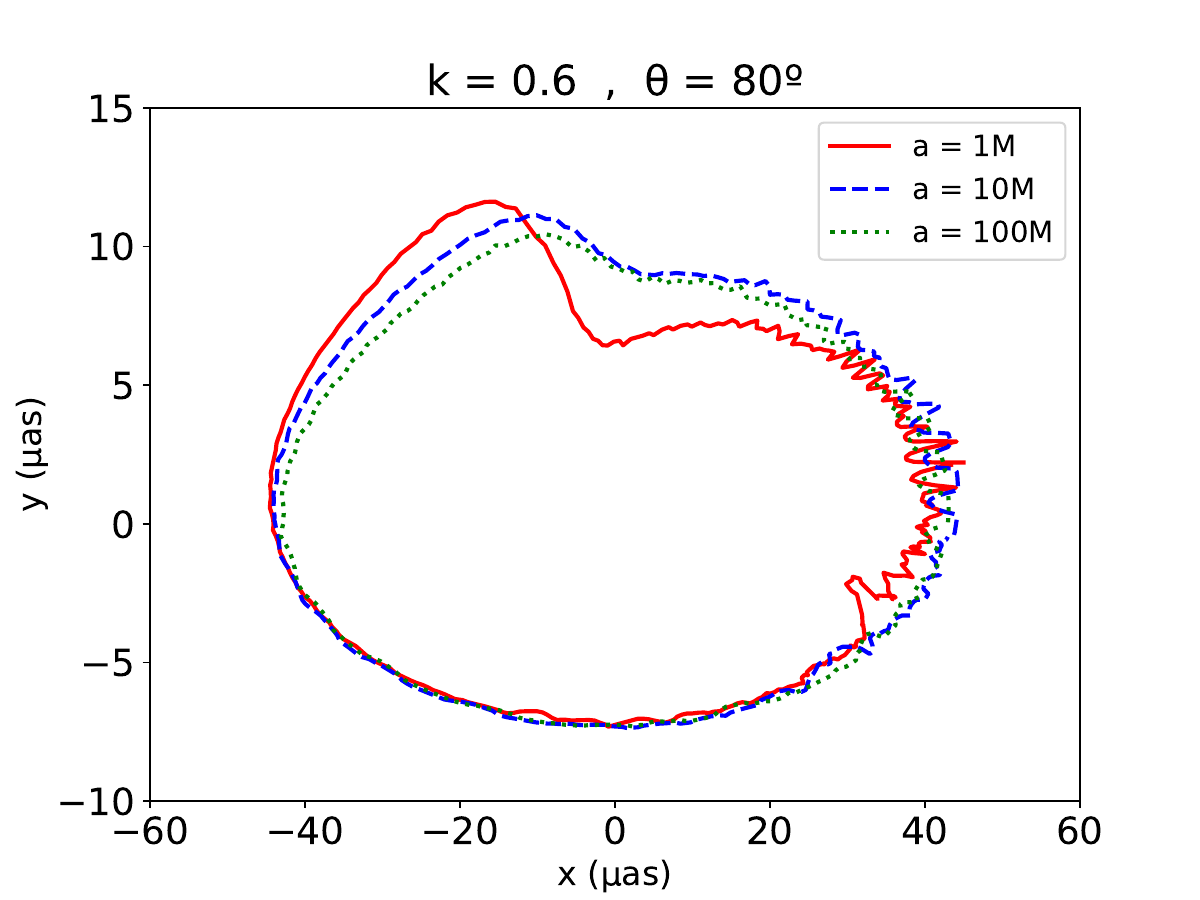}
\includegraphics[scale=0.28]{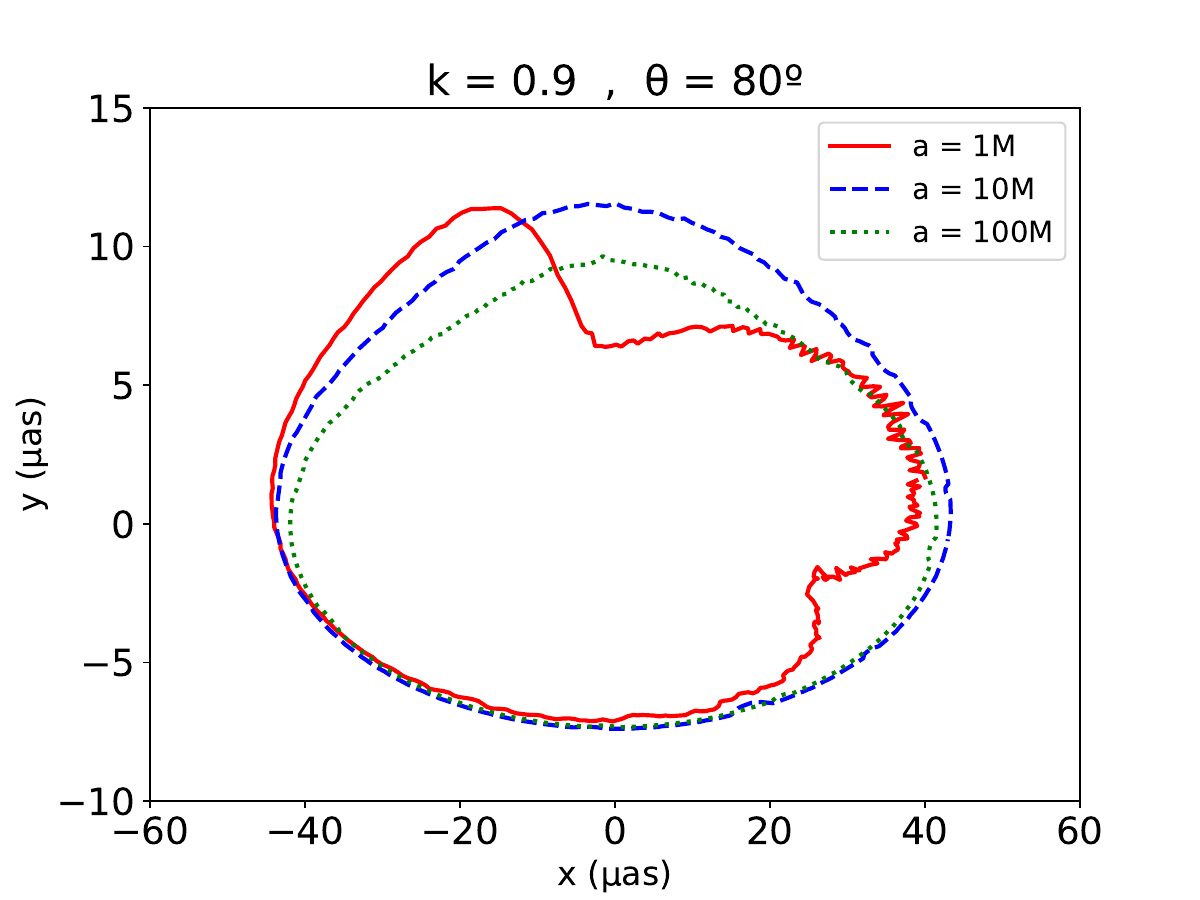}\\
\includegraphics[scale=0.28]{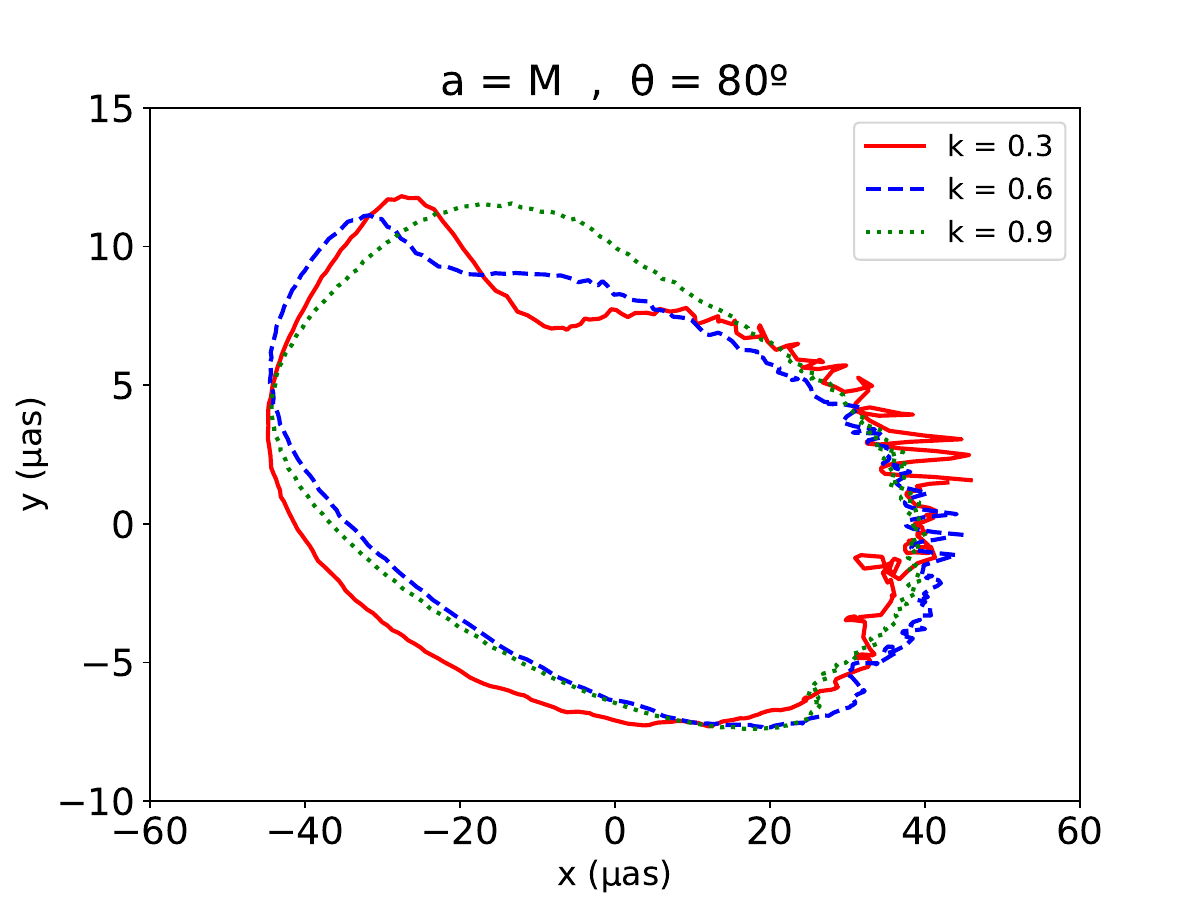}
\includegraphics[scale=0.28]{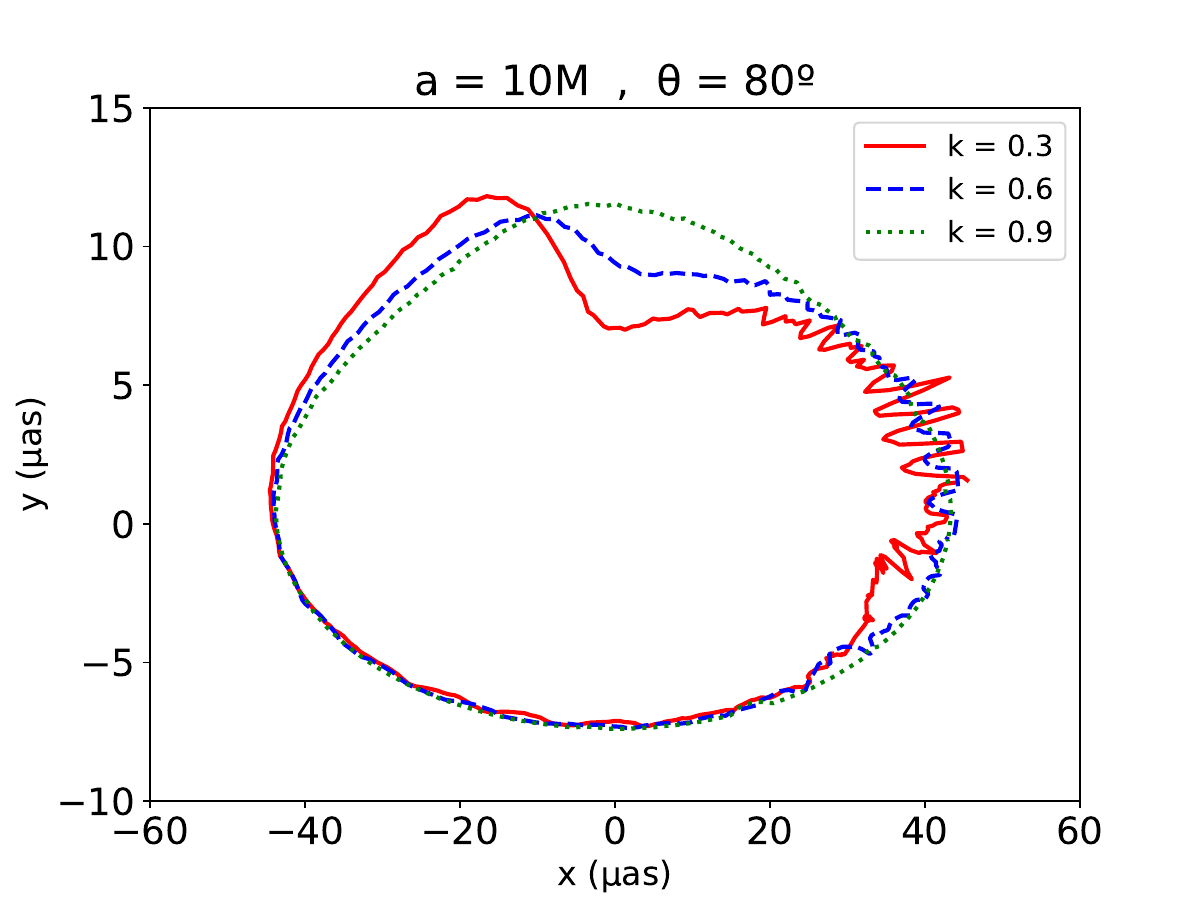}
\includegraphics[scale=0.28]{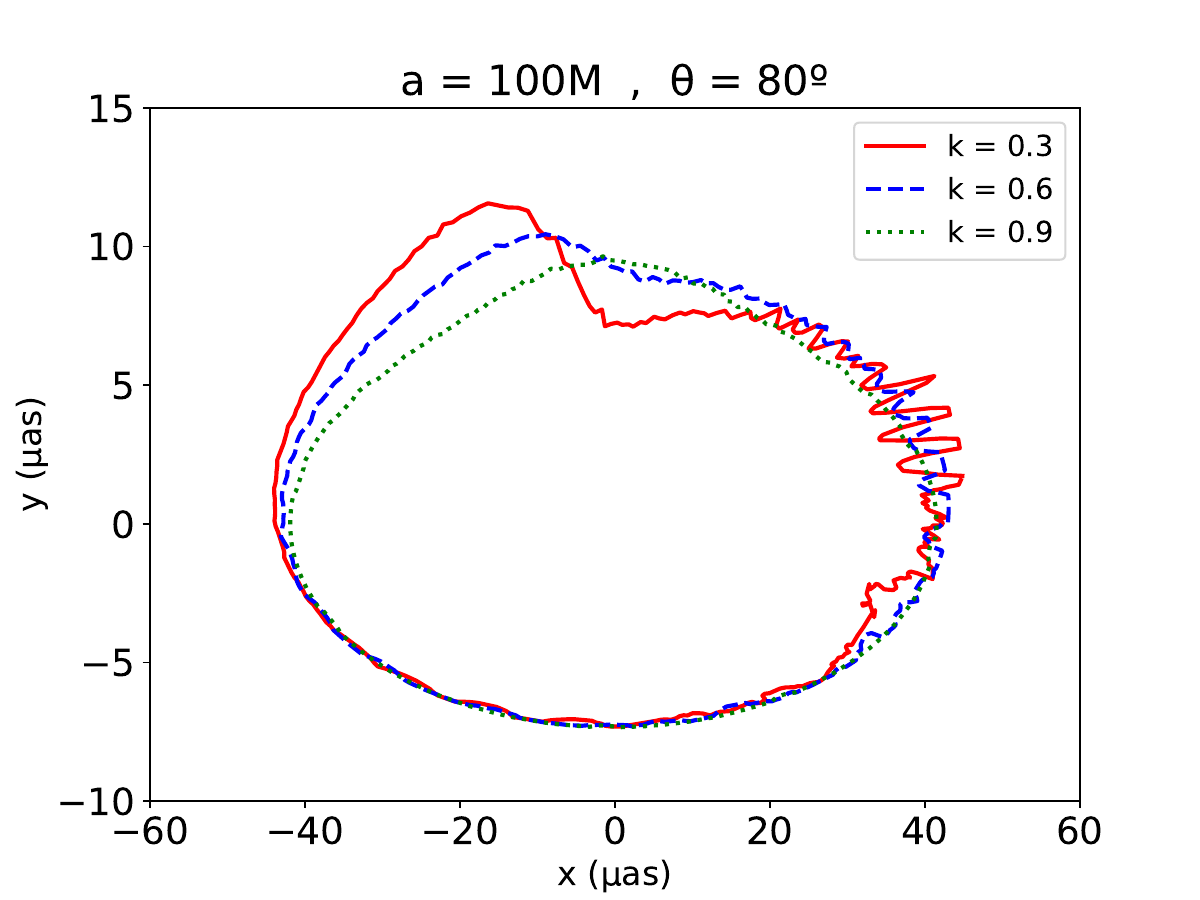}\\
\caption{Temporal centroids $\vec{c}_k$ for an observation angle of $\theta=20^\circ$ (two top rows) and $\theta=80^\circ$ (two bottom rows), for constant values of $k$ and varying $a_0$ (first and third rows), and for constant values of $a_0$ and varying $k$ (second and fourth rows).}
\label{fig:centroid}
\end{figure*}

\begin{figure*}[t!]
\includegraphics[scale=0.28]{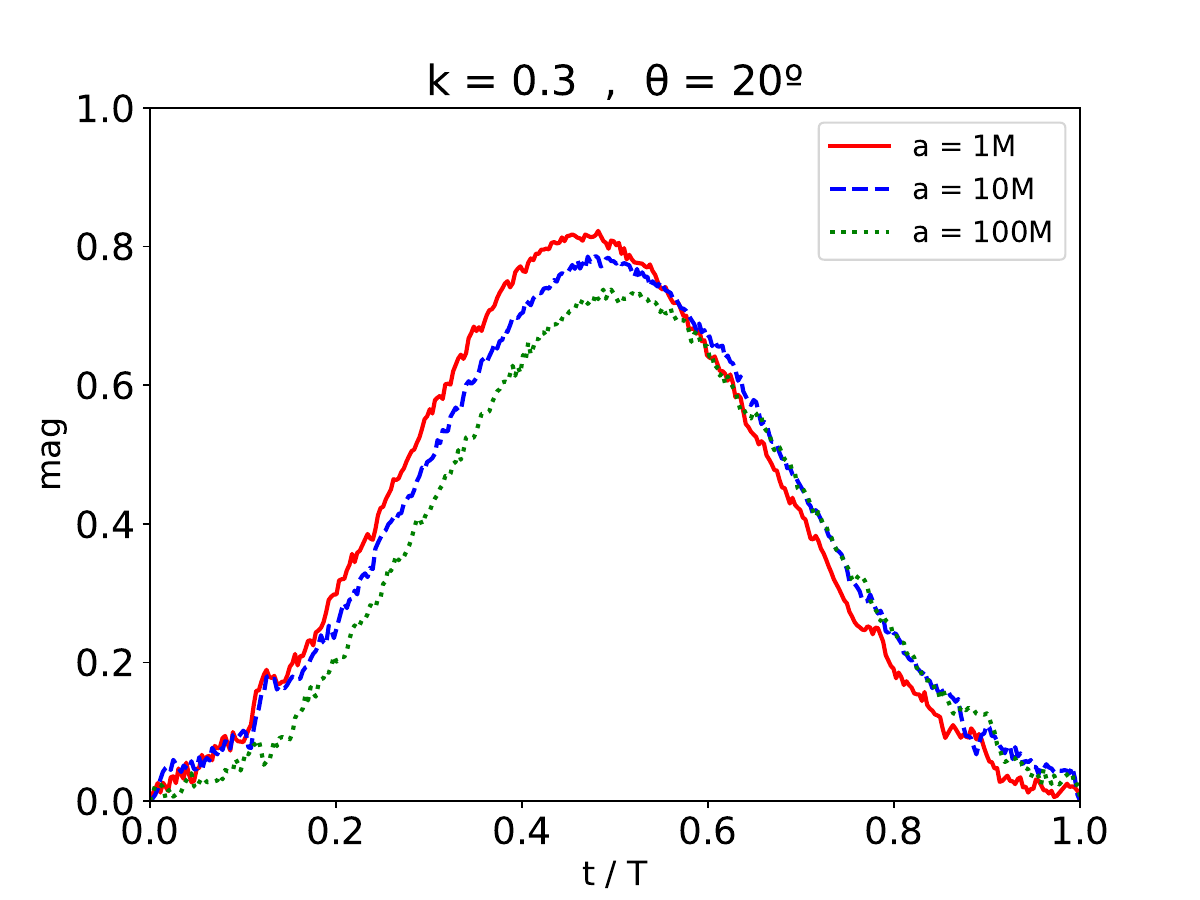}
\includegraphics[scale=0.28]{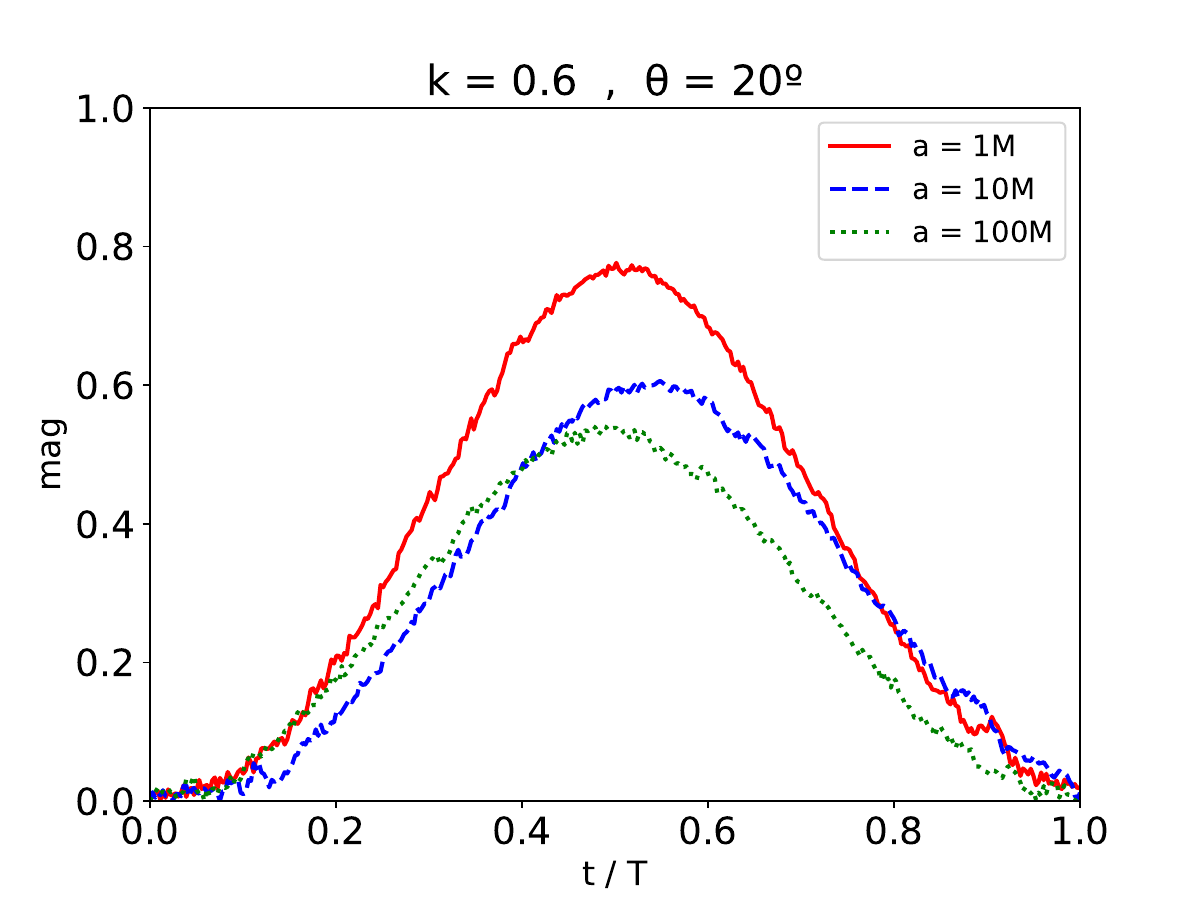}
\includegraphics[scale=0.28]{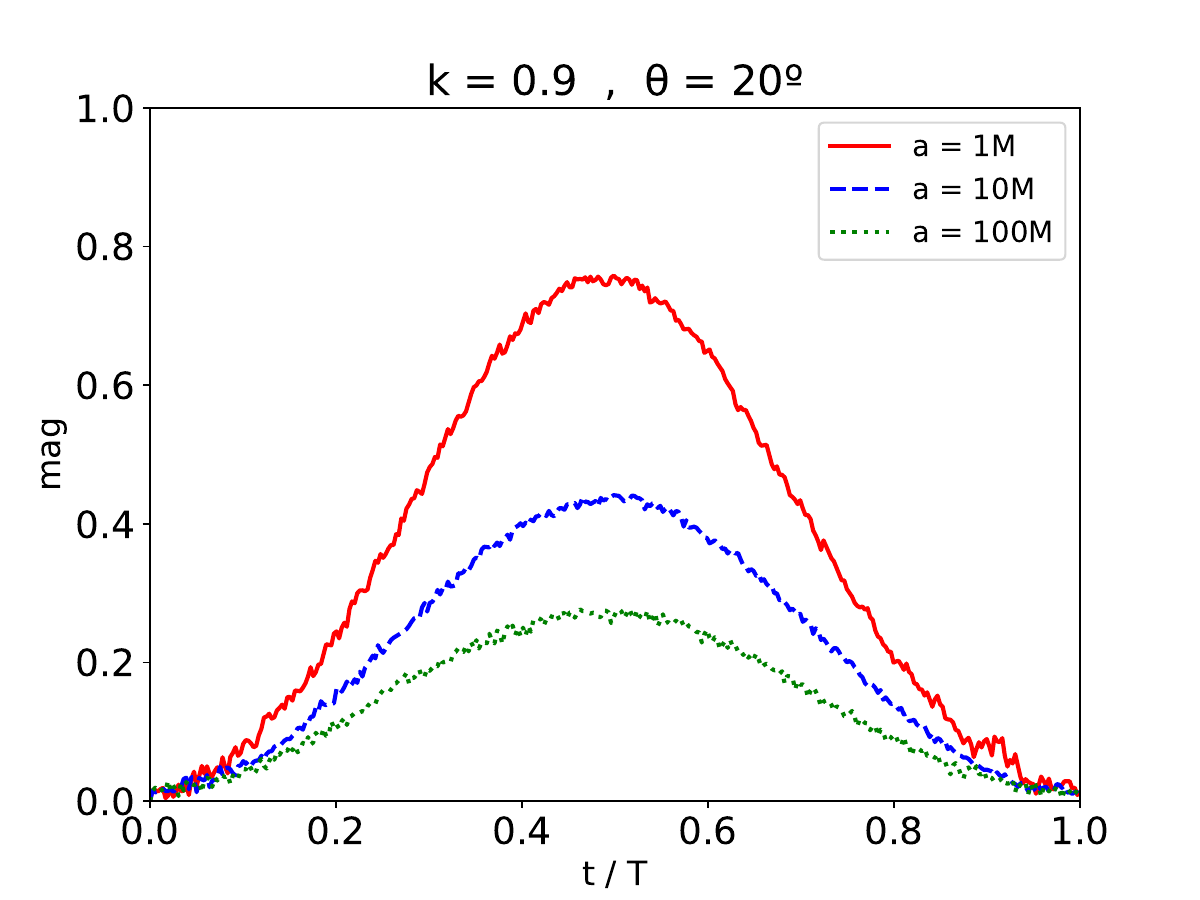}\\
\includegraphics[scale=0.28]{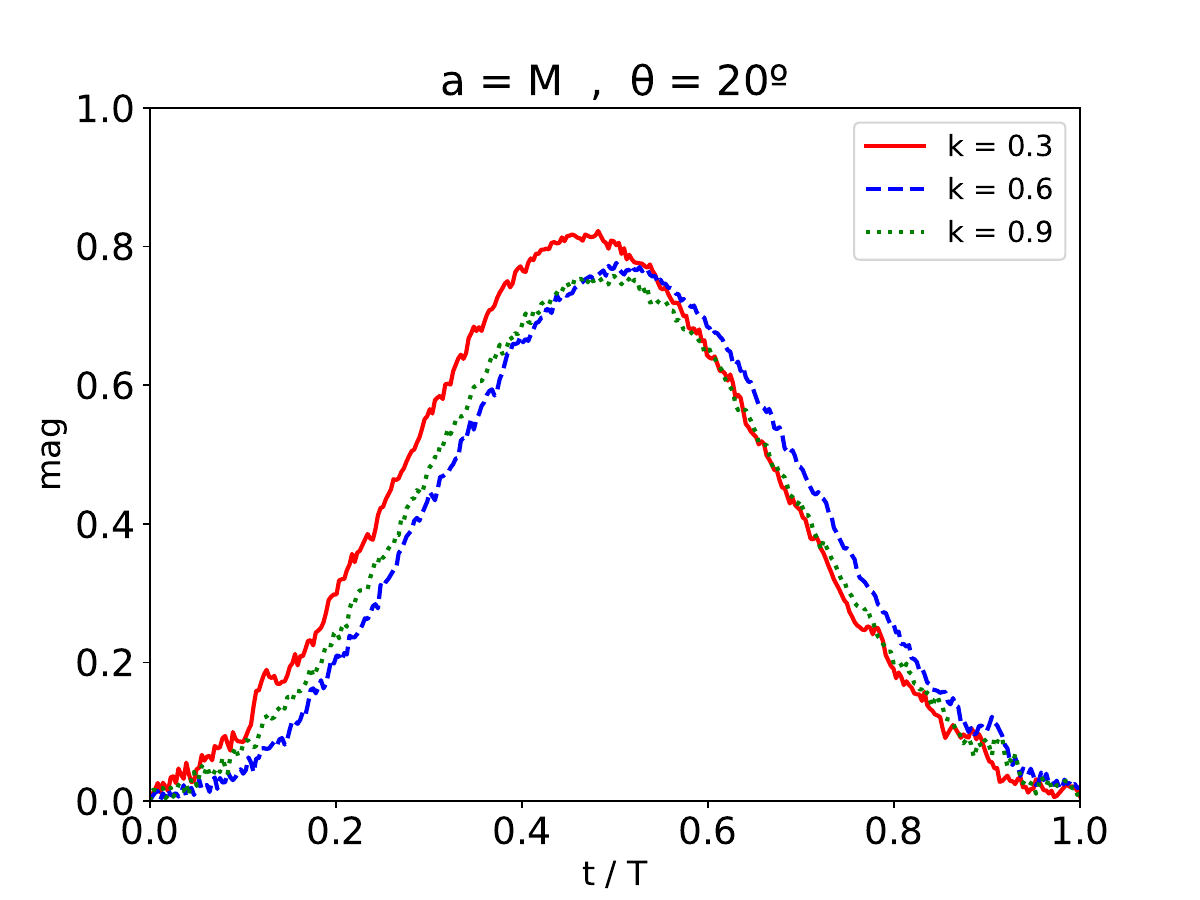}
\includegraphics[scale=0.28]{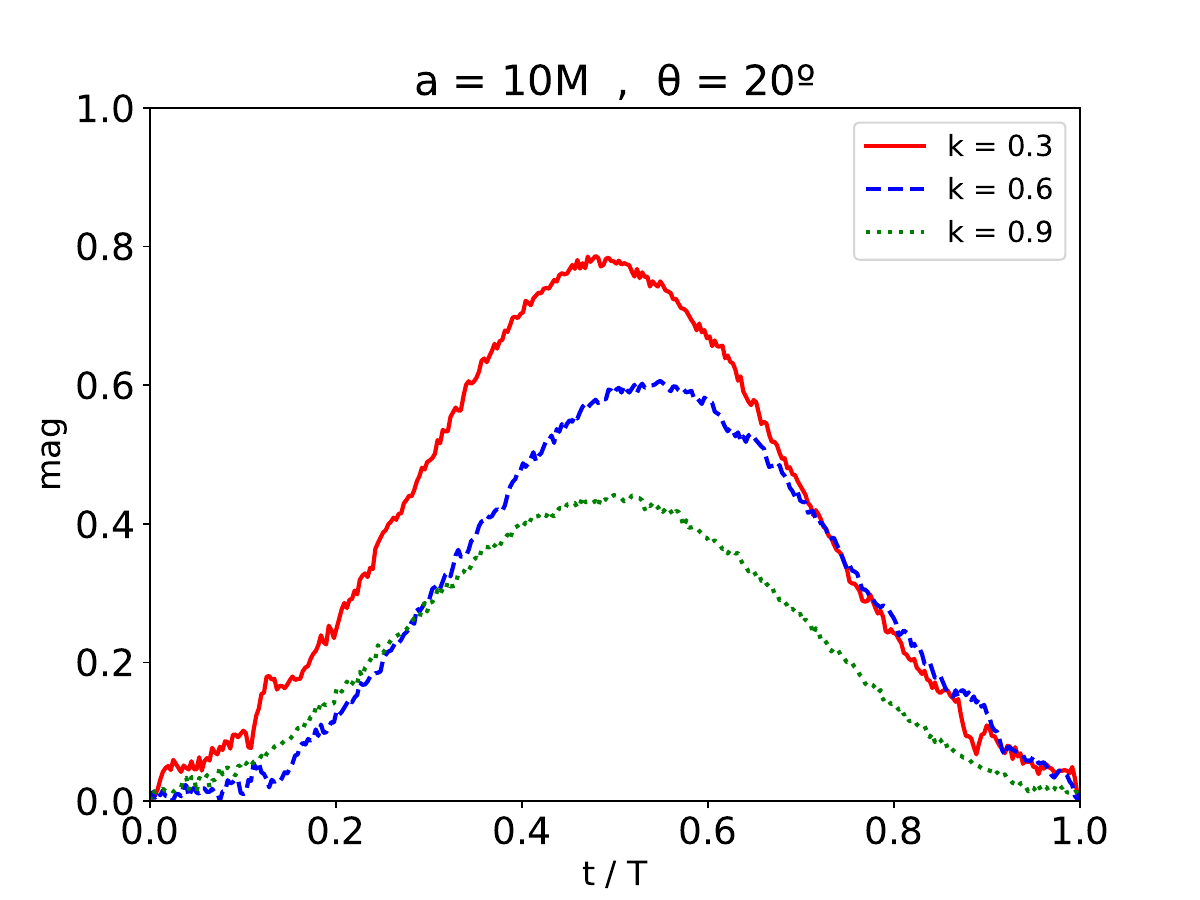}
\includegraphics[scale=0.28]{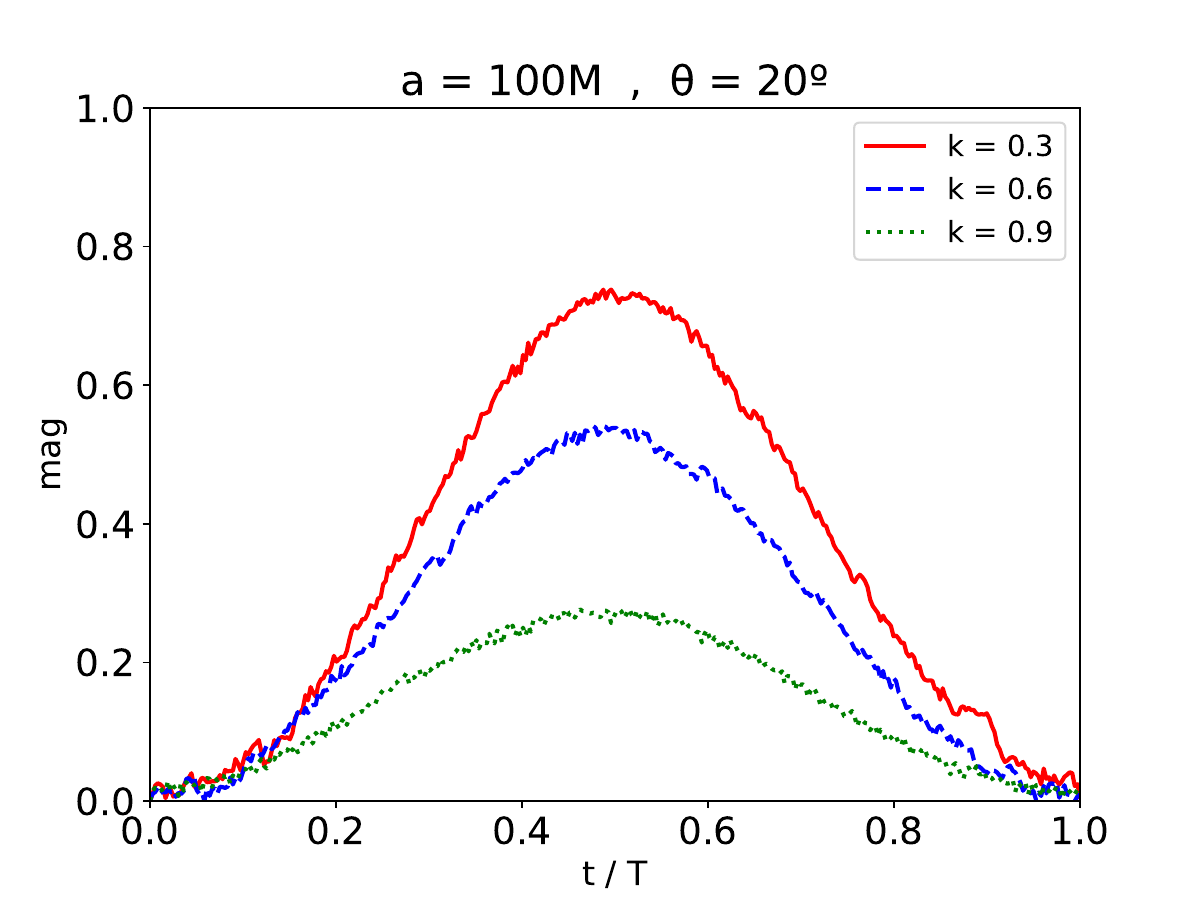}\\
\includegraphics[scale=0.28]{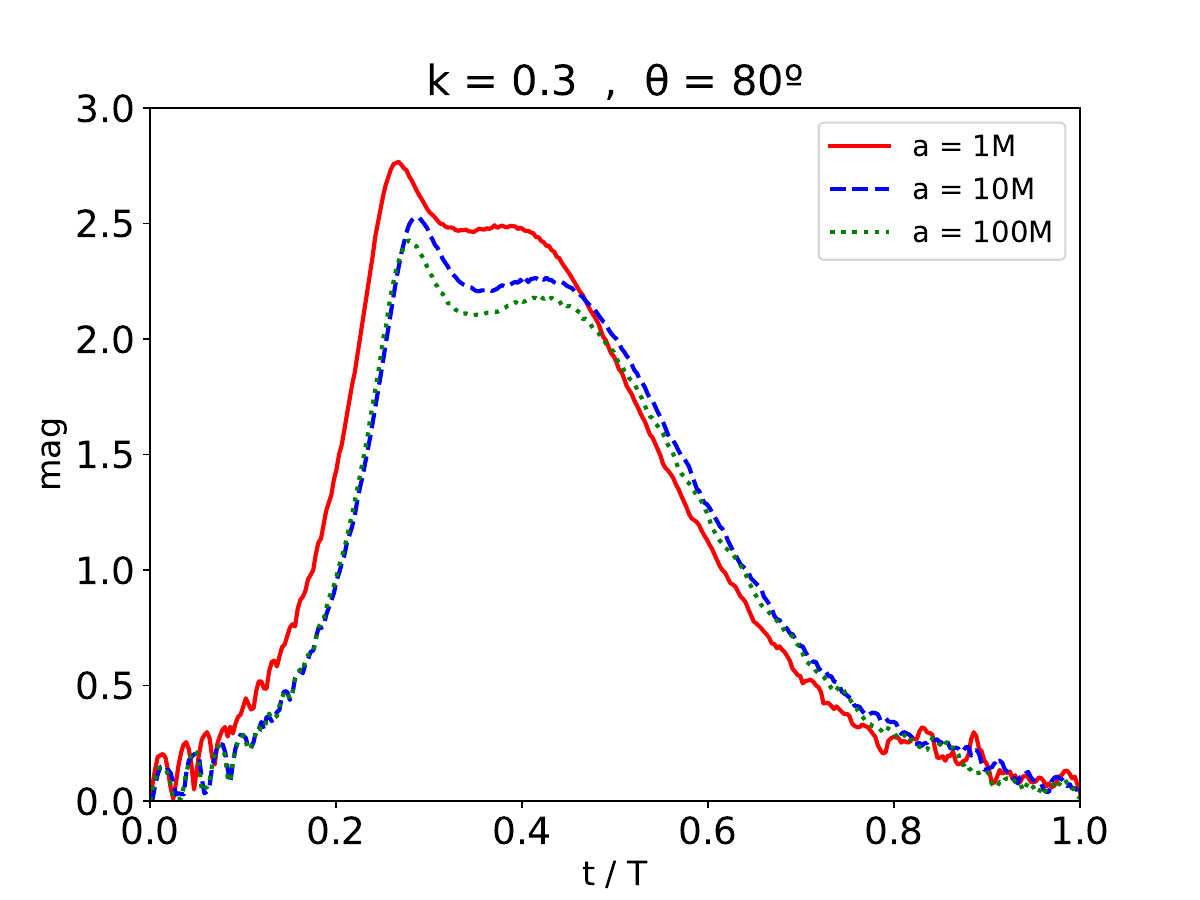}
\includegraphics[scale=0.28]{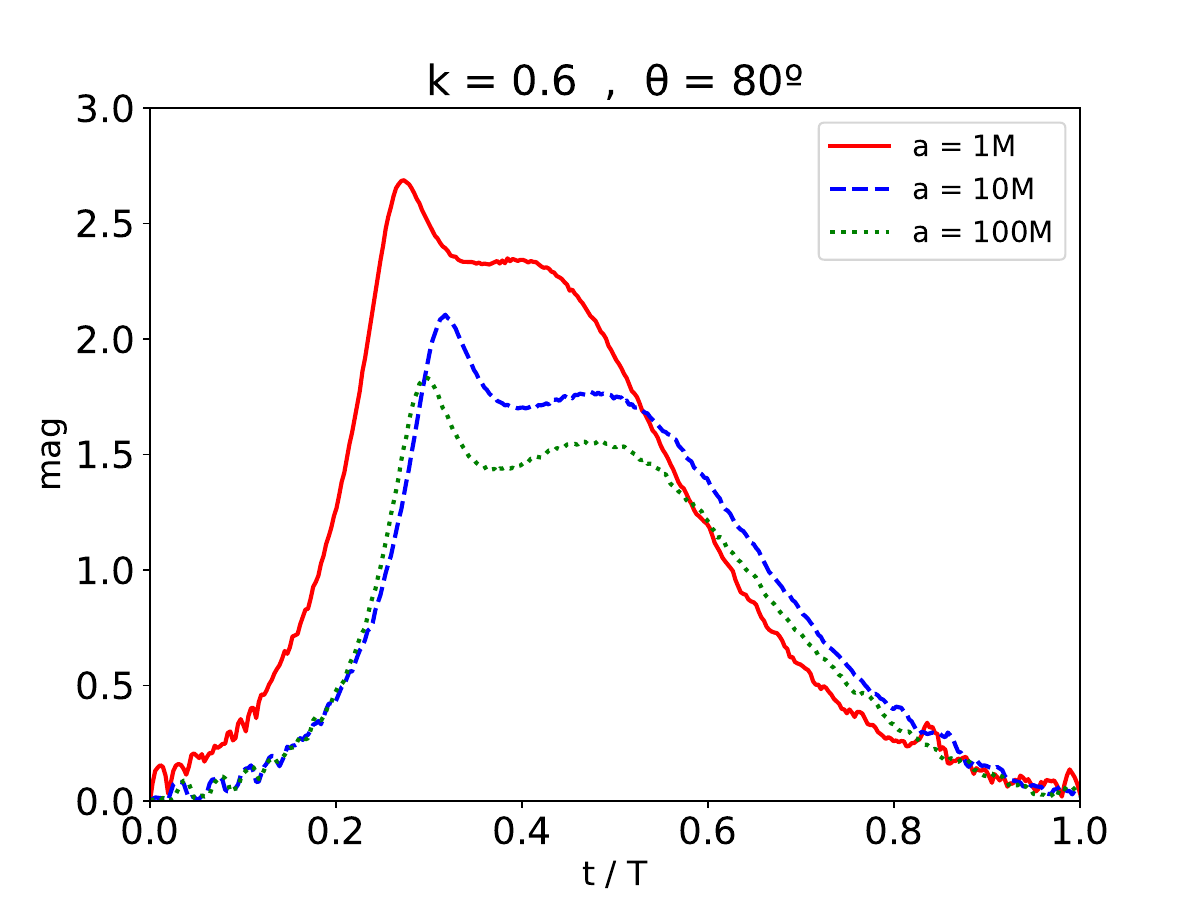}
\includegraphics[scale=0.28]{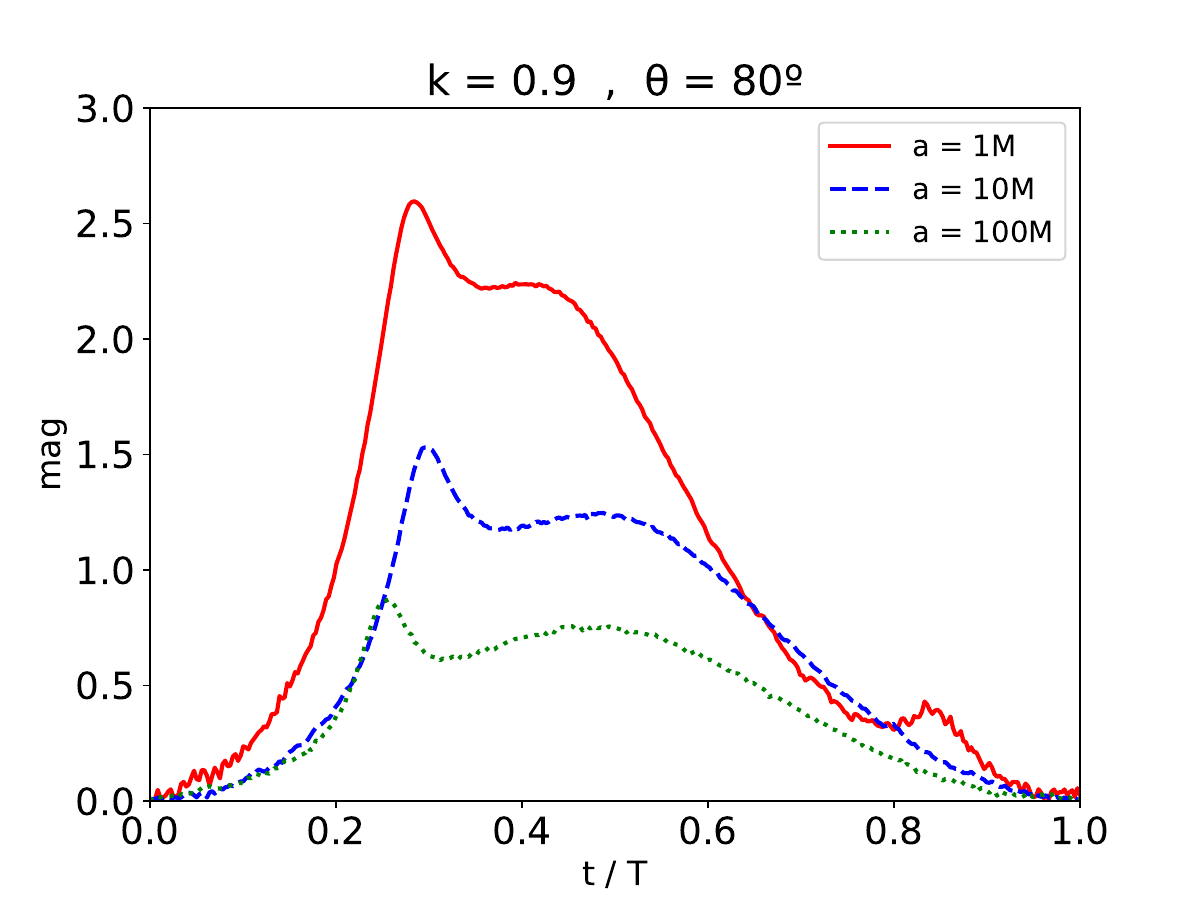}\\
\includegraphics[scale=0.28]{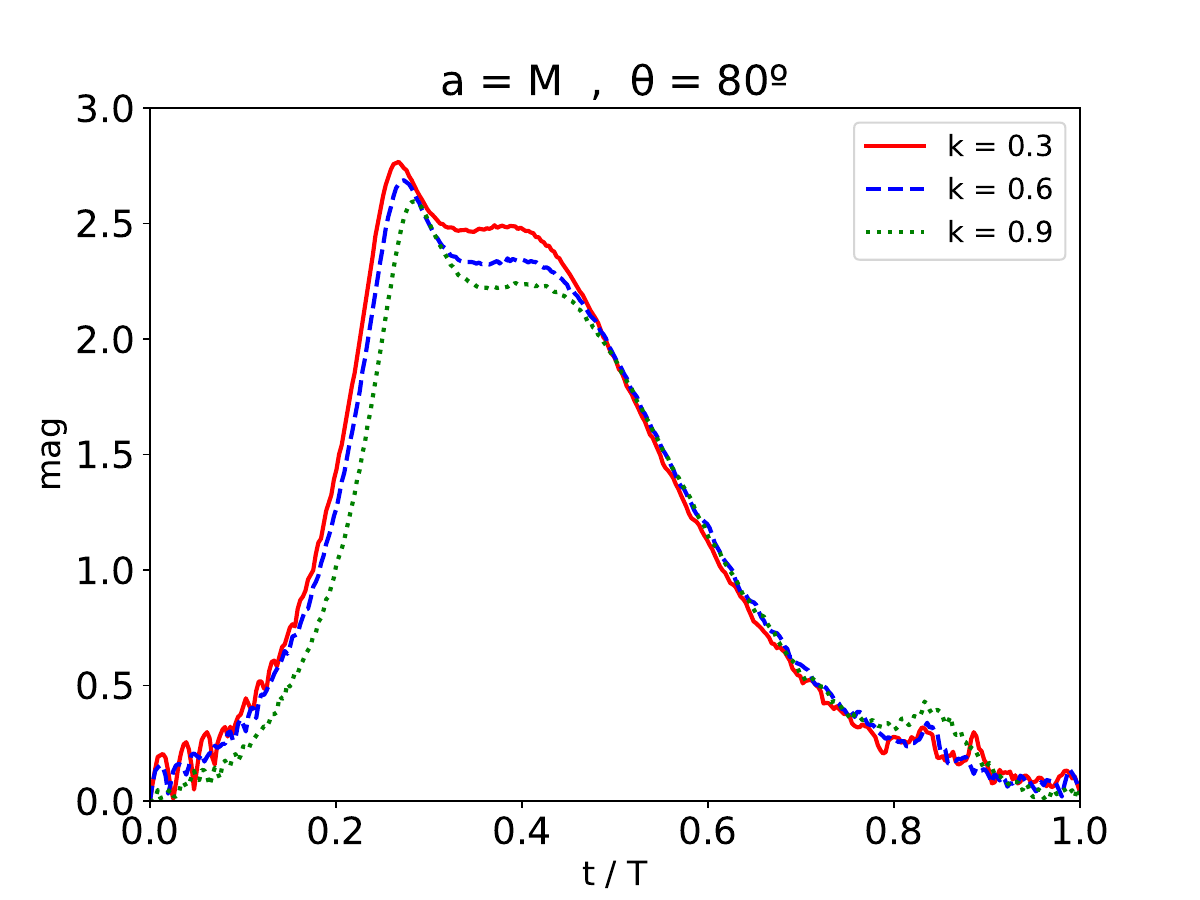}
\includegraphics[scale=0.28]{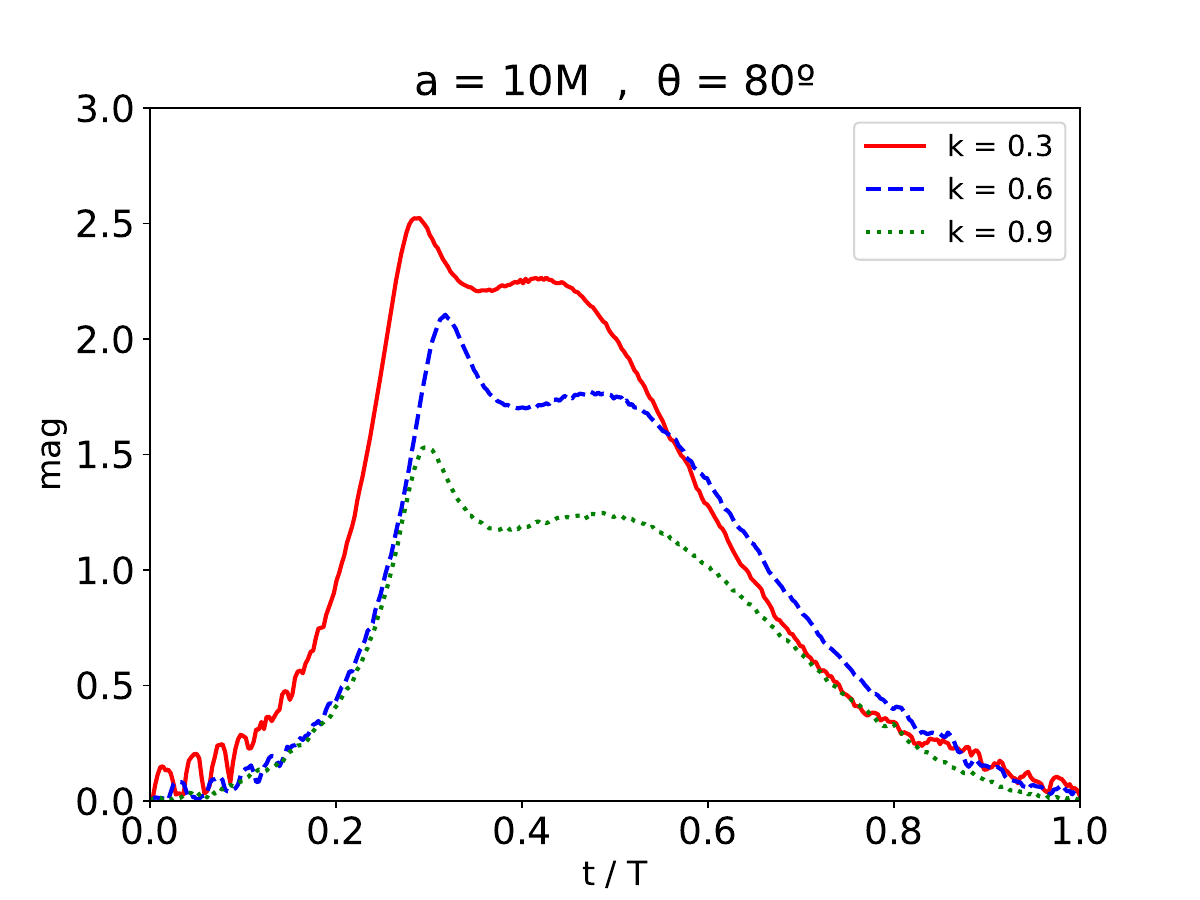}
\includegraphics[scale=0.28]{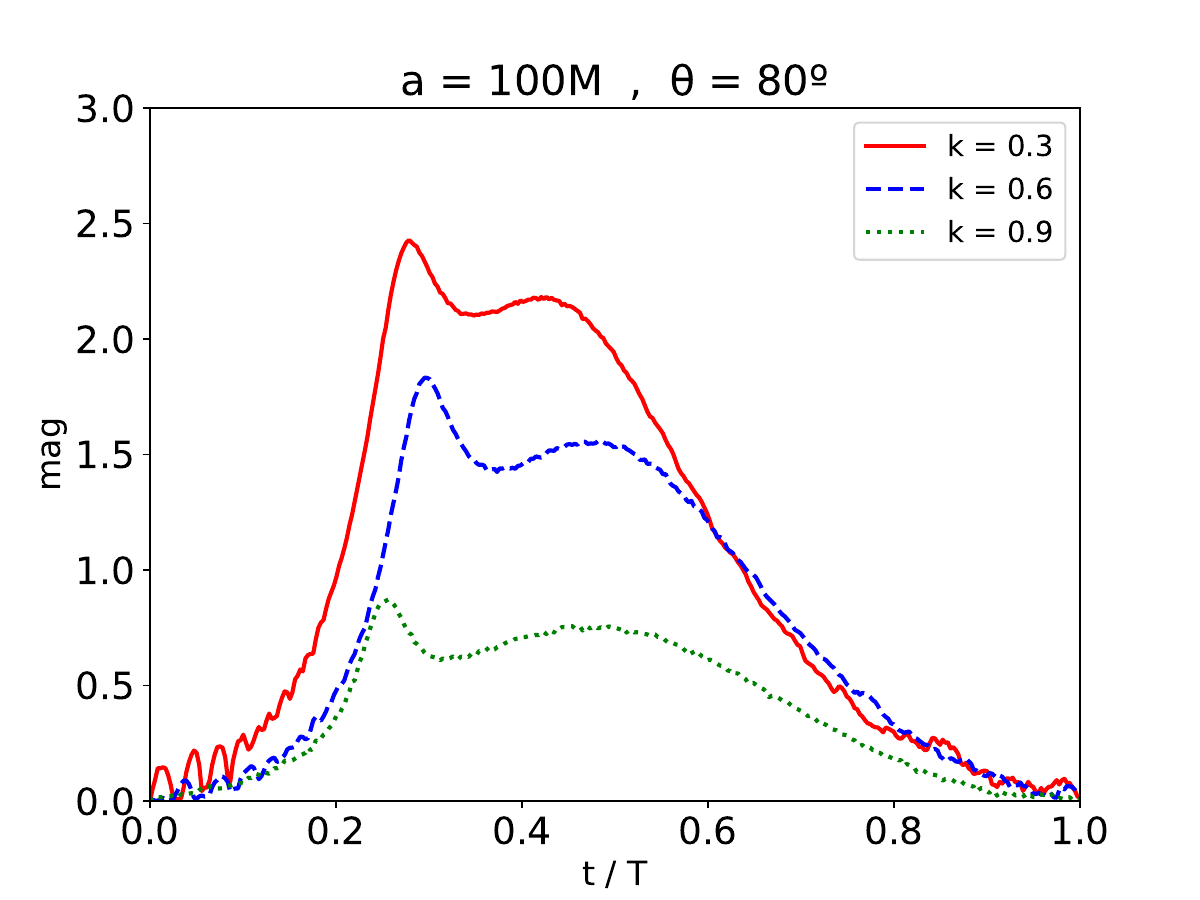}\\
\caption{Temporal magnitudes $m_k$ for an observation angle of $\theta=20^\circ$ (two top rows) and $\theta=80^\circ$ (two bottom rows), for constant values of $k$ and varying $a_0$ (first and third rows), and for constant values of $a_0$ and varying $k$ (second and fourth rows).}
\label{fig:magnitude}
\end{figure*}

The temporal centroids and magnitudes are plotted in Figs. \ref{fig:centroid} and \ref{fig:magnitude}. Similarly to what is observed in the time integrated flux, the effects of the two free parameters of the model are sharper for higher observation inclinations. Indeed, for low inclination, the centroids always follow a slightly distorted ellipse and the magnitudes always follow a single Doppler-shift induced peak, independently of the values of the parameters $k$ and $a_0$, where small variations of these parameters induce slight quantitative variations on the behaviour of these two observables without inducing any qualitative changes. It is, however, noticeable that an increase in both $k$ and $a_0$ slightly decrease the radius of the centroid trajectory, caused by a decrease in the light deflection angles for these variations. As for the magnitude, one observes that an increase in $ a_0$ also induces slight quantitative variations in the height of the Doppler peak, which are particularly noticeable for larger values of both $k$ and $a_0$. 

For larger inclination angles, one additional observational imprint is visible in both the temporal centroid track and the magnitude. Indeed, for small values of $k$ corresponding to the models where most of the mass is concentrated in the central black hole, one observes that the centroid tracks suffer a shifting to the center of the observation, caused by the appearance of a large secondary image. This effect is visible independently of the value of $a_0$. As the value of $k$ increases, which corresponds to a decrease in the size of the secondary and photon ring tracks as observed in the time integrated fluxes, the shifting effect of the centroid becomes smaller, eventually becoming unnoticeable for large values of $k$ and $a_0$. As for the magnitude, the presence of a strong secondary image in the observer's screen, no matter for how short of a time, always induces an additional peak of brightness on top of the primary Doppler-induced peak. Thus, the only noticeable changes with parameter variations correspond to quantitative changes, i.e., the height of the peaks which, similarly to what happens for low inclination, decrease slightly with an increase in both $a_0$ and $k$. 

\subsection{High-compactness regime}

The results obtained for the integrated fluxes, temporal centroids, and temporal magnitudes for the high-compactness configuration considered with $k=0.9375$ and $\bar a_0=0.620525$. For this configuration, an additional pair of photons spheres and MSOs induces by the high compactness of the DM halo arise in the space-time, see Figs. \ref{fig:ddpot} and \ref{fig:potential}. Even though the mass of the black hole in this configuration is small compared to the remaining scenarios analyzed, $M_{\rm BH}=M/16$, one observes that the observational properties are more closely related to those of the $k=0.3$ and $k=0.6$ scenarios, and not with the $k=0.9$ scenario. This is particularly visible for large inclinations, for which the integrated flux closely resembles the ones on the left panels of Fig. \ref{fig:flux80deg}, the centroid tracks closely resemble the ones for $k=0.3$ in the bottom row of Fig. \ref{fig:centroid} (especially for $\bar a_0\geq 10$, and the peaks of magnitude attained reach the same values as those for the left panels of Fig. \ref{fig:magnitude}. Thus, one verifies that if the compactness of the DM halo is large enough, the observational properties of the DMHBH model closely match those of models for which the black hole dominates the mass ratio of the system.

\begin{figure*}[t!]
\includegraphics[scale=0.35]{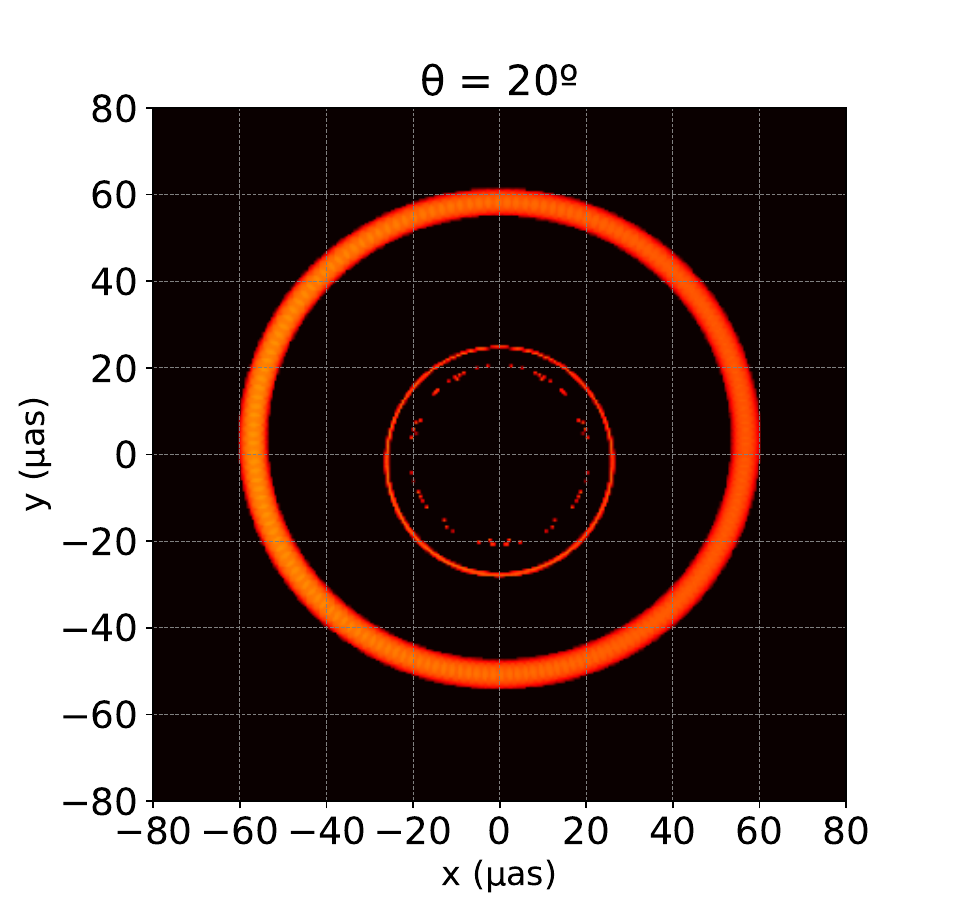}
\includegraphics[scale=0.28]{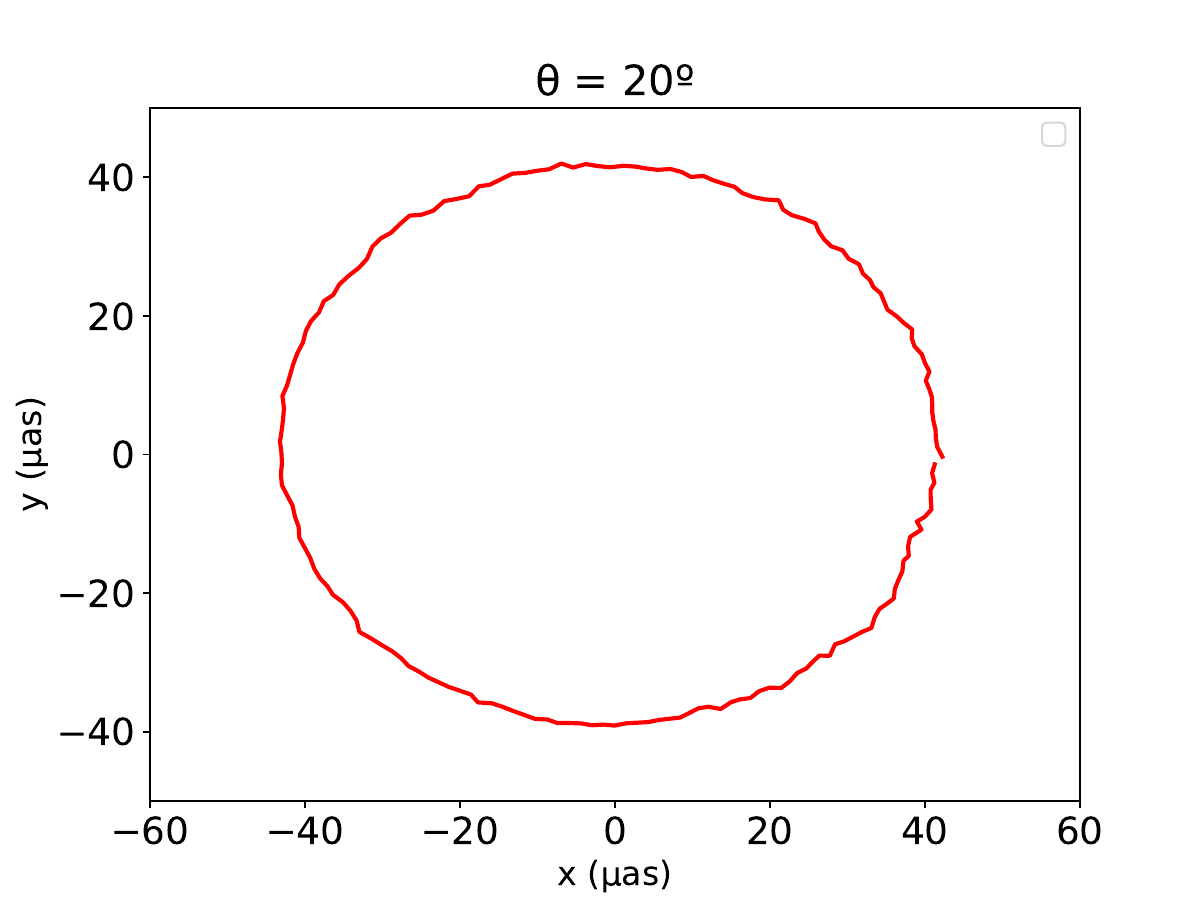}
\includegraphics[scale=0.28]{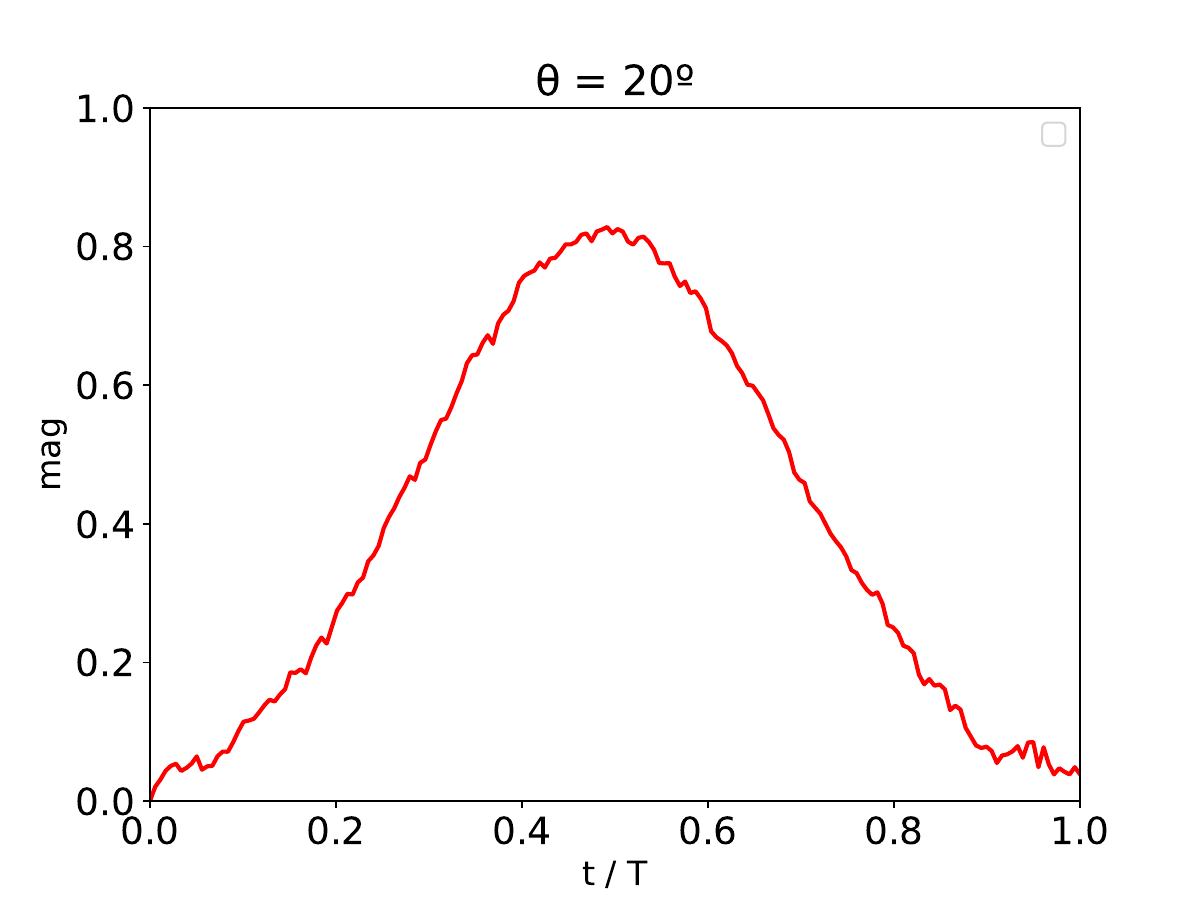}\\
\includegraphics[scale=0.35]{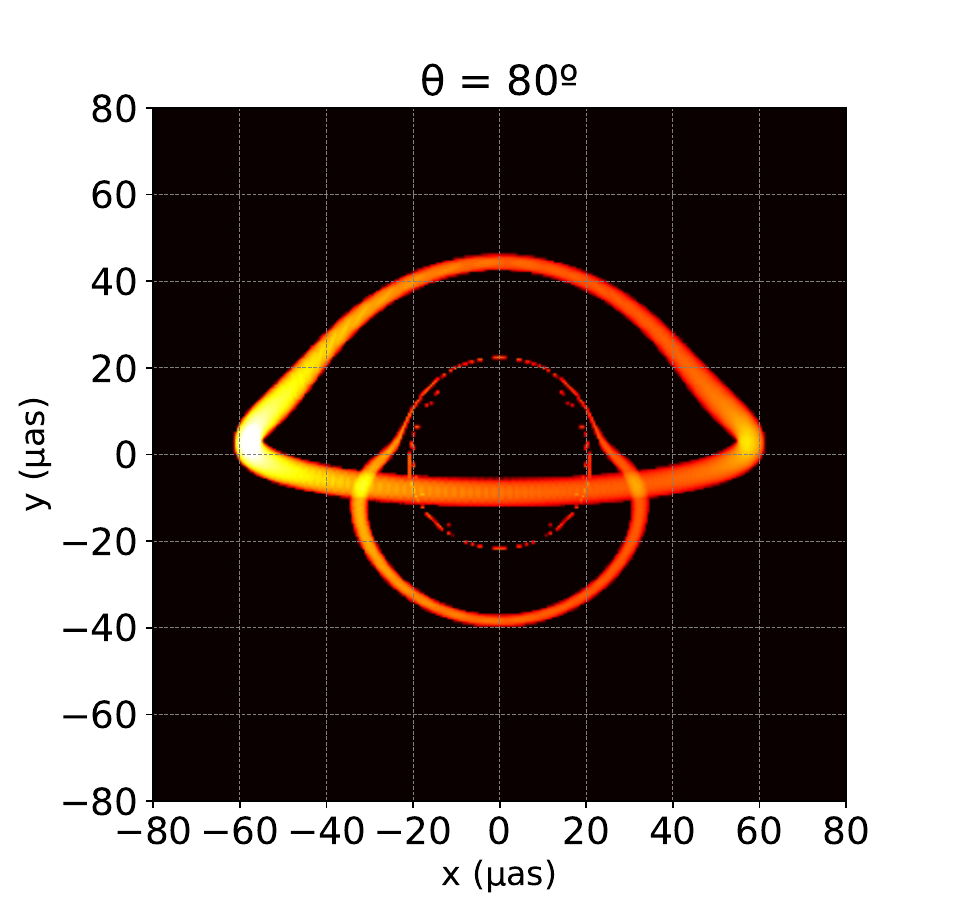}
\includegraphics[scale=0.28]{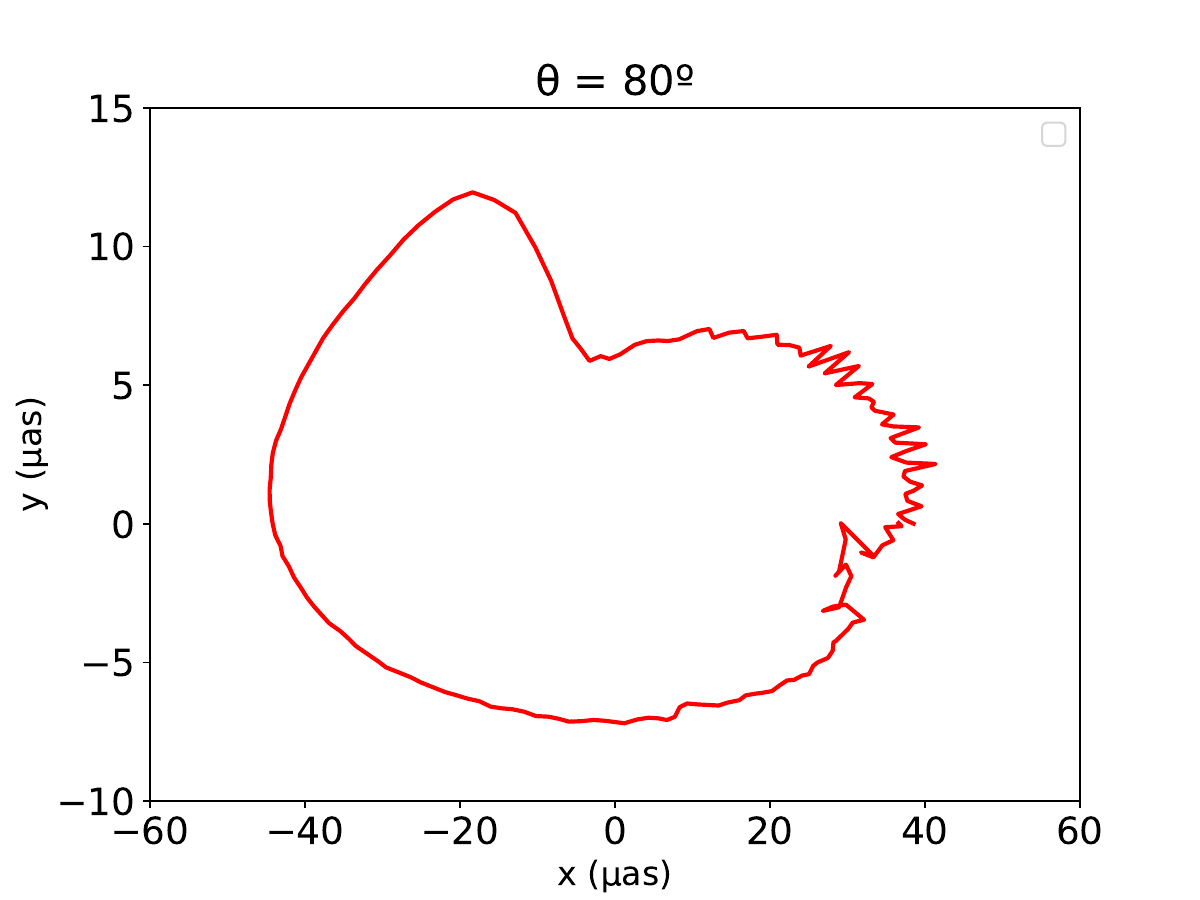}
\includegraphics[scale=0.28]{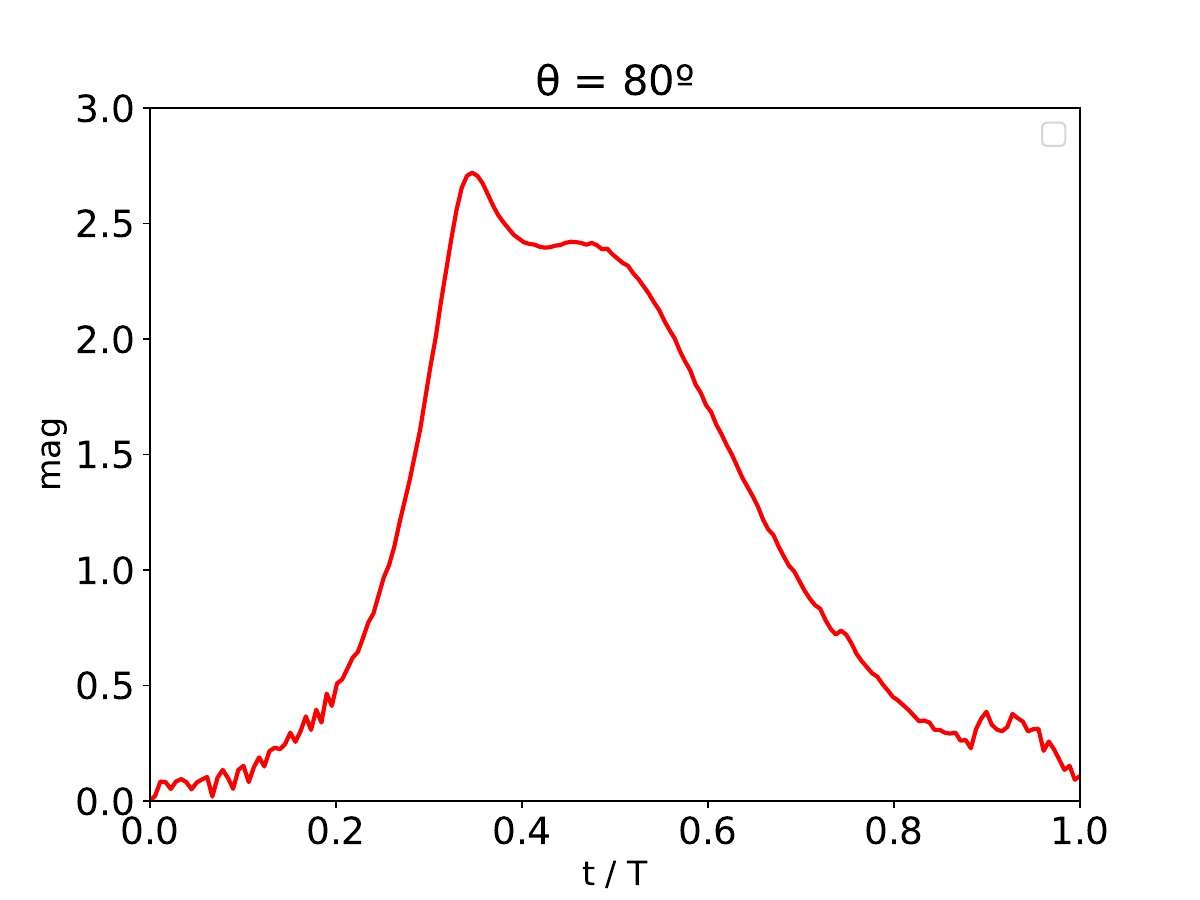}
\caption{Time integrated fluxes $\left<I\right>_{lm}$ (left panel), Temporal centroids $\vec{c}_k$ (middle panel) and temporal magnitudes $m_k$ (right panel) for an observation angle of $\theta=20^\circ$ (top row) and $\theta=80^\circ$ (bottom row), for the high-compactness configuration with $k=0.9375$ and $\bar a_0=0.620525$.}
\label{fig:astro_ultra}
\end{figure*}

\section{Imaging dark matter halo black holes from thin accretion disks}\label{sec:disks}
%%%%%%%%%%%%%%%%%%%%%%%%%%%%%%%%%%%%%%%%%%%%%%%%%%%%

\subsection{Ray-tracing in optically and geometrically thin disk models}

In this section we consider the imaging of DMHBHs from a thin disk. In order to generate the observed images, we perform a backwards ray-tracing procedure in which a congruence of geodesics is integrated from the observer towards the black hole using our own Mathematica-based code. This is done using Eq.(\ref{eq:effp}) suitably rewritten using the conservation of both the energy $E$ and the angular momentum $L$ as
\begin{equation}
\frac{d\phi}{dr}= \mp \frac{b}{r^2} \frac{\sqrt{AB}}{\sqrt{1-\frac{bA}{r^2}}} \ ,
\end{equation}
where the $\mp$ correspond to outgoing/ingoing trajectories from the point of view of the asymptotic observer, so only when a turning point is reached does the $+$ sign manifest; otherwise the light ray is absorbed by the event horizon of the black hole. Furthermore, this equation allows one to classify the different trajectories according to the angular deviation or, alternatively, according to the number of times $n\equiv \frac{\phi}{2\pi}$ the geodesic crosses the equatorial plane, where the (thin) accretion disk is located\footnote{In practical terms, however, one must subtract $\pi$ to account for the fact that an undeflected light ray suffers an angular deviation of $\pi$.}. Photon trajectories that execute $n$ half-loops around the black hole are associated with the number of photon rings appearing on the observer's screen, and in typical black holes there is a one-to-one correspondence: the $n^{th}$ photon ring is created by the photon trajectories that perform $n$ half-loops around the black hole (note that this is not necessarily so for horizonless compact objects). 

We consider an optically thin (zero absorption) and geometrically thin accretion disk emitting, in its own frame of reference, a monochromatic radiation with an intensity profile $I_{\nu_e} \equiv I(r)$. Given that the invariant intensity $I_{\nu}/\nu^3$ is conserved along a geodesic congruence \cite{Gold:2020iql}, in the reference frame of the observer the intensity takes the form $I_{\nu_o}=g^{3} I_{\nu_e}$, where $g \equiv \nu_o^3/\nu_e^3$, with $\nu_o$ ($\nu_e$) the frequency in the observer (emitter) frame. For the line element in Eq.(\ref{eq:le}), this means that $g =A^{1/2}(r)$ (given its asymptotically flat character), and thus every quantity depends solely on the radial coordinate. The total intensity comes from performing an integration over all the observed frequencies, and bearing in mind the number $n$ of half-orbits every photon takes on their winding around the black hole, which yields the result \cite{Gralla:2019xty}
\begin{equation}
I_{o}=\int d\nu_o I_{\nu_o}= \sum_{n=0} \xi_n A^2 I(r) \vert_{r=r_n(b)}
\end{equation}
where $r=r_n(b)$ encodes the information about the radial location at which each light ray of impact parameter $b$ hits the disk (usually known as the ``transfer function"). Additionally we have included a {\it fudge} factor $\xi_n$ which accounts for two main elements: what rings one decides to suppress in the image, and what ``weight" is attributed to the luminosity of each ring. For the first element, we assume all photon ring contributions to the observed profile with $n>2$ to be negligible in their contributions to the image, as shall be justified below. This assumption implies that $\xi_n=0, \forall n>2$. The second ingredient allows to simulate, in a simple way, the effects on non-vanishing thickness, as considered in the pool of simulations of \cite{Paugnat:2022qzy,Paugnat:2022qzy}; below we shall use two choices for it.

Under the considerations outlined above, the observed image is characterized by a single function $I(r)$, whose choice is to some degree arbitrary, depending on the specific features of the accretion flow and/or the background geometry one is interested in analyzing. As a consequence, many such profiles have been considered in the literature to generate their corresponding images. To fix this profile we call upon the adaptation of Standard Unbound profile which was proposed in \cite{Gralla:2020srx} under the form (hereafter called as SU models)
\begin{equation} \label{eq:SU}
J_{SU}(r;\mu,\gamma,\sigma)=\frac{e^{-\frac{1}{2} \left[\gamma +\arcsinh \left(\frac{r-\mu}{\sigma}\right) \right]^2}}{\sqrt{(r-\mu)^2+\sigma^2}}
\end{equation}
which is characterized by solely three parameters: $\mu$ is related to the radius at which the peak of the profile occurs, $\gamma$ to the axisymmetry of the profile, and $\sigma$ to its width. This kind of profile turns out to be flexible enough to reproduce some scenarios of GRMHD simulations of the accretion flow around a Kerr black hole, see e.g. \cite{Paugnat:2022qzy,Cardenas-Avendano:2023dzo} for a discussion on this topic.

\subsection{Shadow's size constraints on DMHBHs}

Black hole images, found via GRMHD or via semi-analytic analysis, are characterized by two main features: an exterior bright ring of radiation, and a central brightness depression. We shall use the latter feature to select the classes of DMHBH configurations to be analyzed, and the former in order to carry out the analysis of their characteristic signatures.

In the canonical interpretation of a black hole shadow by Falcke \cite{Falcke:1999pj}, this is related to those photons that lie inside the photon sphere to intersect the event horizon and thus cannot reach the asymptotic observer.  Assuming the latter to be located at some point $(r_0,\theta_0)$, the celestial coordinates are introduced as \cite{Perlick:2021aok}
\begin{eqnarray}
\alpha&=&-\displaystyle\lim_{r_0 \to \infty}\left(-r_0^2  \sin \theta_0 \frac{d\phi}{dr} \right)\Big\vert_{r=r_0}, \\
\beta &=& -\displaystyle\lim_{r_0 \to \infty}\left(r_0 \frac{d\theta}{dr} \right)\Big\vert_{r=r_0} 
\end{eqnarray}
where $r_0$ is the distance from the black hole to the asymptotic observer, and $\theta_0$ the inclination angle between the axis perpendicular to the equatorial plane  of the black hole and the observer's line of sight. Using Eq.(\ref{eq:effp}), along with the conservation of both $E$ and $L$, and the spherically symmetric and asymptotically flat character of the space-time, one verifies that the radius of the dark circle observed by the asymptotic observer (which is interpreted as the radius of the shadow) is $r_{sh}=\sqrt{\alpha^2 + \beta^2}= b_c$, i.e., it coincides with the critical impact parameter. This fact is of relevance at the light of the reported results by the EHT Collaboration regarding the size of the shadow of Sgr A$^*$; while the latter cannot be directly measured, it can be inferred via a correlation with the size of the outer bright ring, which is observable (although subject to several caveats and assumptions on the behavior of the accretion flow). Such a correlation reports a size of the shadow $r_{sh}$ within the bounds
\begin{equation}
4.21 \lesssim \frac{r_{sh}}{M_{Sch}} \lesssim 5.56,
\end{equation}
where $M_{Sch}$ denotes the mass of a Schwarzschild black hole. If we identify the black hole mass in the DMHBH model $M_{\rm BH}$ with the Schwarzschild mass $M_{Sch}$, then comparing Eq.(\ref{eq:bc}) with the above (upper) bound for the radius of the shadow sets an upper bound for the compactness of approximately $ \mathcal{C} \gtrsim 1/12$\footnote{A more detailed analysis of this issue has been recently carried out in \cite{Myung:2024tkz}.}. 

\begin{table}[t!]
\begin{tabular}{|c|c|c|c|c|c|}
\hline
 & $\frac{M_{\rm DM}}{M_{\rm BH}}$   & $\frac{a_0}{M_{\rm BH}}$ & $\mathcal{C}$  &  $\frac{r_{ps}}{M_{\rm BH}}$ & $\frac{b_c}{M_{\rm BH}}$  \\ \hline
DMHBH-I & 100  & 1500 & 1/15  & 3.00013 & 5.562 \\ \hline
DMHBH-II & 10  & 150  & 1/15  & 3.00128  & 5.555 \\ \hline
DMHBH-III & 1  & 12 & 1/12 & 3.01368 & 5.563 \\ \hline
\end{tabular}
\caption{The features of the three dark matter halo black hole (DMHBH) configurations chosen in our analysis for our first pool of simulations: the halo-to-black hole mass ratio $M_{DM}/M_{BH}$, the parameter $a_0$-to-black hole mass ratio $a_0/M_{BH}$, the compactness parameter $\mathcal{C} \equiv M_{DM}/a_0$, the photon sphere radius $r_{ps}/M_{BH}$, and the critical impact parameter $b_c/M_{BH}$.}
\label{Table:fea}
\end{table}

Following this discussion, for our first pool of simulations we shall consider three DMHBH configurations that saturate the bound above, with roughly the same compactness but exploring different ranges of the parameters $M_{\rm DM}$ and $a_0$ of two orders of magnitude (the black hole mass $M_{\rm BH}$ is normalized to one in every configuration). The corresponding parameters are summarized in Table \ref{Table:fea}. For all of these models, the radii of the photon spheres are very close to their Schwarzschild counterpart (with the relative difference decreasing with an increase in $M_{\rm DM}$ and $a_0$), while the critical impact parameter saturates the EHT constraint for compactness values of $\mathcal{C} \sim \frac{1}{12}-\frac{1}{15}$.

\subsection{Lyapunov exponents of nearly-bound geodesics}

\begin{table*}[t!]
\begin{tabular}{|c|c|c|c|c|}
\hline
Em/Geo & Sch  & DMHBH-I  & DMHBH-II & DMHBH-III \\ \hline
$n=0$ & $b \not \in (5.02,6.17)$ &   $b \not \in (5.37,6.59)$   & $b \not \in (5.36,6.62)$   & $b \not \in (5.36,6.77)$   \\ \hline
$n=1$ & $b\in (5.02,5.23) \cup (5.19,6.17)$ & $b\in (5.37,5.55) \cup (5.59,6.59)$  & $b\in (5.36,5.55) \cup (5.59,6.62)$ & $b\in (5.36,5.55) \cup (5.60,6.77)$ \\ \hline
$n=2$ & $b \in (5.188,5.227)$ & $b \in (5.553,5.595)$ & $b \in (5.546,5.590)$ & $b \in (5.553,5.603)$ \\ \hline
\end{tabular}
\caption{The impact parameter regions of the direct $n=0$ and photon ring $n=1,2$ emissions for the Schwarzschild black holes and three three samples of DMHBHs appearing in Table \ref{Table:fea}.}
\label{Table:r-tr}
\end{table*}

To characterize the photon rings we need to elaborate on nearly bound geodesics, which are defined as those whose minimum radius is arbitrarily near the photon sphere, i.e., $r=r_{ps} + \delta r$, with $\delta r \ll r_{ps}$. This way, perturbing the geodesic equation in Eq.(\ref{eq:effp}) up to second order, imposing the photon sphere radius condition in Eq.(\ref{eq:rm}), and using the conservation of the angular momentum $L$, one arrives at
\begin{eqnarray}
\pi \frac{d \delta r}{d\phi}&=& \gamma_L \delta r, \\
\gamma_L &=& \pi \frac{A_m^{1/2}}{A_m'} \frac{1}{(A_m B_m)^{1/2}} \left[A_m'^2-2A_mA_m''\right]^{1/2},
\end{eqnarray}
where $\gamma_L$ is dubbed as the {\it Lyapunov exponent} and it depends only on the background metric. To interpret the role of $\gamma_L$, consider the integration of the equation above as
\begin{equation}
\delta r_n = \delta r_0 e^{\gamma_L n},
\end{equation}
where we have rewritten the result in terms of the number $n$ of half-orbits. This means that every half-orbit the photon streams to a radius that is a factor $\sim e^{\gamma_L}$ larger than the previous one. As a consequence, the impact parameter window is reduced on each successive half-loop also via a factor $\sim e^{-\gamma_L}$. In Table \ref{Table:r-tr} we classify the impact parameter space for the three DMHBH configurations above (and also of the Schwarzschild black hole for comparison), following the prescription of \cite{Gralla:2019xty} into the direct emission ($n=0$) and first ($n=1$) and second ($n=2$) photon rings, since these are the only ones which we hope to observe in future interferometric projects. Several aspects of the modification induced by the DMHBHs can be inferred: the location of both photon rings is pushed to a significantly larger impact parameter radius despite the fact that the photon sphere radius remains almost unmodified\footnote{One should note that the lower end of the $n=2$ photon ring extends inside the critical curve in the observer's screen: such values correspond to light rays that are present when an accretion disk whose inner edge is located inside the photon sphere emits photons outwards, which can therefore escape from the black hole and reach the asymptotic observer, thus enhancing the corresponding width.}. Furthermore, the width of both the $n=1$ and $n=2$ photon rings is increased as compared to their Schwarzschild counterparts, e.g., a $\sim 25\%$ increase in the DMHBH-III model. 

Another tangible consequence of the Lyapunov exponent is that, should the luminosity collected by the photon on each turn be exactly similar, then the ratio between the observed luminosities of different photon rings should be also be expected to follow a behavior of the form \cite{Cardenas-Avendano:2023dzo}
\begin{equation} \label{eq:Isup}
\frac{I_{n+1}}{I_n} \propto e^{-\gamma_L}.
\end{equation}
Thus, the Lyapunov exponent becomes a universal quantifier of the way the background geometry bends successive photon trajectories, which translates into their corresponding luminosities. However, in practical terms, the features of the emission regions every pack of photons travel in their winding around the black hole (depending on their respective impact parameters) is not exactly the same, a fact that introduces differences in the actual luminosities of the photon rings. This takes us to set our first pool of emission SU models of the accretion disk.

\subsection{Accretion disk models}

In this section we shall choose three intensity profiles to model the accretion disk emission previously employed in other works, which are suitable implementations of the SU model in Eq. \eqref{eq:SU} above and, in particular, of the three profiles introduced in \cite{Gralla:2020srx} by Gralla, Lupsasca and Marrone (GLM), adapted to a non-rotating scenario (where, in particular, no inner horizon is present) as
\begin{eqnarray}
\text{GLM1}&:& \gamma=-\frac{3}{2}, \quad \mu=0,  \quad  \sigma=\frac{M_{BH}}{2} \label{eq:GLM1} \\
\text{GLM2}&:& \gamma=0, \quad \mu=0,  \quad  \sigma=\frac{M_{BH}}{2} \label{eq:GLM2} \\
\text{GLM3}&:& \gamma=0, \quad \mu=\frac{17M_{BH}}{3},  \quad  \sigma=\frac{M_{BH}}{4}  \label{eq:GLM3}
\end{eqnarray}
In our setting we take advantage of the fact that the event horizon of the DMHBH model is located exactly at the Schwarzschild radius (in terms of the black hole mass), such that GLM1/GLM2, both extending inside the event horizon, have their peak of emission located at the exact same radius, something which does not happen in other black hole space-times. Similarly, the intensity profile of the GLM3 model peaks near the radius of the ISCO radius of a Schwarzschild black hole, $r_{ISCO}^S=6M$, and also the one of the DHMBHin the low compactness regime, which is given by
\begin{equation}
r_{ISCO} \approx r_{ISCO}^S \left(1-\frac{32\mathcal{C}M_{BH}}{a_0} \right) 
\end{equation}
where the corrections are strongly suppressed in such a scenario. The ISCO sets the inner boundary of the region where the massive particles composing the accretion disk may undergo stable orbital motion; below this radius, orbital motion is still possible but it is unstable, meaning that any perturbation in the orbit either induces the infall of matter towards the black hole horizon or ejection to the exterior. Therefore, GLM3 model accounts for the first scenario, and GLM1/GLM2 for the other two. Furthermore, DMHBHs are particularly well-suited objects to be tested with minimal differences (again in the low-compactness regime) in their geometrical and emission profiles as compared to their Schwarzschild counterpart.

\begin{figure*}[t!]
\includegraphics[width=4.4cm,height=4.0cm]{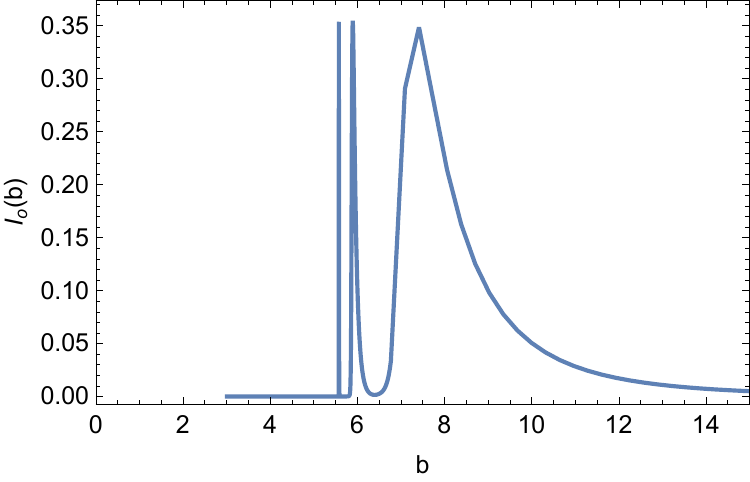}
\includegraphics[width=4.4cm,height=4.3cm]{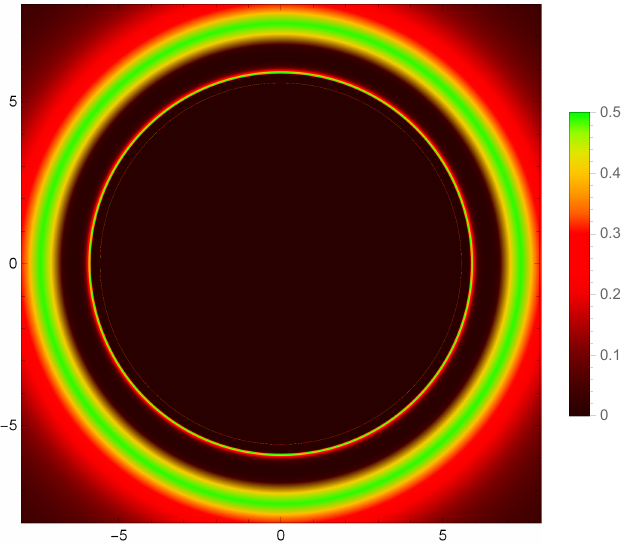}  \\
\includegraphics[width=4.4cm,height=4.0cm]{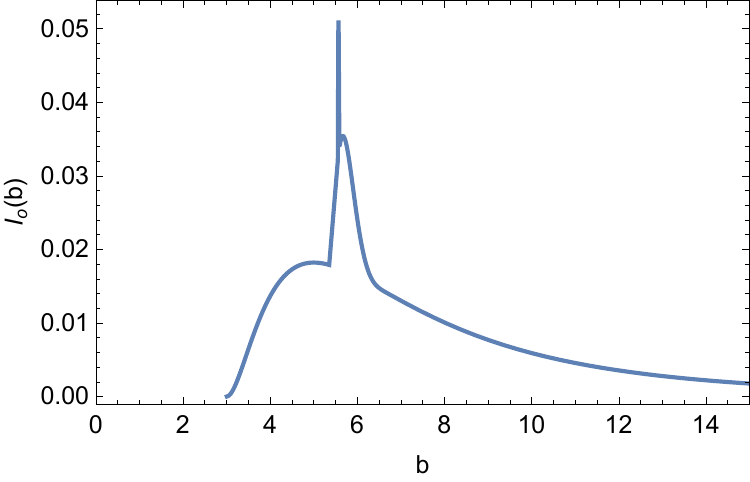}
\includegraphics[width=4.4cm,height=4.3cm]{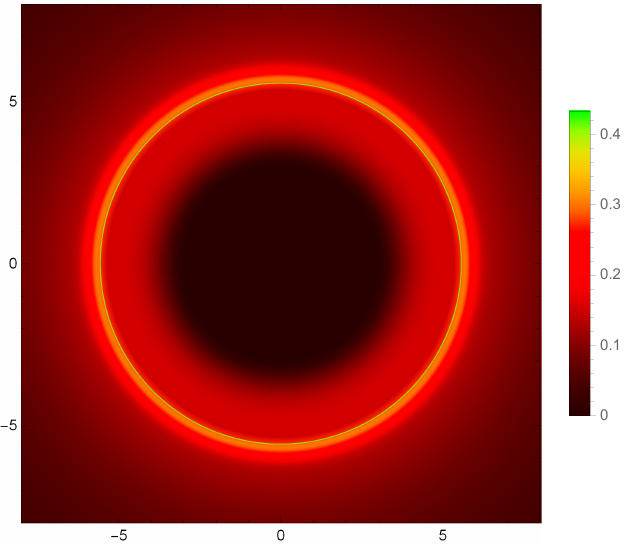}
\includegraphics[width=4.4cm,height=4.0cm]{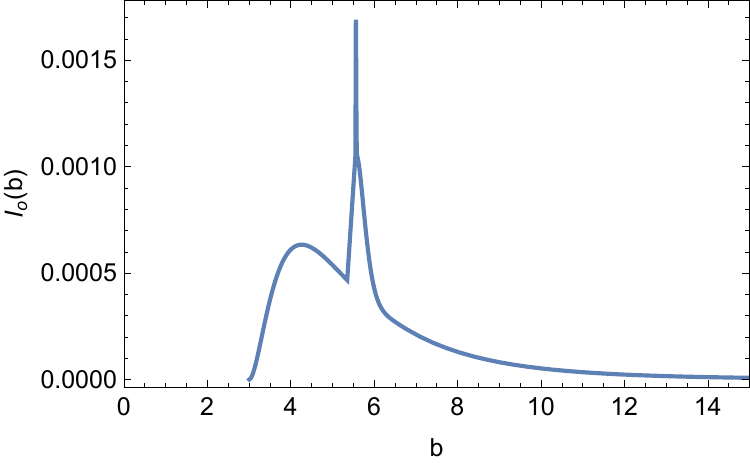}
\includegraphics[width=4.4cm,height=4.3cm]{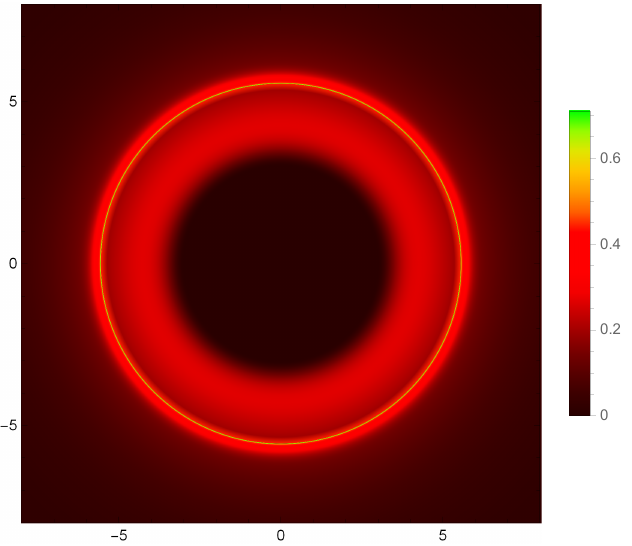} 
\caption{Images of a DMHBH-III configuration for the GLM3 (top), GLM1 (left) and GLM2 (right) emission models, as given by Eqs.(\ref{eq:GLM3}), (\ref{eq:GLM1}), and (\ref{eq:GLM2}), respectively. For each emission model we depict the observed intensity $I_{o}(b)$ as well as the full image. Note the contrast between the GLM3 and GLM1/GLM2 images: in the former the photon rings ($n=1$ and $n=2$, though only the first one is neatly visible) are clearly distinguished from the direct emission $n=0$ ring, while in the latter they are inserted into the $n=0$ ring.   }
\label{fig:images1}
\end{figure*}

We shall consider two examples of fudge factors to describe the relative weight of the photon rings of different order, namely: type I) $\xi_{0,1,2}=1$, corresponding to an infinitely-thin disk, and type II) $\xi_0=1, \xi_{1,2}=1.5$. We run our simulations for the three solutions summarized in Table \ref{Table:fea} for a fudge factor of type I (i.e. same weight for the luminosity of photon rings as compared to direct emission). Our main objects of interest are i) the relative luminosity of the photon rings, and ii) the size of the central brightness depression in the three GLM models. We supply in Fig. \ref{fig:images1} the imaging of the DMHBH-III configuration which, as we justify in what follows, provides the largest modification in comparison with the Schwarzschild solution out of the three DMHBH geometries chosen in this section.

Regarding the luminosity of the photon rings, in Table \ref{Table:lum} we provide the extinction rate $I_1/I_2$ of the relative luminosities between the $n=1$ and $n=2$ photon rings for the three GLM models, as well as the Schwarzschild black hole for comparison, organized in decreasing values, and together with its comparison with the theoretical expectation based on the Lyapunov exponent. One verifies that the latter overestimates the values of these rates for every solution. In any case, extinction rates are slightly lower in the DMHBHs than in the Schwarzschild solution, meaning that they appear as more luminous in their corresponding image. This effect is further exaggerated the lower the values of $M_{\rm DM}$ and $a_0$ are, which places the DMHBH-III model as the one deviating the most from the Schwarzschild solution in our analysis. Thus, relative luminosities of photon rings are hardly a good observational discriminator between the Schwarzschild and DMHBH configurations, at least within this scenario. A better opportunity could be present in the locations of the photon rings given the larger critical impact parameter which manifests a moderate shifting of $\sim 7\%$ as compared to the Schwarzschild black hole, though this signature should also be shared by many modifications of the Schwarzschild solution proposed in the literature.

\begin{table}[t!]
\begin{tabular}{|c|c|c|c|c|}
\hline
 & Sch & DMHBI-I & DMHBH-II  & DMHBH-III  \\ \hline
$e^{\gamma_L}$ & 23.35  & 23.34  & 23.26  & 22.46  \\ \hline
$I_{GLM3}$ & 27.84 & 27.84  & 27.84 & 27.15  \\ \hline
$I_{GLM1}$ & 24.81 & 24.74  & 24.70 & 24.00  \\ \hline
$I_{GLM2}$ & 23.57 & 23.45  & 23.38 & 22.58 \\ \hline
\end{tabular}
\caption{The extinction rate between the $n=1$ and $n=2$ photon rings for the three DMHBH configurations appearing in Table \ref{Table:fea} for the three GLM models (\ref{eq:GLM3}), (\ref{eq:GLM1}) and (\ref{eq:GLM2}), and its comparison with the theoretical Lyapunov-based expectancy and with the Schwarzschild solution.}
\label{Table:lum}
\end{table}

Regarding the central brightness depression, we see neat differences between the GLM3 and the GLM1/GLM2 models: in the GLM1 model the $n=1$ and $n=2$ photon rings are clearly separated in the observed intensity $I_o(b)$ from the main ring of direct radiation, though only the $n=1$ is visible to the naked eye in the optical appearance plot. Furthermore, the innermost $n=2$ photon ring is located around the critical impact parameter $b=b_c$, and thus the shadow in this model fills entirely the critical curve. In contrast, for the GLM2/GLM3 models, the fact that the emission region extends all the way down to the event horizon implies that the photon rings overlap with the direct emission. Furthermore, the outer edge of the shadow is moderately increased: in the Schwarschild black hole the outer edge of the shadow stands at a radius of $\sim 2.88M$, while in the DMHBH-III it stands at a radius of $\sim 3.09M$. These reductions in the size of the central brightness depression from the canonical black hole shadows (as given by the critical impact parameter, as discussed above) in all solutions is consistent with previous results found in the literature: in those models for which the inner edge of the effective region of emission is located at $r_e \gg r_h$, then one observes that the classical black hole shadow fills the critical curve, whereas if $r_e \gtrsim r_h$, then the outer edge of the central brightness depression is strongly reduced and becomes a gravitationally redshifted image of the event horizon, a phenomenon dubbed in \cite{Chael:2021rjo} as the inner shadow.

If we repeat our analysis for a fudge factor of type II, then the luminosity of the photon rings gets boosted by a factor $1/3$ as compared to the direct emission, due to the decrease in their extinction rates by such an amount. The ratio between the $n=1$ and $n=2$, however, remains unaffected by this modification. Therefore, in this scenario one would expect more luminous photon rings but, since the Schwarzschild solution would also be similarly modified and no other significant signatures are introduced, the refinement of the fudge factor in the emission profile provides no additional support to distinguish between DMHBH models and the Schwarzschild solution.

A final comment of our analysis refers to the horizonless configurations discussed in \cite{Xavier:2023exm}. These configurations feature two photon spheres, as well a local minimum of the effective potential (an anti-photon sphere). The fact that the innermost photon sphere corresponds in such a scenario to a larger peak in the effective potential implies that light trajectories whose impact parameters lie in the intermediate region may circulate in the potential well and give rise to further photon rings, as found in other models with similar features \cite{Kar:2023dko}. Furthermore, their relative luminosity does not necessarily follow the exponential decay rule of the Lyapunov exponent. One can criticize such configurations on the grounds that anti-photon spheres are prone to induce instabilities associated to the accumulation of waves within the potential well \cite{Cardoso:2014sna,Cunha:2022gde}, while the additional photon rings typically change the optical appearance in a most drastic way, making them hardly compatible with current black hole images.

\subsection{ISCO and event horizon models}

In this section, we adopt two additional SU intensity profile models to describe the emission of the accretion disk to perform a wider parameter space analysis of the observational properties of the DMHBH model. These SU models are characterized by the parameters:
\begin{eqnarray}
\text{ISCO:} \ \gamma&=&-2, \qquad \mu=6M_{\text BH}, \qquad \sigma=\frac{M}{4},\\
\text{EH:} \ \gamma&=&0, \qquad \mu=0, \qquad \sigma=2M.
\end{eqnarray}
We denote the first model as the ISCO model given that, for dilute dark matter configurations, the radius at which the intensity profile of this model peaks corresponds to the radius of the ISCO. We clarify that this does not correspond, in general, to the ISCO of the models considered (indeed, the exact value of the ISCO changes from model to model). Nevertheless, to avoid over-complicating the modeling of the disk intensity profiles, we take this assumption in every combination of parameters considered.

\begin{figure*}[t!]
\includegraphics[scale=0.45]{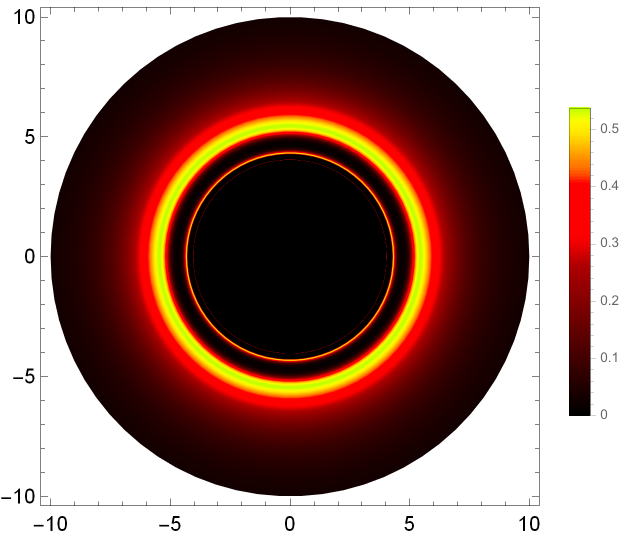}\qquad
\includegraphics[scale=0.45]{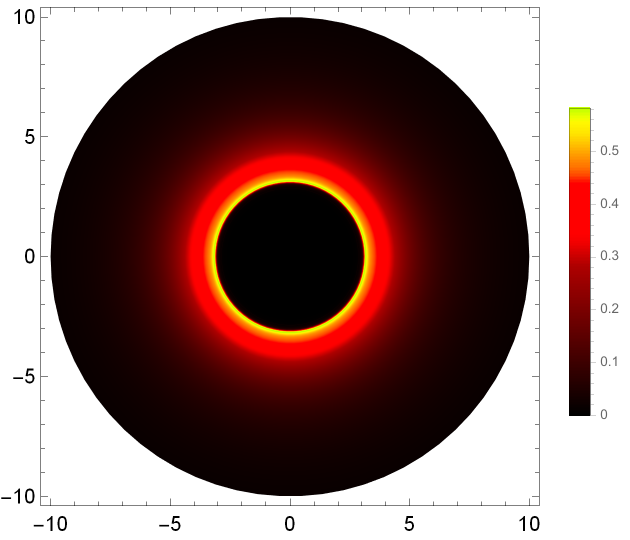}\qquad
\includegraphics[scale=0.45]{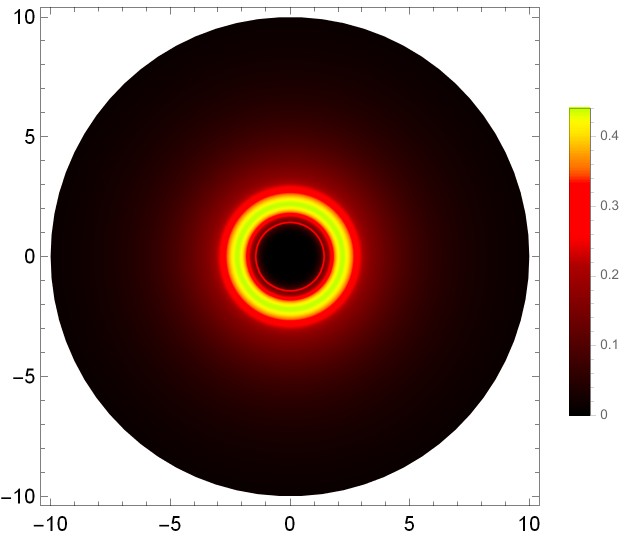}\\
\includegraphics[scale=0.45]{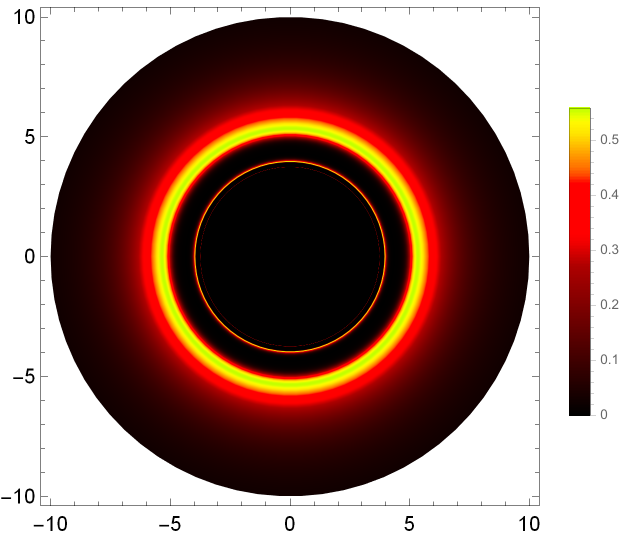}\qquad
\includegraphics[scale=0.45]{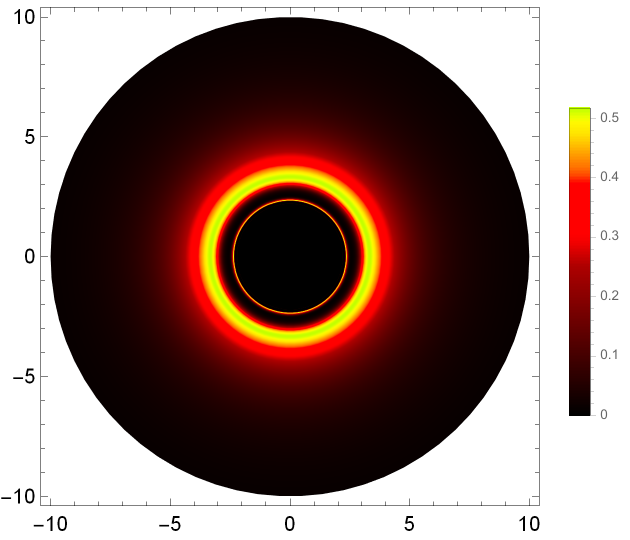}\qquad
\includegraphics[scale=0.45]{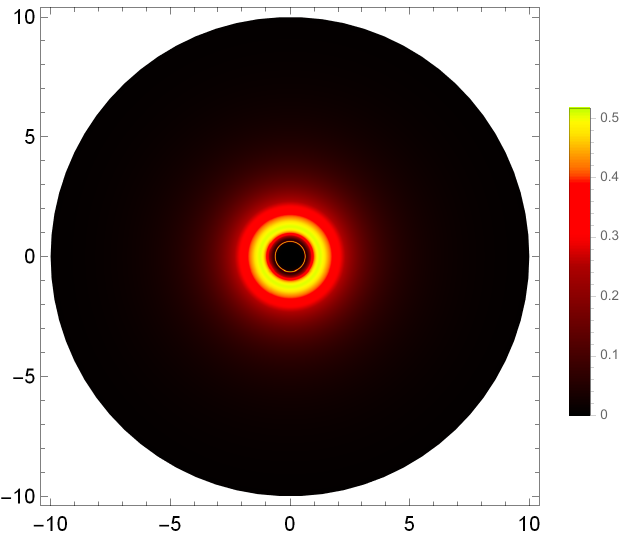}\\
\includegraphics[scale=0.45]{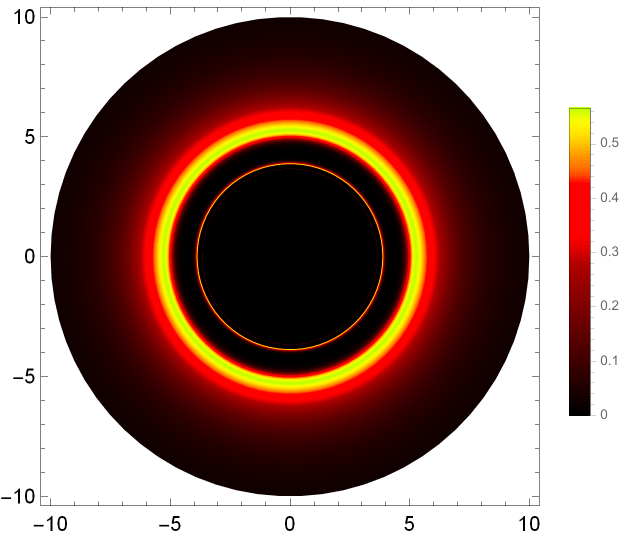}\qquad
\includegraphics[scale=0.45]{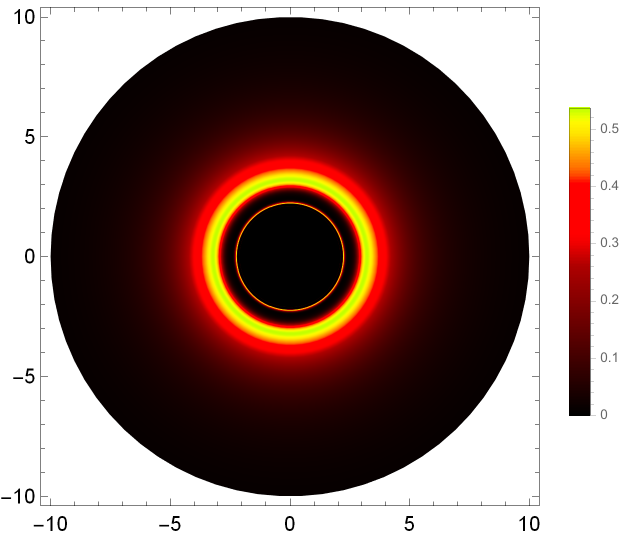}\qquad
\includegraphics[scale=0.45]{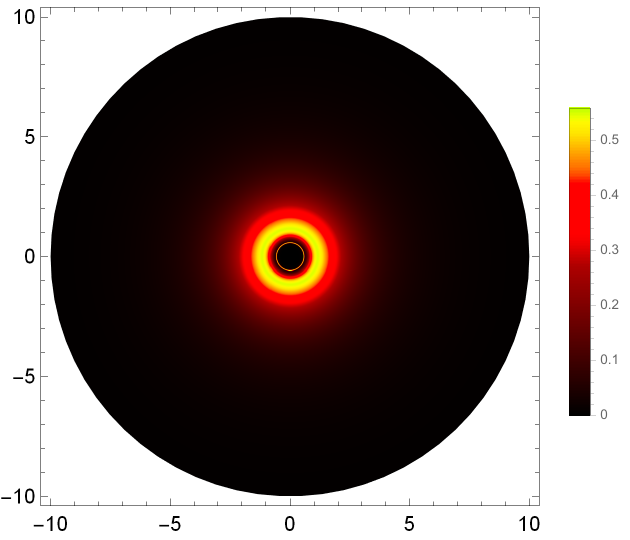}\\
\caption{Observed shadow images for an observation angle of $\theta=0^\circ$ with the ISCO disk model for the DMBH models with $a=1M$ (top row), $a=10M$ (middle row), and $a=100M$ (bottom row), and with $k=0.3$ (left column), $k=0.6$ (middle column), and $k=0.9$ (right column). The axes of the figures are given in units of $r/M$.}
\label{fig:shadowsISCO}
\end{figure*}

\begin{figure*}[t!]
\includegraphics[scale=0.65]{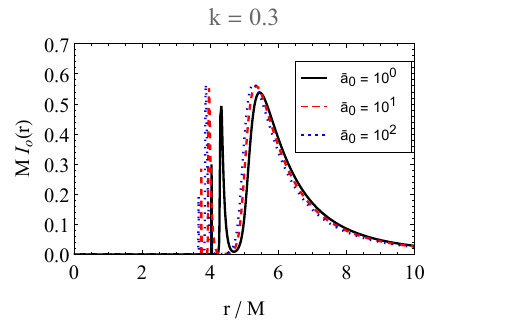}
\includegraphics[scale=0.65]{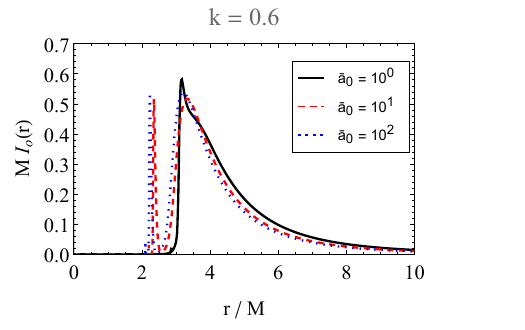}
\includegraphics[scale=0.65]{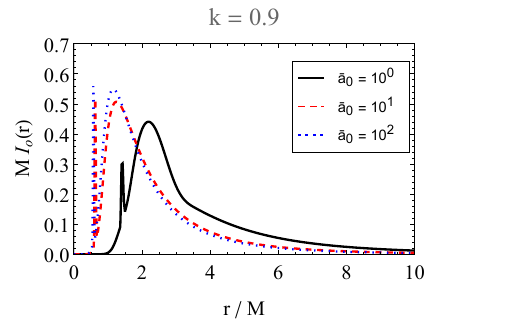}\\
\includegraphics[scale=0.65]{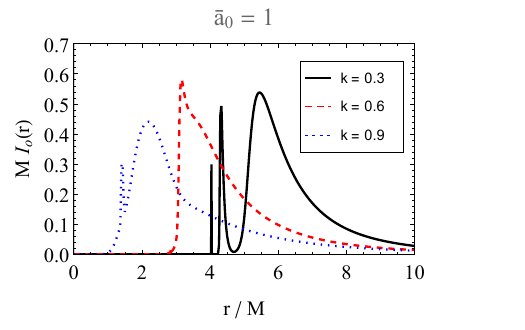}
\includegraphics[scale=0.65]{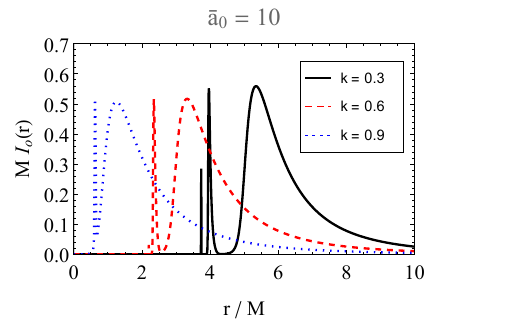}
\includegraphics[scale=0.65]{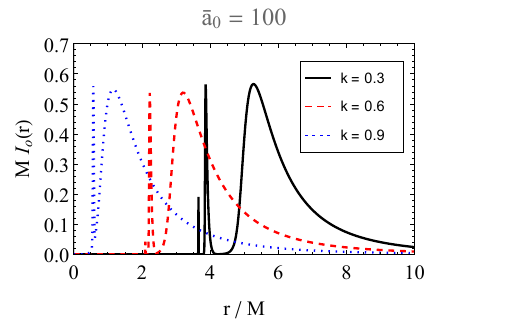}
\caption{Observed intensity profiles with the ISCO disk model for the DMHBH models with fixed $k$ and varying $\bar a_0\equiv a_0/M$ (top row) or fixed $\bar a_0$ and varying $k$ (bottom row).}
\label{fig:intensityISCO}
\end{figure*}

\begin{figure*}[t!]
\includegraphics[scale=0.45]{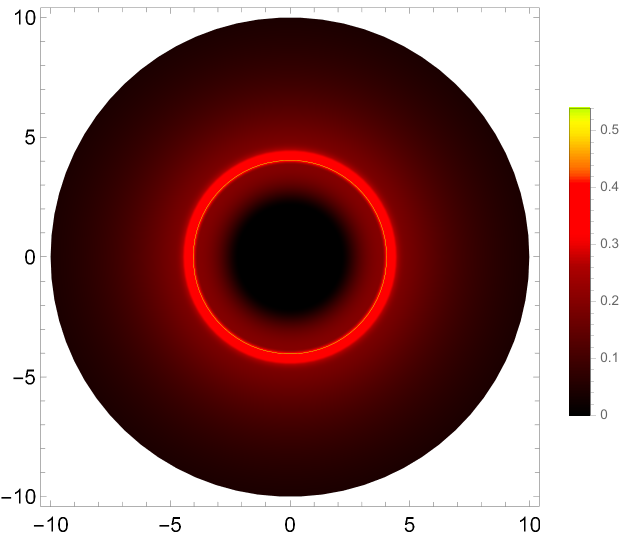}\qquad
\includegraphics[scale=0.45]{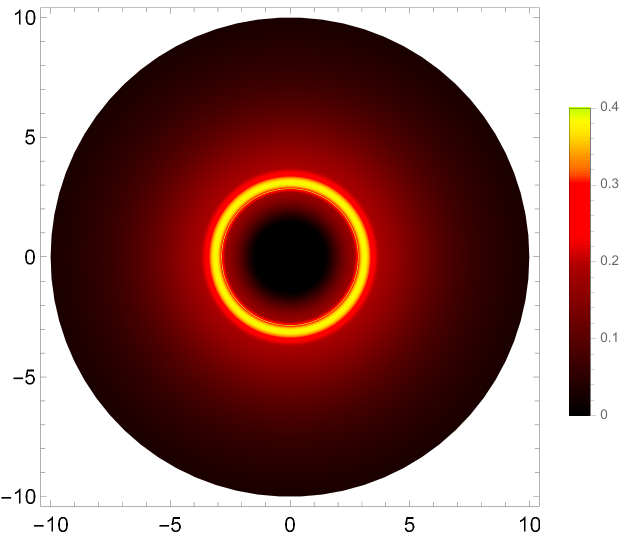}\qquad
\includegraphics[scale=0.45]{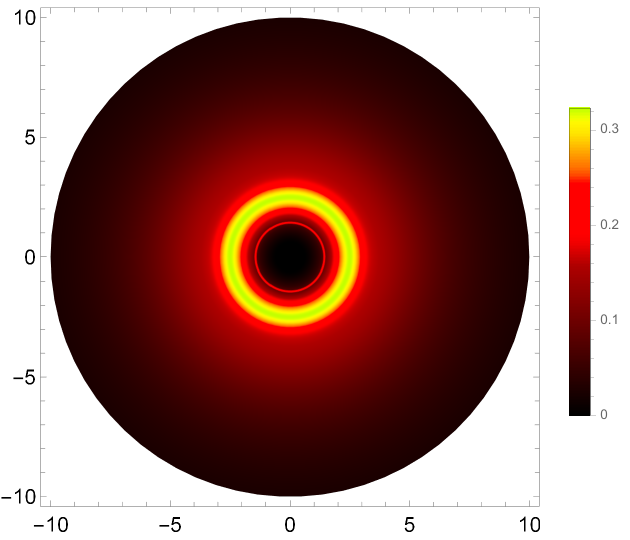}\\
\includegraphics[scale=0.45]{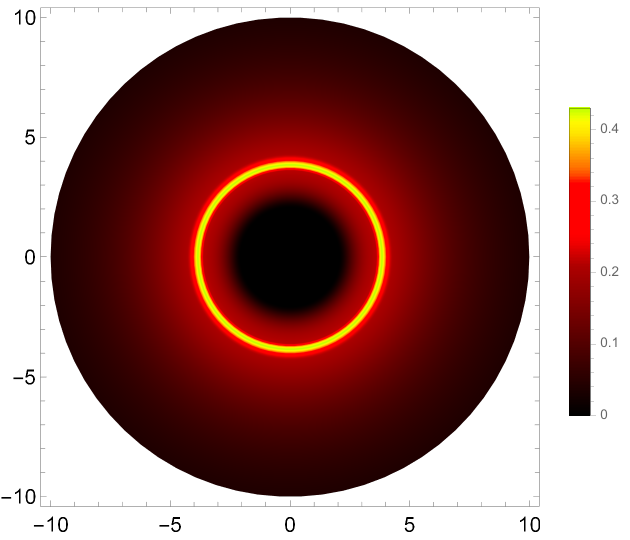}\qquad
\includegraphics[scale=0.45]{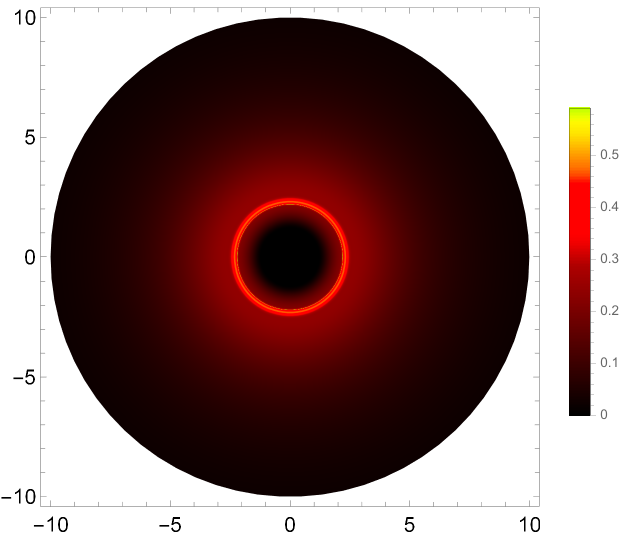}\qquad
\includegraphics[scale=0.45]{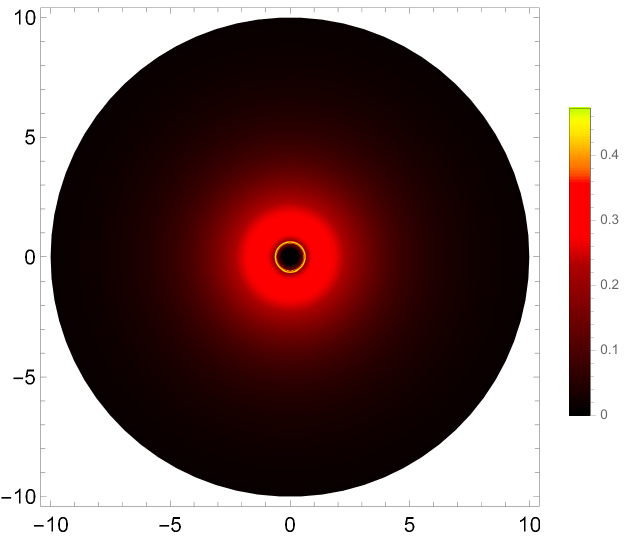}\\
\includegraphics[scale=0.45]{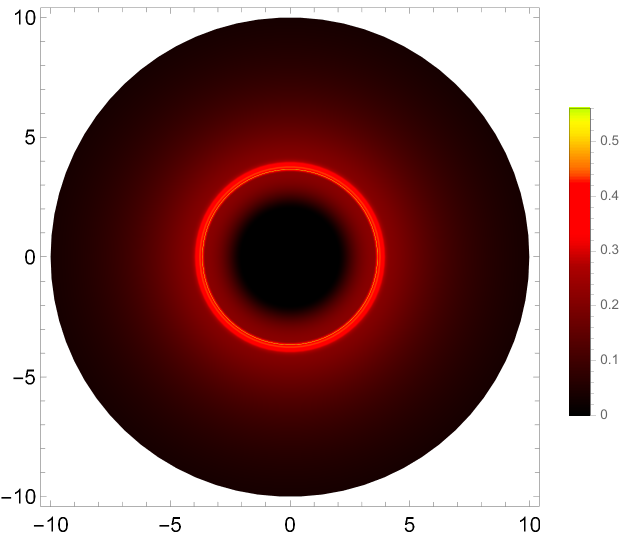}\qquad
\includegraphics[scale=0.45]{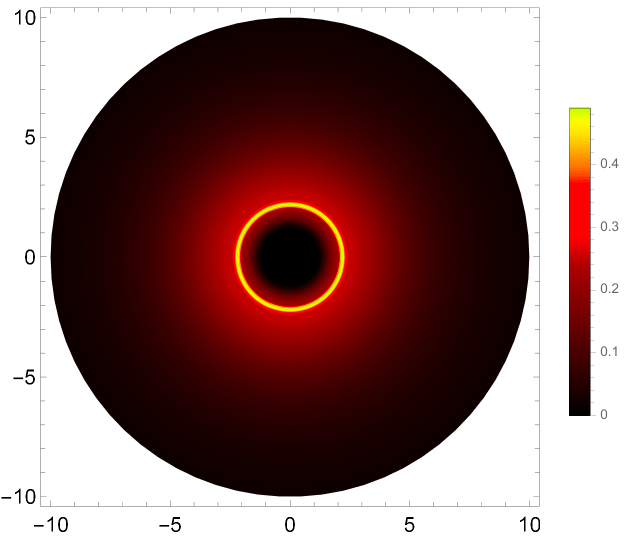}\qquad
\includegraphics[scale=0.45]{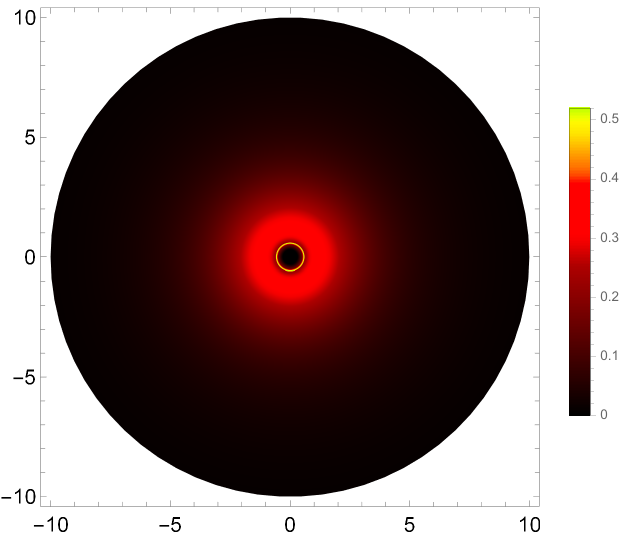}\\
\caption{Observed shadow images for an observation angle of $\theta=0^\circ$ with the EH disk model for the DMHBH models with $a=1M$ (top row), $a=10M$ (middle row), and $a=100M$ (bottom row), and with $k=0.3$ (left column), $k=0.6$ (middle column), and $k=0.9$ (right column). The axes of the figures are given in units of $r/M$.}
\label{fig:shadowsC}
\end{figure*}

\begin{figure*}[t!]
\includegraphics[scale=0.65]{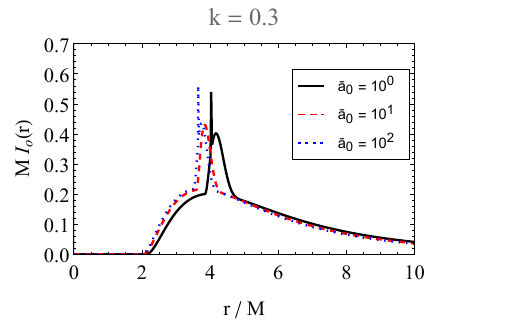}
\includegraphics[scale=0.65]{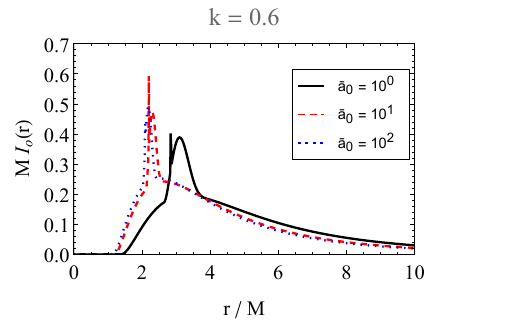}
\includegraphics[scale=0.65]{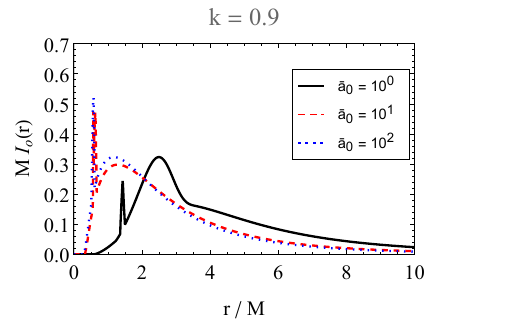}\\
\includegraphics[scale=0.65]{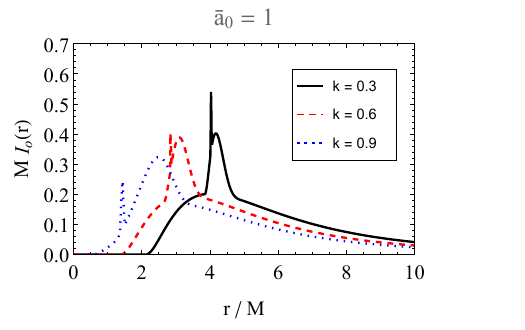}
\includegraphics[scale=0.65]{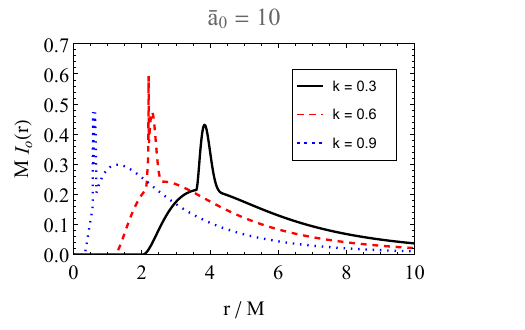}
\includegraphics[scale=0.65]{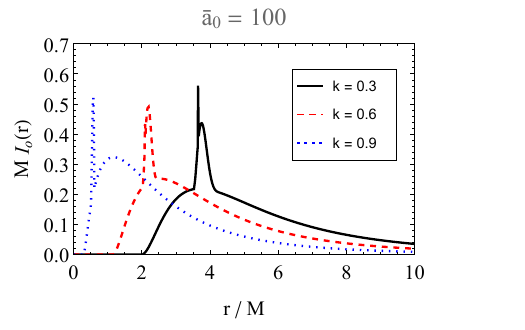}
\caption{Observed intensity profiles with the EH disk model for the DMHBH models with fixed $k$ and varying $\bar a_0\equiv a_0/M$ (top row) or fixed $\bar a_0$ and varying $k$ (bottom row).}
\label{fig:intensityC}
\end{figure*}

Let us now analyze how the two free parameters of the model (in normalized form), namely $k$ and $\bar a_0$, affect the observable properties of optically thin-accretion disks in the background of DMHBH models, pushing the parameter space farther than in the previous analysis. For this purpose, we select nine further DMHBH models to analyze. These models are characterized by the combinations of the parameters $k=\{0.3, 0.6, 0.9\}$ and $\bar a_0=\{10,100,1000\}$. In Figs. \ref{fig:shadowsISCO} and \ref{fig:shadowsC} we show the images for the ISCO and EH disk models, respectively, whereas in Figs. \ref{fig:intensityISCO} and \ref{fig:intensityC} we show the observed intensity profiles for the same models.

For the ISCO disk model, see Figs. \ref{fig:shadowsISCO} and \ref{fig:intensityISCO}, one observes that the properties of the direct image are strongly dependent on the value of $k$. This is so since this parameter is directly related to the mass ratio and, thus, variations of $k$ imply variations of $M_{\text BH}$, which controls the inner radius of the disk. Consequently, an increase in $k$ leads to a decrease in the radius of the observed shadow. On the other hand, the parameter $\bar a_0$ is particularly important for larger compactness, causing larger deviations in the size of the shadow for $\bar a_0\sim 10$ in comparison to the smaller deviations for $\bar a_0\sim 100$.

\begin{figure*}[t!]
\includegraphics[scale=0.45]{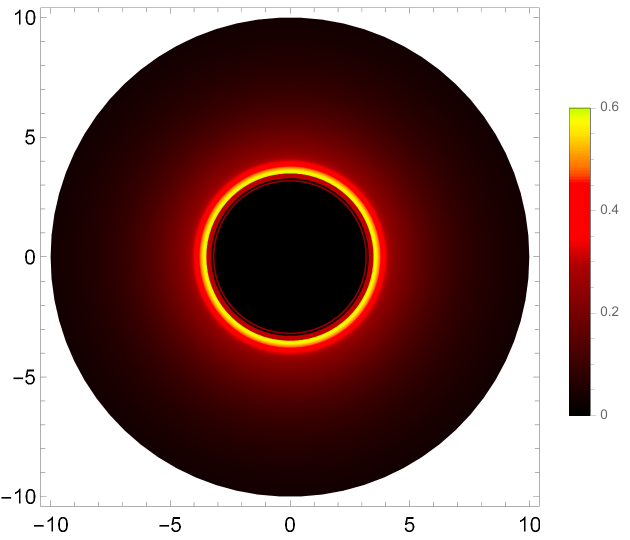}
\includegraphics[scale=0.45]{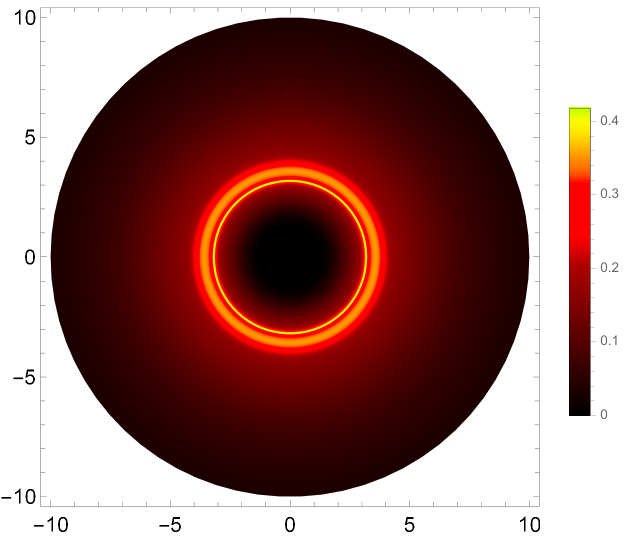}\qquad
\includegraphics[scale=0.75]{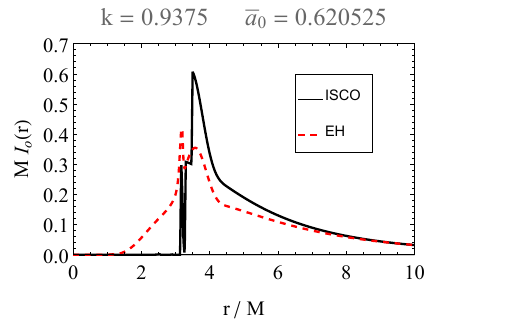}
\caption{Observed shadow images for an observation angle of $\theta=0^\circ$ with the ISCO disk model (left panel) and the EH disk model (center panel), and corresponding observed intensity profiles (right panel) for the DMHBH model in the ultra-compact configuration with $k=0.9375$ and $\bar a_0=0.620525$. The axes of the figures are given in units of $r/M$.}
\label{fig:disk_ultra}
\end{figure*}

Consider now the EH disk model, see Figs. \ref{fig:shadowsC} and \ref{fig:intensityC}. Unlike what happens for the ISCO disk model, one verifies that the observed intensity profiles are strongly dependent on both the values of $k$ and $\bar a_0$. Indeed, given that the emission radius of the EH model and the radius of the photon sphere decrease with an increase in $k$, one verifies that an increase in $k$ allows for the direct component to extend closer to the center of the configuration, thus altering immensely the size of the inner shadow, and also that the $n=1$ and $n=2$ contributions follow the same fate. Changes in the parameter $\bar a_0$ are particularly effective in altering the observational properties of this model for small values of $\bar a_0$ and large values of $k$. Indeed, one observes that the modifications induced by a variation of $\bar a_0=1$ to $\bar a_0=10$ are much more pronounced than further variations to $\bar a_0=100$. Finally, for this model one also observes that the $n=1$ and $n=2$ contributions are superimposed with the direct emission.

\subsection{High-compactness regime}

To finalize the analysis, consider one additional model in the high-compactness regime. This model is characterized by the parameters $k=0.9375$ and $\bar a_0=0.620525$, and it corresponds to a situation for which additional photons spheres and MSO are present in the space-time due to the high-compactness of the DM halo (see Figs. \ref{fig:ddpot} and \ref{fig:potential}). Indeed, for this model, one observes that timelike circular orbits are stable in the regions $0.284375<r/M<0.478125$ and $r/M>2.51313$. To restrict our analysis of the ISCO model to the region where orbits are stable, we have forced the intensity profile of the ISCO model to zero in the regions where orbits are unstable, i.e., regions complementary to those listed above. The observed images for this configuration with the ISCO and the EH disk models, as well as the corresponding observed intensity profiles, are given in Fig. \ref{fig:disk_ultra}. These results indicate that, due to the rise of additional MSOs and photon spheres induced by the DM halo, which causes an increase in the critical impact parameter $b_c$ (see Sec. \ref{sec:framework}), the sizes of the shadows for this configuration are similar to those of models for which the black hole dominates the mass ratio, see e.g. the results for $k=0.3$ in Fig. \ref{fig:shadowsISCO}, even though $M_{\rm BH}=M/16$ is much smaller in this case. Furthermore, in the ISCO model, even though there exists emission from the inner region $0.284375<r/M<0.478125$, these photons are strongly redshifted and consequently undetectable in the reference frame of the observer.

\section{Conclusion}\label{sec:conclusion}

In this work we have considered a recently introduced family of black hole configurations surrounded by a dark matter halo distribution with the objective of analyzing their observational properties in suitable astrophysical scenarios, namely, when orbited by isotropically emitting light sources, and  when surrounded by a (geometrically and optically thin) accretion disk. The background geometry is characterized by two parameters, $a_0$ and $M_{DM}$, associated to the halo's typical length scale and mass. The total - ADM - mass of the space-time is thus given by the sum of the black hole and dark matter halo masses. The geometry built this way has strong resemblances to the Schwarschild black hole in the key quantities for the generation of observables. Indeed, the event horizon location is exactly the same as for a Schwarzschild black hole of equal mass, while the corrections to the unstable critical null curve and the innermost stable circular orbit are strongly suppressed in the astrophysically relevant range. The critical impact parameter, though, is less suppressed.

We first considered the orbits of isotropically emitting sources. Our results indicate that observable differences between a bunch of background geometries covering a relevant range of the parameters space of $\{a_0,M_{BH}\}$  are particularly noticeable for large observation inclinations. Indeed, an increase in the mass of the DM halo induces a strong decrease in the light deflection and sizes of the secondary and photon ring tracks, which leads to a reduction of the centroid shifting effect caused by the secondary image. Furthermore, an increase in both the mass and the radius of the DM halo induces a decrease in the total luminosity of the observations which, although being a solely quantitative effect, is large enough to produce an observable imprint. These simulated astrometrical properties can be directly compared with the experimental observations of the GRAVITY instrument in order to impose constraints on both free parameters of the model.

We next turned our attention upon the optical appearance of our DMHBH solutions when illuminated by an optically and geometrically thin accretion disk emitting monochromatically in the reference frame of the disk and with an intensity given by suitable adaptations of the Standard Unbound profile peaking at the ISCO (GLM3 model) and the event horizon (GLM1/GLM2 models), which have been employed previously in the literature for the sake of matching the results of specific scenarios of GRMHD simulations in a simplified setting.  Furthermore, we used the recent observations of the EHT Collaboration on the calibrated size of the shadow's size of Sgr A$^*$ to choose three samples of such black hole geometries consistent with such observations. Our results point towards the strong resemblance of the features of the images of DMHBHs and the canonical Schwarzschild black hole even in those scenarios deviating the most each other. Indeed, photon rings suffer only minor modifications in their relative luminosities and locations, though moderated in their widths (up to a $\sim 25\%$ in the most extreme case). As for the size of the central brightness depression, there are tiny differences in the GLM3 model, given the fact that in such a case the outer edge of the shadow fills the critical impact parameter, which is different in the Schwarzschild and DMHBH models considered here, while in the GLM1/GLM2 models such an outer edge also manifest small but non-vanishing differences given by the lensed image of the event horizon in the different gravitational field of each background geometry. 

In order to dwell further into these differences and resemblances between images, we pushed further the space of parameters by considering nine additional configurations resulting from the combination of parameters characterizing the ratio between dark matter mass and ADM mass, $k$, and the halo length scale, $a_0$, combined with two additional SU-type models, peaking near the ISCO and at the event horizon, respectively. We verified that large variations in both parameters introduce significant changes in the images, with particular prominence for $k$, and which manifests in strong modifications to the locations of the photon rings and, consequently, on the size of the central brightness depression. A particularly interesting result regards configurations for which the DM halo presents a large compactness, where our results demonstrate that the size of the observed shadow increases when the DM halo develops a pair of photon rings and additional MSOs, thus indicating that smaller black holes immersed in DM halos can produce observables similar to those of larger black holes.

To conclude, DMHBHs, up to the astrophysical constraints, particularly those of the EHT Collaboration and the GRAVITY instrument, mimic quite closely the expected features of a Schwarzschild black hole, while also introducing slight possibly detectable modifications. From this point of view, it poses the opposed problem as other alternatives available in the literature (as horizonless compact objects) that modify too strongly the expected images, namely, that here we find a difficulty in having clear cut observational discriminators, unless the parameter space is pressed sufficiently upwards. A chance could be present via precise measurements of the features of the $n=1$ and $n=2$ photon rings in future upgrades of very long baseline interferometry, like those forecast in the next-generation Event Horizon Telescope \cite{Ayzenberg:2023hfw}.

\begin{acknowledgements}
CFBM thanks Fundação Amazônia de Amparo a Estudos
	e Pesquisas (FAPESPA), Conselho Nacional de Desenvolvimento Científico e Tecnológico (CNPq) and Coordenação de
	Aperfeiçoamento de Pessoal de Nível Superior (CAPES) -
	Finance Code 001, from Brazil, for partial financial support;
	JLR acknowledges the European Regional Development Fund and the programme Mobilitas Pluss for financial support through Project No. MOBJD647, and project
	No. 2021/43/P/ST2/02141 co-funded by the Polish National
	Science Centre and the European Union Framework Programme for Research and Innovation Horizon 2020 under
	the Marie Sklodowska-Curie grant agreement No. 94533.;
	DRG is supported by Grant PID2022-138607NB-I00, funded
	by MCIN/AEI/10.13039/501100011033 (FEDER, UE). 
\end{acknowledgements}

\end{document}